\documentstyle[namedreferences]{kapproc}

\font\gothic=eufm10 scaled \magstep1

\def\LL{\mathbin{\rm\vphantom{X}%       %right contraction
                        \hskip 1.8pt%
                        \vrule width 0.5pt%
                        \hskip -0.7pt%
                        \_\hskip -0.3em\_}
                        \hskip 0.8pt}

\def\JJ{\mathbin{\rm\vphantom{X}%      %left contraction
                       \hskip 1.1pt%
                       \hskip -0.5pt%
                       \_\hskip -0.3em\_%
                       \vrule width 0.5pt%
                       \hskip 1.7pt}}

\def\w{\wedge}
\def\bigw{\bigwedge}

  \let\R=\BR

  \let\C=\BC

\def\BH{{\rm\hskip 0.1pt%
                       I\hskip -2.15pt H}}

  \def\BZ{{\bf Z}}

\def\cB{{\cal B}}\def\cC{{\cal C}}
\def\cE{{\cal E}}\def\cF{{\cal F}}
\def\cG{{\cal G}}\def\cH{{\cal H}}
\def\cL{{\cal L}}
\def\cM{{\cal M}}
\def\cR{{\cal R}}

\let\g=\gamma
\let\al=\alpha          
          \let\om=\omega
    \let\fii=\varphi

  \def\ba{{\bf a}}
  \def\bb{{\bf b}}
\def\bC{{\bf C}}  
\def\bD{{\bf D}}

  \def\bi{{\bf i}}
\def\bJ{{\bf J}}

\def\bM{{\bf M}}

\def\bR{{\bf R}}  
  
\def\bT{{\bf T}}

\def\bX{{\bf X}}    \def\x{{\bf x}}
    \def\y{{\bf y}}

\def\cl{C\kern -0.3em\ell}     %Clifford algebra
\def\Cl{\cC \kern -0.1em \ell}   %Clifford bundle 
\def\Spin{{\bf Spin}}            \def\Pin{{\bf Pin}}
\def\DSpin{{\it \rlap{$\kern 0.7pt |\kern -0.7pt$}Spin}}
\def\Lip{{\bf\Gamma}}
\def\DLip{{\it \rlap{$\kern 0.7pt |\kern -0.7pt$}S\Gamma}}

         \def\SL{\mathop{\rm SL}}
\def\Or{\mathop{\rm O}}          \def\SO{\mathop{\rm SO}}

\def\End{\mathop{\rm End}}
\def\Aut{\mathop{\rm Aut}}
        
\def\Char{\mathbin{\rm char}\,}
\def\Ad{\mathop{\rm Ad}}

\def\inv#1{#1^{-1}}

\def\bil#1{\mathopen{<} #1\mathclose{>}}
\def\dgr#1{\langle #1 \rangle}

\def\bpartial{\hbox{\boldmath$\partial$}}

\def\ut#1{{\setbox0=\hbox{$#1$}\mathsurround=0pt
       \rlap{\raisebox{-0.8\dp0}{\raisebox{-0.8ex}
       {\kern -0.15ex\hbox{$\tiny\sim$}\kern 0.15ex}}}#1}}
\def\uti#1{{\setbox0=\hbox{$#1$}\mathsurround=0pt
       \rlap{\raisebox{-0.8\dp0}{\raisebox{-0.8ex}
       {\kern -0.3ex\hbox{$\tiny\sim$}\kern 0.3ex}}}#1}}

\def\ub#1{{\rlap{\raisebox{-0.6ex}
      {$ \hskip 0.02em \_ \hskip -0.02em$}}#1}}
\def\ubi#1{{\rlap{\raisebox{-0.8ex}
   {$ \hskip -0.1em \_  \hskip 0.1em $}}#1}}

\def\dis{\displaystyle}

\def\Ad{\mathop{\rm Ad}}
\def\Aut{\mathop{\rm Aut}}
\def\bgamma{\mbox{\boldmath$\gamma$}}
\def\bomega{\mbox{\boldmath$\omega$}}
\def\got#1{\hbox{\gothic#1}}
\def\BM{{\rm I\!M}}
\def\Lie{\mbox{\rm \pounds}}
\let\R=\BR
\let\C=\BC
\def\bP{{\bf P}}
\def\squarecdot{\mathbin{\rule[1pt]{1mm}{1mm}}}

\runningtitle{}
\runningauthor{}

\begin{opening}

\title{DIRAC-HESTENES SPINOR FIELDS IN RIEMANN-CARTAN SPACETIME$^*$}

\author{W. A. Rodrigues, Jr.}
\author{Q. A. G. de Souza}
\author{J. Vaz, Jr.} 

\institute{Departamento de Matem\'atica Aplicada\\
Universidade Estadual de Campinas\\
13081--970, Campinas, SP, Brazil\\
email: walrod@ime.unicamp.br}

\author{P. Lounesto}

\institute{Department of Mathematics\\
Helsinki University of Technology\\
SF-02150 Espoo, Finland}

\end{opening}

\begin{document}

\begin{abstract}
In this paper we study Dirac-Hestenes spinor fields~(DHSF) on a
four-di\-men\-sional Riemann-Cartan spacetime~(RCST). We prove that
these fields must be defined as certain equivalence classes of even
sections of the Clifford bundle (over the RCST), thereby being certain
particular sections of a new bundle named Spin-Clifford bundle~(SCB).
The conditions for the existence of the SCB are studied and are
shown to be equivalent to the famous Geroch's theorem concerning to
the existence of spinor structures in a Lorentzian spacetime.
We introduce also the covariant and algebraic Dirac spinor fields and
compare these with DHSF, showing that all the three kinds of spinor
fields contain the same mathematical and physical information. We
clarify also the notion of (Crumeyrolle's) amorphous spinors
(Dirac-K\"ahler spinor fields are of this type), showing that they
cannot be used to describe fermionic fields. 
We develop a rigorous theory for the covariant derivatives of
Clifford fields (sections of the Clifford bundle (CB)) and of
Dirac-Hestenes spinor fields. We show how to generalize the original
Dirac-Hestenes equation in Minkowski spacetime for the case of a
RCST. Our results are obtained from a variational principle
formulated through the multiform derivative approach to Lagrangian
field theory in the Clifford bundle.
\end{abstract}

\footnotetext{$^*$ accepted for publication, {\it Int. J. Theor. Physics},
1996}

\section{Introduction}

In the following we study the theory of
Dirac-Hestenes spinor fields~(DHSF) and the theory of their covariant
derivatives on a Riemann-Cartan spacetime~(RCST). We also show how to
generalize the 
so-called Dirac-Hestenes equation -- originally introduced
in (\citeauthor{REF-2}, \citeyear{REF-2};\citeyear{REF-3}) for the
formulation of Dirac theory of the 
electron using the spacetime algebra $\cl_{1,3}$ in Minkowski
spacetime -- 
for an arbitrary Riemann-Cartan spacetime. We use an approach based
on the multiform derivative formulation of Lagrangian field theory to
obtain the above results. They are important for the study of spinor
fields in gravitational theory and  are essential for an understanding
of the relationship between Maxwell and Dirac theories and quantum
mechanics (\citeauthor{REF-4}, \citeyear{REF-4}; \citeyear{REF-MD}). 

In order to achieve our goals we start clarifying many misconceptions
concerning the usual presentation of the theory of covariant, 
algebraic and Dirac-Hestenes spinors. Section~2 is dedicated to this
subject and we must say that it improves over other presentations 
-- e.g.,~\cite{REF-4,REF-5,REF-6,REF-7,REF-8,REF-9,REF-10,REF-11,REF-12} --
introducing a new and important 
fact, namely that all kind of spinors referred above must be defined
as  special equivalence classes  in appropriate Clifford algebras.
The hidden geometrical meaning of the covariant Dirac Spinor is
disclosed and the physical and geometrical meaning of the famous
Fierz identities \cite{REF-8,REF-9,REF-13,REF-14} becomes obvious.

In Section~3 we study the Clifford bundle of a Riemann-Cartan
spacetime \cite{REF-13a} and its irreducible module representations.
This permit us 
to define Dirac-Hestenes spinor fields~(DHSF) as certain equivalence
classes of even sections of the Clifford bundle. DHSF are then
naturally identified with sections of a new bundle which we call the
Spin-Clifford bundle.

We discuss also the concept of amorphous spinor fields (ASF) -- a name 
introduced by Crumeyrolle \shortcite{REF-15}. The so-called
Dirac-K\"ahler spinors \cite{REF-16} discussed by Graf \shortcite{REF-17}
and used in presentations of field theories in the
lattice \cite{REF-18,REF-19} are examples of ASF. We prove that they
cannot be used to describe fermion fields because they cannot be used
to properly formulate the Fierz identities.

In Section~4 we show how the Clifford and Spin-Clifford bundle
techniques permit us to give a simple presentation of the concept of
covariant derivative for Clifford fields, algebraic Dirac Spinor
Fields and for the DHSF. We show that our elegant
theory agrees with the standard one developed for the so-called
covariant Dirac spinor fields as developed, e.g.,
in (\citeauthor{REF-21}, \citeyear{REF-21}, \citeyear{REF-22};
\citeauthor{REF-23}, \citeyear{REF-23}). 

In Section~5 we introduce the concepts of Dirac and Spin-Dirac
operators acting respectively on sections of the Clifford and
Spin-Clifford bundles. We show how to use the Spin-Dirac operator
on the representatives of DHSF on the Clifford bundle.

In Section~6 we present the multiform derivative approach to
Lagrangian field theory and derive the Dirac-Hestenes equation on a
RCST \cite{REF-23}. We compare our results with
some others that appear in the literature for the covariant Dirac
Spinor field \cite{REF-24,REF-25} and also for Dirac-K\"ahler
fields \cite{REF-16,REF-17,REF-26}. 

Finally in Section~7 we present our conclusion.

\section{Covariant, Algebraic and Dirac-Hestenes Spinors}

\subsection{Some General Features about Clifford Algebras}

In this section we fix the notations to be used in this paper
and introduce the main ideas concerning the theory of Clifford
algebras necessary for the intelligibility of the paper. We
follow with minor modifications the conventions used
in~\cite{REF-8,REF-9}.

\subsubsection*{Formal Definition of the Clifford Algebra $\cl(V,Q)$}

Let $K$ be a field, $\Char K \neq 2$,\footnote{In our applications in
this paper, $K$ will be $\R$ or $\C$, respectively the real or complex
field. The quaternion ring will be denoted by $\BH$.}
$V$ a vector space of finite dimension $n$ over $K$, and 
$Q$ a nondegenerate quadratic form over $V$. Denote by
\begin{equation}
\x \cdot \y = {\textstyle\frac{1}{2}}( Q(\x+\y) - Q(\x) - Q(\y))
\end{equation}
the associated {\sl symmetric\/} bilinear form on $V$ and define
the {\sl left contraction\/} $\JJ : \bigw V \times \bigw V \to
\bigw V$ and the {\sl right contraction\/} $\LL : \bigw V \times
\bigw V \to \bigw V$ by the rules
\begin{enumerate}
\item $\x\JJ\y = \x \cdot \y$\\
      $\x\LL\y = \x \cdot \y$
\item $\x\JJ(u\w v) = (\x\JJ u) \w v + \hat u \w (\x\JJ v)$ \\
      $(u\w v) \LL \x = u \w (v\LL \x) + (u\LL\x)\w \hat v$
\item $(u\w v)\JJ w = u \JJ (v\JJ w)$ \\
      $u \LL (v \w w) = (u\LL v)\LL w$
\end{enumerate}
where $\x,\y\in V$, $u,v,w\in \bigw V$, and $\hat{\ }$ is the grade involution
in the algebra $\bigw V$. The notation
$\ba\cdot\bb$ will be used for contractions when it is clear from
the context which factor is the contractor and which factor is
being contracted. When just one of the factors is homogeneous, it is
understood to be the contractor. When both factors are homogeneous,
we agree that the one with lower degree is the contractor, so that
for $\ba\in\bigw^r V$ and $\bb\in\bigw^s V$, we have
$\ba\cdot\bb = \ba \JJ \bb$ if $r\leq s$ and 
$\ba\cdot\bb = \ba \LL \bb$ if $r\geq s$.

Define the (Clifford) product of $\x\in V$ and $u\in\bigw V$ by
\begin{equation}
\x u = \x \w u + \x \JJ u
\end{equation}
and extend this product by linearity and associativity to all of
$\bigw V$. This provides $\bigw V$ with a new product, and provided
with this new product $\bigw V$ becomes isomorphic to the {\sl
Clifford algebra\/} $\cl(V,Q)$.

We recall that $\bigw V = T(V)/I$ where $T(V)$ is the tensor
algebra of $V$ and $I\subset T(V)$ is the bilateral ideal
generated by the elements of the form $\x\otimes\x$, $\x\in V$. It
can also be shown that the Clifford algebra of $(V,Q)$ is
$\cl(V,Q)=T(V)/I_Q$, where $I_Q$ is the bilateral ideal generated by
the elements of the form $\x\otimes\x-Q(\x)$, $\x\in V$.
The Clifford algebra so constructed is an associative algebra with
unity. Since $K$ is a field, the space $V$ is naturally embedded in
$\cl(V,Q)$
\begin{eqnarray}
&&V \stackrel{i}{\hookrightarrow} T(V) \stackrel{j}{\rightarrow}
T(V)/I_Q = \cl(V,Q)\nonumber\\
&&I_Q = j \circ i \ \ {\rm and} \ \ V \equiv i_Q(V) \subset 
\cl(V,Q)
\end{eqnarray}
Let $\cl^{+}(V,Q)$ [resp., $\cl^-(V,Q)]$ be the $j$-image
of ${\displaystyle\oplus^\infty_{i=0}} T^{2i}(V)$ [resp.,
${\displaystyle\oplus^\infty_{i=0}} T^{2i+1}(V)]$ in
$\cl(V,Q)$. The elements of $\cl^+(V,Q)$ form a
sub-algebra of $\cl(V,Q)$ called the even sub-algebra of
$\cl(V,Q)$.

$\cl(V,Q)$ has the following property: If $A$ is an
associative $K$-algebra with unity then all linear mappings
$\rho:V \to A$ such that $(\rho(x))^2 = Q(x)$, $x \in V$, can be
extended in a unique way to an algebra homomorphism
$\rho: \cl(V,Q) \to A$. 

In $\cl(V,Q)$ there exist three linear mappings which are
quite natural. They are extensions of the mappings

\medbreak
\noindent{\bf Main involution:} an automorphism
$\hat{} :\cl(V,Q) \to \cl(V,Q)$, extension of
$\al: V \to T(V)/I_Q, \al(x)= -i_Q(x) = - x$, $\forall x \in V$. 

\medskip
\noindent{\bf Reversion:} an anti-automorphism
$\tilde{} : \cl(V,Q) \to \cl(V,Q)$, extension of
$^{t}: T^r(V) \to T^r(V)$; 
$T^r(V) \ni x = x_{i_1} \otimes \ldots \otimes x_{i_r} \mapsto x^t =
x_{i_r}	\otimes \ldots \otimes x_{i_1}$. 

\medskip
\noindent{\bf Conjugation:} $\bar{}: \cl(V,Q) \to \cl(V,Q)$,
defined by the composition of the main involution $\hat{}$
with the reversion $\tilde{}$, i.e., if $x \in \cl(V,Q)$ then
$\overline{x} = (\hat{x})^{\tilde{}} = (\tilde{x})^{\hat{}}$.

\medbreak
$\cl(V,Q)$ can be described through its generators, i.e., if
$\Sigma =\{E_i\}$  $(i=1,2, \ldots, n)$ is a $Q$-orthonormal basis of $V$, then
$\cl(V,Q)$ is generated by 1 and the $E_i$'s are subjected to the
conditions
\begin{eqnarray}
&&E_i E_i = Q(E_i)\nonumber\\
&&E_iE_j + E_jE_i = 0, \qquad 
i\neq j; \ i, j = 1,2, \ldots, n\nonumber \\
&&E_1 E_2 \cdots E_n \neq \pm 1. 
\end{eqnarray}

\subsubsection*{The Real Clifford Algebra $\cl_{p,q}$}

Let $\R^{p,q}$ be a real vector space of dimension $n=p+q$ endowed
with a nondegenerate metric $g: \R^{p,q} \times \R^{p,q} \to \R$.
Let $\Sigma = \{E_i\}$, $(i=1,2,\ldots,n)$ be an
orthonormal basis of $\R^{p,q}$,
\begin{equation}
g(E_i, E_j) = g_{ij} = g_{ji} = \left\{\begin{array}{rl}
+1, & i=j=1,2,\ldots p\\
-1, & i=j=p+1,\ldots,p+q=n\\
0,  & i \neq j
\end{array}\right.
\end{equation}

The Clifford algebra $\cl_{p,q} = \cl(\R^{p,q}, Q)$ is the Clifford
algebra over $\R$, generated by 1 and the $\{E_i\}$, $(i =1,2,\ldots,n)$
such that $E_{i}^{2} = Q(E_i) = g(E_i, E_i)$, $E_{i}E_{j} =
-E_{j}E_{i}$ ($i \neq j$), and \cite{REF-27} $E_1 E_2\ldots E_n \neq \pm
1$. \ $\cl_{p,q}$ is obviously of dimension 2$^n$ and as
a vector space it is the direct sum of vector spaces $\bigw^k \R^{p,q}$
of dimensions $\left(^n_k\right), 0 \leq k\leq n$.  The canonical
basis of $\bigw^k \R^{p,q}$ is given by the elements $e_A =
E_{\al_1} \ldots 
E_{\al_k}$, $1 \leq \al_1 <\ldots <\al_k \leq n$. The element $c_J = E_1
\ldots E_n \in \bigw^n \R^{p,q}$ commutes ($n$ odd) or anticommutes ($n$
even) with all vectors $E_1, \ldots,E_n \in \bigw^{1} \R^{p,q} \equiv
\R^{p,q}$. The center of $\cl_{p,q}$ is $\bigw^{0} \R^{p,q} \equiv
\R$ if $n$ is even and its is the direct sum $\bigw^{0} \R^{p,q} \oplus
\bigw^n \R^{p,q}$ if $n$ is odd.

All Clifford algebras are semi-simple. If $p + q =n$ is even,
$\cl_{p,q}$ is simple and if $p+q =n$ is odd we have the following
possibilities:
\begin{enumerate}
\item $\cl_{p,q}$ is simple $\leftrightarrow c^2_J = -1
\leftrightarrow p-q \neq 1$ (mod 4) $\leftrightarrow$ center of
$\cl_{p,q}$ is isomorphic to $\C$
\item $\cl_{p,q}$ is not simple (but is a direct sum of
two simple algebras) $\leftrightarrow c^2_J = +1 \leftrightarrow p-q
=1$ (mod 4) $\leftrightarrow$ center of $\cl_{p,q}$ is isomorphic to
$\R \oplus \R$.
\end{enumerate}
All these semi-simple algebras are direct sums of two simple
algebras.

If $A$ is an associative algebra on the field $K, K
\subseteq A$, and if $E$ is a vector space, a homomorphism $\rho$
from $A$ to $\End E$ ($\End E$ is the endomorphism algebra of $E$)
which maps the unit element of $A$ to
Id$_E$ is a called a {\it representation} of $A$ in $E$. The
dimension of $E$ is called the degree of the representation. The
addition in $E$ together with the mapping $A \times E \to E$,
$(a, x) \mapsto \rho(a) x$ turns $E$ in an $A$-module, the {\it
representation module}. 

Conversely, $A$ being an algebra over $K$ and $E$ being an $A$-module,
$E$ is a vector space over $K$ and if $a \in A$, the mapping
$\g: a \to \g_a$ with $\g_a(x) = ax$, $x \in E$, is a homomorphism
$A \to \End E$, and so it is a representation of $A$ in $E$.
The study of $A$ modules is then equivalent to the study of the
representations of $A$. A representation $\rho$ is {\it faithful\/}
if its kernel is zero, i.e., $\rho(a) x = 0$,
$\forall x \in E \Rightarrow a = 0$. The
kernel of $\rho$ is also known as the annihilator of its module.
$\rho$ is said to be {\it simple} or irreducible if the only invariant
subspaces of $\rho(a), \ \forall a \in A$, are $E$ and $\{0\}$. Then
the representation module is also simple, this meaning that it has no
proper  submodule. $\rho$ is said to be {\it semi-simple}, if it is
the direct sum of simple modules, and in this case $E$ is the direct
sum of subspaces which are globally invariant under $\rho(a), \forall
a \in A$. When no confusion arises $\rho(a) x$ will be denoted 
by $a \mathbin{\hbox{\small$\bullet$}} x$, $a \squarecdot x$ or $ax$.
Two $A$-modules $E$ and $E'$ (with the exterior multiplication being
denoted respectively by \hbox{\small$\bullet$} and $\squarecdot$) are
{\it isomorphic} if there exists a bijection $\fii: E \to E'$ such that,
\begin{eqnarray}
&\fii(x + y) = \fii(x) + \fii(y), \quad \forall x, y
\in E , \nonumber\\
&\fii(a \mathbin{\hbox{\small$\bullet$}} x) = a \squarecdot \fii(x),
\quad \forall a \in A,
\end{eqnarray}
and we say that representations $\rho$ and $\rho'$ of $A$ are
equivalent if their modules are isomorphic. This implies the
existence of a $K$-linear isomorphism $\fii:E \to E'$ such
that $\fii \circ \rho(a) = \rho'(a) \circ \fii,\ \forall a \in A \
{\rm or} \ \rho'(a) = \fii \circ \rho(a) \circ \fii^{-1}$. If
$\dim E = n$, then $\dim E'=n$. We shall need:

\proclaim Wedderburn Theorem. \cite{REF-28} If $A$ is
simple algebra then $A$ is equivalent to $F(m)$, where $F(m)$ is a
matrix algebra with entries in $F$, $F$ is a division algebra and $m$
and $F$ are unique (modulo isomorphisms).

\subsection{Minimal Left Ideals of $\cl_{p,q}$}

The minimal left (resp., right) ideals of a semi-simple algebra $A$
are of the type $A e$ (resp., $e A$), where $e$ is a primitive
idempotent of $A$, i.e., $e^2 =e$ and $e$ cannot be written as a sum
of two non zero annihilating (or orthogonal) idempotents, i.e, $e
\neq e_1 + e_2$, where $e_1 e_2 = e_2 e_1 = 0$, $e^2_1 = e_1$, $e^2_2 =
e_2$.

\proclaim Theorem. The maximum number of pairwise annihilating
idempotents in $F(m)$ is $m$.

The decomposition of $\cl_{p,q}$ into minimal ideals is then
characterized by a spectral set $\{ e_{pq,i}\}$ of idempotents of
$\cl_{p,q}$ satisfying
(i)~$\sum_{i} e_{pq,i} =1$;
(ii)~$e_{pq,i} e_{pq,j} = \delta_{ij} e_{pq,i}$;
(iii)~rank of $e_{pq,i}$ is minimal $\neq 0$, i.e., $e_{pq,i}$ is
primitive $(i=1,2,\ldots,m)$

By rank of $e_{pq,i}$ we mean the rank of the $\bigw \R^{p+q}$-morphism
$e_{pq,i}: \psi \mapsto \psi e_{pq,i}$ and $\bigw \R^{p,q} =
\oplus_{k=0}^{n}  
\bigw^k(\R^{p,q})$ is the exterior algebra of $\R^{p,q}$. Then
$\cl_{p,q} = \sum_{i} I^i_{p,q}$, $I^i_{p,q} = \cl_{p,q} e_{pq,i}$ and
$\psi \in I^i_{p,q}$ is such that $\psi e_{pq,i} = \psi$.
Conversely any element $\psi \in I^i_{p,q}$ can be characterized by
an idempotent $e_{pq,i}$ of minimal rank $\neq 0$ with $\psi e_{pq,i}
=\psi$. We have the following

\proclaim Theorem. \cite{REF-29} A minimal left ideal
of $\cl_{p,q}$ is 
of the type $I_{p,q} = \cl_{p,q} e_{pq}$ where $e_{pq} =
\frac{1}{2}(1+ e_{\al_1})\ldots \frac{1}{2}
(1 + e_{\al_k})$ is a primitive idempotent of $\cl_{p,q}$ and
are $e_{\al_1},\ldots, e_{\al_k}$ commuting elements of
the canonical basis of $\cl_{p,q}$ such that $(e_{\al_i})^2 =1$,
$(i=1,2,\ldots,k)$ that generate a group of order $2^k$,
$k=q - r_{q-p}$ and $r_i$ are the Radon-Hurwitz numbers,
defined by the recurrence formula $r_{i+8} = r_i + 4$ and
\begin{center}
\begin{tabular}{rrrrrrrrrr}
$i$   & $\;$ & 0 & 1 & 2 & 3 & 4 & 5 & 6 & 7 \\ \hline
$r_i$ & $\;$ & 0 & 1 & 2 & 2 & 3 & 3 & 3 & 3
\end{tabular}
\end{center}

\smallbreak
If we have a linear mapping $L_a: \cl_{p,q} \to \cl_{p,q}$, 
$L_a(x) = ax$, $x \in \cl_{p,q}$, $a \in \cl_{p,q}$, then since
$I_{p,q}$ is invariant under left multiplication with arbitrary
elements of $\cl_{p,q}$ we can consider
$L_a|_{I_{p,q}}: I_{p,q} \to I_{p,q}$ and taking into account
Wedderburn theorem we have

\proclaim Theorem. If $p+q =n$ is even or odd with
$p-q \neq 1$ (mod 4) then 
$$
\cl_{p,q} \simeq \ {\rm End}_F(I_{p,q}) \simeq F(m)
$$
where $F=\R$ or $\C$ or $\BH$, $\End_F(I_{p,q})$ is the algebra of linear
transformations in $I_{p,q}$ over the field $F, m =$dim$_F(I_{p,q})$
and $F \simeq e F(m) e$, $e$ being the representation of $e_{pq}$ in
$F(m)$.

If $p+q=n$ is odd, with $p-q=1$ (mod 4) then
$$
\cl_{p,q} = \ {\rm End}_F(I_{p,q}) \simeq F(m) \oplus F(m)
$$
and $m = \dim_F(I_{p,q})$ and $e_{pq} \cl_{p,q} e_{pq} \simeq \R \oplus
\R$ or $\BH \oplus \BH$. 

\medbreak
Observe that $F$ is the set
$$
F = \{T \in \ {\rm End}_F (I_{p,q}), TL_a = L_a T, \,  \ \forall \ a \in
\cl_{p,q}\}
$$

\proclaim Periodicity Theorem. \cite{REF-28}
For $n = p + q \geq 0$ there exist the following isomorphisms
\begin{eqnarray}
\cl_{n+8,0} \simeq \cl_{n,0} \otimes \cl_{8,0} \qquad \cl_{0,n+8} \simeq
\cl_{0,n} \otimes \cl_{0,8}\\
\cl_{p+8,q} \simeq \cl_{p,q} \otimes \cl_{8,0} \qquad \cl_{p,q+8} \simeq
\cl_{p,q} \otimes \cl_{0,8}\nonumber
\end{eqnarray}

We can find, e.g., in~\cite{REF-28,REF-5,REF-6} tables giving the
representations of all algebras $\cl_{p,q}$ as matrix algebras. For
what follows we need
\begin{eqnarray}
\mbox{complex numbers}  &&   \cl_{0,1} \simeq \ \C\nonumber\\
\mbox{quarternions}  &&  \cl_{0,2} \simeq \ \BH\nonumber\\
\mbox{Pauli algebra} &&  \cl_{3,0} \simeq \ M_2(\C)\nonumber\\
\mbox{spacetime algebra}  &&  \cl_{1,3}  \simeq \ M_2(\BH)\nonumber\\
\mbox{Majorana algebra}  &&  \cl_{3,1}  \simeq \ M_4(\R)\\
\mbox{Dirac algebra}  && \cl_{4,1} \ \simeq \ M_4(\C)\nonumber
\end{eqnarray}

\noindent We also need the following 

\proclaim Proposition. $\cl^+_{p,q} = \cl_{q,p-1}$, 
for $p > 1$ and  $\cl^+_{p,q} = \cl_{p,q-1}$ for $q > 1$.

From the above proposition we get the following particular results
that we shall need later
\begin{equation}
\cl^+_{1,3} \simeq \cl^+_{3,1} = \cl_{3,0} \qquad  \cl^+_{4,1} \simeq
\cl_{1,3} , 
\end{equation}
\begin{equation}
\cl_{4,1} \simeq \C \otimes \cl_{3,1} \qquad  \cl_{4,1} \simeq \C
\otimes \cl_{1,3}, 
\end{equation}
which means that the Dirac algebra is the complexification of both the
spacetime or the Majorana algebras.

\subsubsection*{Right Linear Structure for $I_{p,q}$}

We can give to the ideal
$I_{p,q} = \cl_{p,q}e$ (resp. $I_{pq} = e \cl_{pq})$ a right
(resp. left) linear structure over the field $F( \cl_{p,q} \simeq
F(m)$ or $\cl_{p,q} \simeq F(m) \oplus F(m))$. A right linear
structure, e.g, consists of an additive group (which is $I_{p,q}$) and
the mapping
$$
I \times F \to  I; \quad (\psi, T) \mapsto \psi T
$$
such that the usual axioms of a linear vector space structure are
valid, e.g., we have\footnote{For $\cl_{3,0}$,
$I = \cl_{3,0} \frac{1}{2} (1+\sigma_3)$ is a minimal left ideal.
In this case it is also possible to give a left linear structure
for this ideal. See~\cite{REF-4,REF-5} }
$(\psi T)T' = \psi(TT') $. 

From the above discussion it is clear that the minimal (left or right)
ideals of $\cl_{p,q}$ are representation modules of $\cl_{p,q}$. In
order to investigate the equivalence of these representations we must
introduce some groups that are subsets of $\cl_{p,q}$. As we shall
see, this is the key for the definition of algebraic and
Dirac-Hestenes spinors.

\subsection{The Groups: $\cl^\star_{p,q}$, Clifford, Pinor and
Spinor}

The set of the invertible elements of
$\cl_{p,q}$ constitutes a non-abelian group which we denote by
$\cl^\star_{p,q}$. It acts naturally on 
$\cl_{p,q}$ as an algebra homomorphism through its adjoint
representation
\begin{equation}
{\rm Ad}: \cl^\star_{p,q} \to \Aut(\cl_{p,q}); \ u \mapsto
{\rm Ad}_{u}, \ {\rm with} \ {\rm Ad}_{u}(x) = uxu^{-1}  .
\end{equation}

\noindent The Clifford-Lipschitz group is the set
\begin{equation}
\Lip_{p,q} = \{u \in \cl^\star_{p,q} \, | \,\,\, \forall x \in
\R^{p,q}, u x \hat{u}^{-1} \in \R^{p,q}\} . 
\end{equation}

\noindent The set $\Lip^+_{p,q} = \Lip_{p,q} \cap
\cl^+_{p,q}$ is called special Clifford-Lipschitz group. 

\smallskip
Let $N:\cl_{p,q} \to \cl_{p,q}$, $N(x) = \dgr{\tilde x x}_{0}$
($\dgr{\;}_0$ means the scalar part of the Clifford number).
We define further: 

\medbreak
\noindent The {\it Pinor group} $\Pin(p,q)$ 
is the subgroup of $\Lip_{p,q}$ such that
\begin{equation}
\Pin (p,q) = \{ u \in \Lip_{p,q} | N(u) = \pm 1\} . 
\end{equation} 

\noindent The {\it Spin group\/} $\Spin (p,q)$ is the set
\begin{equation}
\Spin(p,q) = \{ u \in \Lip^+_{p,q} | N(u) = \pm 1\}  . 
\end{equation}

\noindent The $\Spin_+(p,q)$ group is the set
\begin{equation}
\Spin_+(p,q) = \{ u \in \Lip^+_{p,q} | N(u) = + 1\}  . 
\end{equation}

\proclaim Theorem. ${\rm Ad}_{|\Pin(p,q)}: \Pin(p,q)
\to \Or(p,q)$ is 
onto with kernel $\BZ_2$. ${\rm Ad}_{|\Spin(p,q)}:
\Spin(p,q) \to \SO(p,q)$ is onto with kernel $\BZ_2$. 

$\Or(p,q)$ is the pseudo-orthogonal group of the vector space
$\R^{p,q}$,  $\SO(p,q)$ is the special pseudo-orthogonal group of
$\R^{p,q}$. We also denote by $\SO_+(p,q)$ the connected component
of $\SO(p,q)$. $\Spin_+(p,q)$ is connected for all pairs $(p,q)$ with
the exception of $\Spin_+(1,0) \simeq \Spin_+(0,1) \simeq \{ \pm 1 \}$
and $\Spin_+(1,1)$.  We have,
$$
\Or(p,q) = \frac{\Pin(p,q)}{\BZ_2}  \quad
\SO(p,q) = \frac{\Spin(p,q)}{\BZ_2} \quad
\SO_+(p,q)= \frac{\Spin_+(p,q)}{\BZ_2}.
$$
In the following the group homomorphism between $\Spin_+(p,q)$ and 
$\SO_+(p,q)$ will be denoted
\begin{equation}
\cH: \Spin_+ (p,q) \to \SO_+(p,q)  .
\end{equation}

\noindent We also need the important result:

\proclaim Theorem. \cite{REF-28} For $p+q \leq 5$,
$\Spin_+(p,q) = \{ u \in \cl^+_{p,q} | u \tilde u = 1\}$.

\subsubsection*{Lie Algebra of $\Spin_{+}(1,3)$}

It can be shown that for each $u\in \Spin_{+}(1,3)$ it holds
\begin{equation} \label{2.17}
u = \pm e^F, \qquad F\in \bigw^2\R^{1,3}\subset \cl_{1,3}
\end{equation}
and $F$ can be chosen in such a way to have a positive sign in
Eq.~\ref{2.17}, except in the particular case $F^2=0$ when
$u=-e^F$. From Eq.~\ref{2.17} it follows immediately that the Lie
algebra of $\Spin_+(1,3)$ is generated by the bivectors
$F\in\bigw^2\R^{1,3}\subset \cl_{1,3}$ through the commutator
product.

\subsection{Geometrical and Algebraic Equivalence of the
Representation Modules $I_{p,q}$ of Simple Clifford Algebras
$\cl_{p,q}$}

Recall that $\cl_{p,q}$ is a ring. We already said that the minimal
lateral ideals of $\cl_{p,q}$ are of the form
$I_{p,q} = \cl_{p,q} e_{pq}$
(or $e_{pq} \cl_{p,q}$) where $e_{pq}$ is a primitive idempotent.
Obviously the minimal lateral ideals are modules over  the ring
$\cl_{p,q}$, they are representation modules. According to the
discussion of Section~2.1, given two ideals $I_{p,q} = \cl_{p,q}
e_{pq}$ and $I'_{p,q} = \cl_{p,q} e'_{pq}$ they are by definition
isomorphic if there exists a bijection $\fii: I_{p,q} \to
I'_{p,q}$ such that,
\begin{equation}
\fii(\psi_1 + \psi_2) = \fii(\psi_1) + \fii(\psi_2) \ ; \ \fii(a
\psi) = a \fii(\psi) \ , \ \forall a \in \cl_{p,q}, \forall \psi_1,
\psi_2 \in I_{p,q}
\end{equation} 	   

Recalling the Noether-Skolem theorem, which says that all
automorphism of a simple algebra are inner automorphism, we have:

\proclaim Theorem. When $\cl_{p,q}$ is simple, its
automorphisms are given by inner automorphisms $x\mapsto
ux\inv u$, $x\in\cl_{p,q}$, $u\in\cl^\star_{p,q}$.

\noindent We also have:

\proclaim Proposition. When $\cl_{p,q}$ is simple, all its
finite-dimensional irreducible representations are equivalent
(i.e., isomorphic) under inner automorphisms.

\noindent We quote also the

\proclaim Theorem. \cite{REF-15} $I_{p,q}$ and $I'_{p,q}$ are
isomorphic if and only if $I'_{p,q}=I_{p,q}X$ for non-zero $X\in
I'_{p,q}$.

We are thus lead to the following definitions:
\begin{enumerate}
\item The ideals $I_{p,q} = \cl_{p,q} e_{pq}$ and
$I'_{p,q} = \cl_{p,q} e'_{pq}$ are said to be {\it geometrically
equivalent\/} if, for some $u\in \Lip_{p,q}$,
\begin{equation} \label{2.19}
e'_{pq} =  u e_{pq} u^{-1}. 
\end{equation}
\item $I_{p,q}$ and $I'_{p,q}$ are said to be {\it
algebraically equivalent\/} if
\begin{equation}
e'_{pq} = u e_{pq} u^{-1}  , 
\end{equation}
for some $u \in \cl^\star_{p,q}$, but $u \not \in \Lip_{p,q}$. 
\end{enumerate}

It is now time to specialize the above results for $\cl_{1,3} \simeq
M_2(\BH)$ and to find a relationship between the Dirac algebra
$\cl_{4,1} \simeq M_4(\C)$ and $\cl_{1,3}$ and their respective
minimal ideals.

Let $\Sigma_0 = \{E_0,E_1,E_2,E_3\}$ be an orthogonal basis of
$\R^{1,3} \subset \cl_{1,3}$, $E_\mu E_\nu + E_\nu E_\mu = 2 \eta_{\mu\nu} 
$, $\eta_{\mu\nu} = {\rm diag}(+1,-1,-1,-1)$. Then, the elements
\begin{equation}
e= {\textstyle\frac{1}{2}} (1 + E_0) \qquad 
e' = {\textstyle\frac{1}{2}} (1 + E_3 E_0) \qquad 
e'' = {\textstyle\frac{1}{2}} (1 + E_1 E_2 E_3)  , 
\end{equation}
are easily verified to be primitive idempotents of $\cl_{1,3}$. The
minimal left ideals, $I = \cl_{1,3} e$, $I' = \cl_{1,3} e'$, $I'' =
\cl_{1,3} e''$ are right two dimensional linear spaces over the
quaternion field (e.g., $\BH e = e\BH = e \cl_{1,3} e$). According to
the definition (ii) above these ideals are algebraically equivalent.
For example, $e' = u e u^{-1}$, with $u = (1 + E_3) \not\in
\Lip_{1,3}$

The elements $\Phi \in \cl_{1,3} {\textstyle\frac{1}{2}}(1 + E_0)$
will be called {\it mother\/} spinors (\citeauthor{REF-9}, 
\citeyear{REF-9}; \citeyear{REF-10}). We can
show \cite{REF-5} that each $\Phi$ can be written
\begin{equation} \label{2.22}
\Phi = \psi_1 e + \psi_2 E_3E_1 e + \psi_3 E_3 E_0 e 
+ \psi_4 E_1 E_0 e = \sum_{i} \psi_i s_i  , 
\end{equation}
\begin{equation}
s_1 = e, \ \ s_2 = E_3 E_1e, \ \ s_3 = E_3E_0 e, \ \ s_4 = E_1 E_0 e
\end{equation}
and where the $\psi_i$ are formally complex numbers, i.e., each $\psi_i
= (a_i + b_iE_2 E_1)$ with $a_i ,  b_i \in \R$.

We recall that $\Pin(1,3)/\BZ_2 \simeq \Or(1,3)$,
$\Spin(1,3)/\BZ_2 \simeq \SO(1,3)$, $\Spin_+(1,3)/\BZ_2
\simeq \SO_+(1,3)$, 
$\Spin_+(1,3) \simeq \SL(2, \C)$ the universal covering group of
$\Lie^\uparrow_+\equiv \SO_+(1,3)$, the restrict 
Lorentz group. 

In order to determine the relation between
$\cl_{4,1}$ and $\cl_{1,3}$ we proceed as follows: let $\{F_0, F_1,
F_2, F_3, F_4\}$ be an orthogonal basis of $\cl_{4,1}$ with $-F^2_0 =
F^2_1=F^2_2 =F^2_3 =F^2_4=1$, $F_AF_B = -F_B F_A$ ($A \neq B$; $A,B =
0,1,2,3,4$).  Define the pseudo-scalar
\begin{equation} \label{2.24}
\bi = F_0 F_1 F_2 F_3 F_4 \qquad
\bi^2 = -1  \qquad
\bi F_A = F_A \bi \qquad A = 0,1,2,3,4
\end{equation}
Define
\begin{equation} \label{2.25}
\cE_\mu = F_\mu F_4
\end{equation}
We can immediately verify that $\cE_\mu\cE_\nu + \cE_\nu
\cE_\mu = 2 \eta_{\mu\nu}$. Taking into account that $\cl_{1,3}
\simeq \cl^+_{4,1}$ we can explicitly exhibit here this isomorphism
by considering the map $g: \cl_{1,3} \to \cl^+_{4,1}$
generated by the linear extension of the map $g^{\#}:\R^{1,3}
\to \cl^+_{4,1}, g^{\#}(E_\mu) = \cE_\mu = F_\mu F_4$,
where $E_\mu$, $(\mu = 0,1,2,3)$ is an orthogonal basis of $\R^{1,3}$.
Also $g(1_{\cl_{1,3}}) = 1_{\cl^+_{4,1}}$,
where $1_{\cl_{1,3}}$ and $ 1_{\cl^+_{4,1}}$ are the identity elements
in $\cl_{1,3}$ and $\cl^+_{4,1}$.
Now consider the primitive idempotent of
$ \cl_{1,3} \simeq \cl^+_{4,1}$,
\begin{equation}
e_{41} = g(e) = {\textstyle\frac{1}{2}} (1 + \cE_0)
\end{equation}
and the minimal left ideal $I^+_{4,1} = \cl^+_{4,1} e_{4,1}$. The
elements $Z_{{\scriptscriptstyle \Sigma_{0}}} \in I^+_{4,1}$ can be
written in an analogous way to
$\Phi \in \cl_{1,3} \frac{1}{2} (1 + E_0)$ (Eq.~\ref{2.22}), i.e.,
\begin{equation} \label{2.27}
Z_{{\scriptscriptstyle \Sigma_{0}}} = \Sigma \ z_i \bar{s}_i
\end{equation}
where
\begin{equation}
\bar{s}_1 = e_{41}, \ \ \bar{s}_2 = -\cE_1 \cE_3  e_{41}, \ \
\bar{s}_3 = \cE_3 \cE_0 e_{41}, \ \ \bar{s}_4 = \cE_1 \cE_0
e_{41}, 
\end{equation}
and
$$
z_i = a_i + \cE_2 \cE_1 b_i,
$$
are formally complex numbers, $a_i,b_i\in\R$. 

Consider now the element
$f_{{\scriptscriptstyle \Sigma_{0}}} \in \cl_{4,1}$,
\begin{eqnarray}
f_{{\scriptscriptstyle \Sigma_{0}}} &=&
e_{41} {\textstyle\frac{1}{2}} (1 + \bi \cE_1
\cE_2)\nonumber\\
&=& {\textstyle\frac{1}{2}} (1 + \cE_0)
{\textstyle\frac{1}{2}} (1 + \bi \cE_1 \cE_2),  \label{2.29}
\end{eqnarray}
with $\bi$ given by Eq.~\ref{2.24}.

Since $f_{{\scriptscriptstyle \Sigma_{0}}} \cl_{4,1}
f_{{\scriptscriptstyle \Sigma_{0}}} = \C f_{{\scriptscriptstyle
\Sigma_{0}}}  =
f_{{\scriptscriptstyle \Sigma_{0}}} \C$ it follows that
$f_{{\scriptscriptstyle \Sigma_{0}}}$ is a primitive 
idempotent of $\cl_{4,1}$. We can easily show that each
$\Phi_{{\scriptscriptstyle \Sigma_{0}}} \in I_{{\scriptscriptstyle
\Sigma_{0}}} = \cl_{4,1} f_{{\scriptscriptstyle \Sigma_{0}}}$ can 
be written
$$
\Psi_{{\scriptscriptstyle \Sigma_{0}}} = \sum_{i} \psi_i f_i , \ \
\psi_i \in \C 
$$
\begin{equation} \label{2.30}
f_1 = f_{{\scriptscriptstyle \Sigma_{0}}} \ , \ f_2 = - \cE_1 \cE_3
f_{{\scriptscriptstyle \Sigma_{0}}} \ , 
\ f_3 = \cE_3 \cE_0 f_{{\scriptscriptstyle \Sigma_{0}}},
f_4 = \cE_1 \cE_0
f_{{\scriptscriptstyle \Sigma_{0}}}
\end{equation}
with the methods described in~\cite{REF-4,REF-5} we find the
following representation in $M_4(\C)$ for the generators
$\cE_\mu$ of $\cl^+_{4,1} \simeq \cl_{1,3}$  
\begin{equation} \label{2.31}
\cE_0 \mapsto \ub\g_0 = \left(\begin{array}{ll}
1_2 & \ \ \ 0\\
0 & -1_2\end{array}\right)  \ \leftrightarrow \ \cE_i \mapsto
\ub\g_i = 
\left(\begin{array}{ll}
0 & -\sigma_i\\
\sigma_i & \ \ 0\end{array}\right) 
\end{equation}
where $1_2$ is the unit $2 \times 2$ matrix and $\sigma_i$, $(i
= 1,2,3)$ are the standard Pauli matrices. We immediately recognize the
$\ub\g$-matrices in Eq.~\ref{2.31} as the standard ones appearing,
e.g., in~\cite{REF-30}.

The matrix representation of $\Psi_{{\scriptscriptstyle \Sigma_{0}}}
\in I_{{\scriptscriptstyle \Sigma_{0}}}$ will 
be denoted by the same letter without the index, i.e.,
$\Psi_{{\scriptscriptstyle \Sigma_{0}}} \mapsto \Psi \in M_4 (\C) f$,
where
\begin{equation} \label{2.34}
f = {\textstyle\frac{1}{2}} (1 + \ub\g_0)
{\textstyle\frac{1}{2}} (1 + i \ub\g_1 \ub\g_2) \qquad 
i = \sqrt{-1}. 
\end{equation}
We have
\begin{equation}
\Psi = \left(\begin{array}{llll}
\psi_1 & 0 & 0 & 0\\
\psi_2 & 0 & 0 & 0\\
\psi_3 & 0 & 0 & 0\\
\psi_4 & 0 & 0 & 0
\end{array}\right) \qquad \psi_i \in \C  . 
\end{equation}
Eqs.(~\ref{2.22}, \ref{2.27}, \ref{2.30}) are enough to prove
that there are bijections between the elements of the ideals
$\cl_{1,3} \frac{1}{2} (1 + E_0)$,
$\cl^+_{4,1} \frac{1}{2} (1 + \cE_0)$ and
$\cl_{4,1} \frac{1}{2} ( 1 + \cE_0) \frac{1}{2} (1 + \bi\cE_1 \cE_2)$.

We can easily find that the following relation
exists between $\Psi_{{\scriptscriptstyle \Sigma_{0}}} \in \cl_{4,1}
f_{{\scriptscriptstyle \Sigma_{0}}}$ and 
$Z_{{\scriptscriptstyle \Sigma_{0}}} \in \cl_{4,1}^+
\frac{1}{2} (1 + \cE_0)$,  
\begin{equation}
\Psi_{{\scriptscriptstyle \Sigma_{0}}} = Z_{{\scriptscriptstyle
\Sigma_{0}}} {\textstyle\frac{1}{2}} (1 + \bi\cE_1 \cE_2) . 
\end{equation}
Decomposing $Z_{{\scriptscriptstyle \Sigma_{0}}}$ into even and odd
parts relative to the 
$\BZ_2$-graduation of $\cl^+_{4,1}\simeq \cl_{1,3},
Z_{{\scriptscriptstyle \Sigma_{0}}} = 
Z^+_{{\scriptscriptstyle \Sigma_{0}}} + Z^{-}_{{\scriptscriptstyle
\Sigma_{0}}}$ we obtain \ $Z^{+}_{\scriptscriptstyle \Sigma_{0}} 
= Z^{-}_{\scriptscriptstyle \Sigma_{0}} \cE_0$ which clearly shows
that all information 
of $Z_{\scriptscriptstyle \Sigma_{0}}$ is contained in
$Z^{+}_{\scriptscriptstyle \Sigma_{0}}$. Then, 
\begin{equation} \label{2.37}
\Psi_{\scriptscriptstyle \Sigma_{0}} = Z^+_{\scriptscriptstyle
\Sigma_{0}} {\textstyle\frac{1}{2}}(1 + \cE_0) {\textstyle\frac{1}{2}} 
(1+\bi\cE_1 \cE_2). 
\end{equation}

Now, if we take into account \cite{REF-5} that $\cl^{++}_{4,1}
\frac{1}{2} (1 + \cE_0) = \cl^+_{4,1} \frac{1}{2} (1 + \cE_0)$
where the symbol $\cl^{++}_{4,1}$ means
$\cl^{++}_{4,1} \simeq \cl^+_{1,3} \simeq
\cl_{3,0}$ we see that each
$Z_{\scriptscriptstyle \Sigma_{0}} \in \cl^+_{4,1} 
\frac{1}{2} (1 + \cE_0)$ can be written
\begin{equation} \label{2.38}
Z^{-}_{\scriptscriptstyle \Sigma_{0}} = \psi_{\scriptscriptstyle
\Sigma_{0}} {\textstyle\frac{1}{2}} ( 1 + \cE_0) 
\qquad \psi_{\scriptscriptstyle \Sigma_{0}} \in (\cl^+_{4,1})^+ \simeq
\cl^+_{1,3}. 
\end{equation}
Then putting $Z^{+}_{\scriptscriptstyle \Sigma_{0}} =
\psi_{\scriptscriptstyle \Sigma_{0}}/2$, Eq.~\ref{2.37} can be written
\begin{eqnarray}
\Psi_{\scriptscriptstyle \Sigma_{0}}&=& \psi 
{\textstyle\frac{1}{2}} (1 + \cE_0)
{\textstyle\frac{1}{2}} (1 + \bi \cE_1\cE_2)\nonumber\\
&=& Z_{\scriptscriptstyle \Sigma_{0}}
{\textstyle\frac{1}{2}} (1 + \bi \cE_1\cE_2) . \label{2.39}
\end{eqnarray}

The matrix representations of $Z_{\scriptscriptstyle \Sigma_{0}}$ and
$\psi_{\scriptscriptstyle \Sigma_{0}}$ in 
$M_4(\C)$ (denoted by the same letter without index) in the
spinorial basis given by Eq.~\ref{2.30} are
\begin{equation}
\Psi = \left(\begin{array}{llll}
\psi_1 & -\psi^*_2 & \psi_3 & \ \psi^*_4\\
\psi_2 & \ \ \psi^*_1 & \psi_4 & -\psi^*_3\\
\psi_3 & \ \ \psi^*_4 & \psi_1 & -\psi^*_2\\
\psi_4 & -\psi^*_3 & \psi_2 & \ \psi^*_1
\end{array}\right) \ , \ \  Z = \left(\begin{array}{llll}
\psi_1 & -\psi^*_2 & 0 & 0\\
\psi_2 & \ \ \psi^*_1 & 0 & 0\\
\psi_3 & \ \ \psi^*_4 & 0 & 0\\
\psi_4 & -\psi^*_3 & 0 & 0
\end{array}\right)  . 
\end{equation}

\subsection{Algebraic Spinors for $\R^{p, q}$}

Let ${\cal B}_{\scriptscriptstyle \Sigma}=\{\Sigma_0, \dot{\Sigma},
\ddot{\Sigma}, \ldots \}$ be the 
set of all ordered orthonormal basis for $\R^{p, q}$, i.e., each
$\Sigma\in {\cal B}_{\scriptscriptstyle \Sigma}$ is the set
$\Sigma=\{E_1, \ldots, E_p, 
E_{p+1}, \ldots, 
E_{p+q}\}$, $E^2_1= \ldots=E^2_p=1$, $E^2_{p+1}=\ldots=E^2_{p+q}=-1$,
$E_r E_s=-E_s E_r$, ($r\neq s; \ r, s=1,2, \ldots, p+q=n$).
Any two basis, 
say, $\Sigma_0, \dot\Sigma\in {\cal B}_{\scriptscriptstyle \Sigma}$ are
related by an element of the 
group $\Spin_+ (p, q) \subset \Gamma_{p q}$. We write,
\begin{equation}
\dot{\Sigma}=u \Sigma_0 u^{-1}, \, \, \, u \in \Spin_+(p, q) .
\end{equation}
A primitive idempotent determined in a given basis $\Sigma\in {\cal
B}_{\scriptscriptstyle \Sigma}$ 
will be denoted $e_{\scriptscriptstyle \Sigma}$. Then, the idempotents
$e_{\scriptscriptstyle \Sigma_{0}}, 
e_{\dot{\Sigma}}, e_{\ddot{\Sigma}}$, etc., such that, e.g.,
\begin{equation}
e_{\dot{\Sigma}}=ue_{\scriptscriptstyle \Sigma_{0}}u^{-1}, \, \, \,
u\in \Spin_+(p, q) ,  
\end{equation}
define ideals $I_{\scriptscriptstyle \Sigma_{0}}, I_{\dot{\Sigma}},
I_{\ddot{\Sigma}}$, 
etc., 
that are geometrically equivalent according to the definition given
by Eq.~\ref{2.19}. We have, 
\begin{equation} \label{2.43}
I_{\dot{\Sigma}}=u I_{\scriptscriptstyle \Sigma_{0}}u^{-1} \qquad 
u\in \Spin_+(p, q) 
\end{equation}
but since $uI_{\scriptscriptstyle \Sigma_{0}} \equiv
I_{\scriptscriptstyle \Sigma_{0}}$, Eq.~\ref{2.43} can also be 
written
\begin{equation} \label{2.44}
I_{\dot{\Sigma}}=I_{\scriptscriptstyle \Sigma_{0}} u^{-1}. 
\end{equation}
Eq.~\ref{2.44} defines a new correspondence for the elements of the
ideals, $I_{\scriptscriptstyle \Sigma_{0}}, I_{\dot{\Sigma}},
I_{\ddot{\Sigma}}$, etc. This 
suggests the 

\proclaim Definition. An algebraic spinor for $\R^{p, q}$ is an
equivalence class of the quotient set $\{I_{\scriptscriptstyle
\Sigma}\}/ R$, where 
$\{I_{\scriptscriptstyle \Sigma}\}$ is the set of all geometrically
equivalent ideals, and 
$\Psi_{\scriptscriptstyle \Sigma_{0}}\in I_{\scriptscriptstyle
\Sigma_{0}}$ and $\Psi_{\dot{\Sigma}}\in 
I_{\dot{\Sigma}}$ are equivalent, $\Psi_{\dot{\Sigma}} \simeq
\Psi_{\scriptscriptstyle \Sigma_{0}}$ (mod $R)$ if and only if
\begin{equation}
\Psi_{\dot{\Sigma}}= \Psi_{\scriptscriptstyle \Sigma_{0}}u^{-1} .
\end{equation}
$\Psi_{\scriptscriptstyle \Sigma}$ will be called the representative
of the algebraic spinor 
in the basis $\Sigma\in {\cal B}_{\scriptscriptstyle \Sigma}$.
Recall that $\dot{\Sigma}=u \Sigma u^{-1} = L \Sigma$, $u\in
\Spin_+(1, 3)$, $L \in \Lie^\uparrow_+$.

\subsection{What is a Covariant Dirac Spinor (CDS)}

As we already know $f_{\scriptscriptstyle \Sigma_{0}}= 
\frac{1}{2}(1+ \cE_0)(1+ \bi
\cE_1 \cE_2)$  (Eq.~\ref{2.29}) is a primitive idempotent of
$\cl_{4,1} \simeq M_4 (\C)$. If $u\in\Spin_+ (1,3) \subset \Spin_+
(4,1)$ then all ideals $I_{\dot{\Sigma}}=I_{\scriptscriptstyle
\Sigma_{0}}u^{-1}$ are geometrically 
equivalent to $I_{\scriptscriptstyle \Sigma_{0}}$. Since
$\Sigma_0=\{\cE_0, \cE_1,  
\cE_2, \cE_3\}$ is a basis for $\R^{1,3} \subset \cl_{4,1}^+$, 
the meaning of $\dot{\Sigma}= u\Sigma_0 u^{-1}$ is clear. From
Eq.~\ref{2.30} we can write
\begin{equation}
I_{\scriptscriptstyle \Sigma_{0}}\ni
\Psi_{\scriptscriptstyle \Sigma_{0}}=\sum \psi_i f_i,
\qquad {\rm and} \qquad
I_{\dot{\Sigma}}\ni \Psi_{\dot{\Sigma}}=\sum \dot{\psi}_i \dot{f}_i,
\end{equation}
where
$$
f_1=f_{\scriptscriptstyle \Sigma_{0}},  \quad
f_2=-\cE_1 \cE_3 f_{\scriptscriptstyle \Sigma_{0}}, \quad
f_3=\cE_3\cE_0 f_{\scriptscriptstyle \Sigma_{0}} \quad
f_4=\cE_1\cE_0 f_{\scriptscriptstyle \Sigma_{0}}
$$
and
$$
\dot{f}_1=f_{\dot\Sigma}, \quad
\dot{f}_2=-\bar{\dot{\cE}}_1 \bar{\dot{\cE}}_3 f_{\dot{\Sigma}}, \quad
\dot{f}_3=\bar{\dot{\cE}}_3\bar{\dot{\cE}}_0 f_{\dot{\Sigma}}, \quad
\dot{f}_4=\bar{\dot{\cE}}_1\bar{\dot{\cE}}_0 f_{\dot\Sigma}
$$
Since $\Psi_{\dot{\Sigma}}=\Psi_{\scriptscriptstyle \Sigma_{0}}u^{-1}$,
we get
$$
\Psi_{\dot{\Sigma}}=\sum_{i} \psi_i u^{-1} \dot{f}_i=
\sum_{i,k} S_{ik}(u^{-1}) 
\psi_i \dot{f}_k= \sum_k \dot{\psi}_k \dot{f}_k . 
$$
Then
\begin{equation} \label{2.47}
\dot{\psi}_k= \sum_{i} S_{ik}(u^{-1}) \psi_i ,
\end{equation}
where $S_{ik}(u^{-1})$ are the matrix components of the representation
in
$M_4 (\C)$ of $u^{-1} \in \Spin_+(1,3)$. As proved in~\cite{REF-4,REF-5}
the matrices $S(u)$ correspond to the representation  
$D^{(1/2,0)}\oplus D^{(0, 1/2)}$ of $SL(2, \C) \simeq
\Spin_+(1,3)$.

We remark that all the elements of the set $\{I_{\scriptscriptstyle
\Sigma}\}$ of the 
ideals geometrically equivalent to $I_{\scriptscriptstyle
\Sigma_{0}}$ under the action of 
$u\in \Spin_+(1,3) \subset \Spin_+ (4,1)$ have the same image $I=M_4
(\C)f$ where $f$ is given by Eq.~\ref{2.34}, i.e., 
$$
f={\textstyle\frac{1}{2}}(1+ \ub\g_0)(1+ i \ub\g_1 \ub\g_2)
\qquad  i= \sqrt{-1}, 
$$
where $\ub\g_\mu$, $\mu=0,1,2,3$ are the Dirac matrices given by
Eq.~\ref{2.31}.

Then, if
\begin{equation}
\begin{array}{cll}
\g: \cl_{4,1} &\to& M_4(\C) \equiv \End (M_4 (\C) f)\\
x &\mapsto& \g(x): M_4(\C) f\to M_4 (\C)f
\end{array}
\end{equation}
it follows that $\g(\cE_\mu)=\g({\dot{\cE}_\mu})=\ub\g_\mu$, 
$\g(f_{\scriptscriptstyle \Sigma_{0}})=\g(f_{\dot{\Sigma}})=f$ for all
$\cE_\mu,
{\dot{\cE}}_\mu$ such that ${\dot{\cE}}_\mu=u \cE_\mu u^{-1}$ for some
$u\in \Spin_+(1,3)$. Observe that all the information concerning the
orthonormal frames $\Sigma_0, \dot{\Sigma}$, etc., disappear in the
matrix representation of the ideals $I_{\scriptscriptstyle
\Sigma_{0}}, I_{\dot{\Sigma}}, \ldots$ 
in $M_4 (\C)$ since all these ideals are mapped in the same ideal
$I=M_4 (\C)f$. 

With the above remark and taking into account Eq.~\ref{2.47} we are then
lead to the following

\proclaim Definition. A Covariant Dirac Spinor (CDS) for
$\R^{1,3}$ is an
equivalent class of triplets $(\Sigma, S(u), \Psi)$, $\Sigma$ being an
orthonormal basis of $\R^{1,3}, S(u)\in D^{(1/2,0)} \oplus D^{(0, 
1/2)}$ representation of $\Spin_+(1,3), u\in \Spin_+(1,3)$ and
$\Psi \in M_4(\C)f$ and 
$$
(\Sigma, S(u), \Psi) \sim (\Sigma_0, S(u_0), \Psi_0)
$$
if and only if
\begin{equation}
\Psi= S(u)S^{-1}(u_0)\Psi_0, \quad \cH(u u_0^{-1})=L \Sigma_0,
\quad L\in \Lie^\uparrow_+, \quad  u\in \Spin_+(1,3). 
\end{equation}

The pair $(\Sigma, S(u))$ is called a spinorial frame. Observe that the
CDS just defined depends on the choice of the original spinorial
frame $(\Sigma_0, u_0)$ and obviously, to different possible choices
there correspond isomorphic ideals in $M_4(\C)$. For simplicity we
can fix $u_0=1, S(u_0)=1$.

The definition of CDS just given agrees with that given by
Choquet-Bruhat \shortcite{REF-31} except for the irrelevant fact 
that Choquet-Bruhat uses as the space of representatives of
a CDS the complex
four-dimensional vector space $\C^4$ instead of $I=M_4(\C)f$.
We see that Choquet-Bruhat's definition is well justified from the
point of view of the theory of algebraic spinors presented above.

\subsection{Algebraic Dirac Spinors (ADS) and Dirac-Hestenes
Spinors (DHS)}

We saw in  Section~2.4 that there is bijection between
$\psi_{\scriptscriptstyle \Sigma_{0}}\in \cl^{++}_{4,1} \simeq
\cl^+_{1,3}$ and 
$\Psi_{\scriptscriptstyle \Sigma_{0}}\in I_{\scriptscriptstyle
\Sigma_{0}}= \cl^+_{4,1} f_{\scriptscriptstyle \Sigma_{0}}$, namely 
(Eq.~\ref{2.39}), 
$$
\Psi_{\scriptscriptstyle \Sigma_{0}}= \psi_{\scriptscriptstyle
\Sigma_{0}} {\textstyle\frac{1}{2}}(1 + \cE_0)
{\textstyle\frac{1}{2}}(1+ \bi \cE_1 \cE_2)
$$
Then, as we already said, all information contained in
$\Psi_{\scriptscriptstyle \Sigma_{0}}$ (that is the representative in
the basis $\Sigma_0$ 
of an algebraic spinor for $\R^{1,3}$) is also contained in
$\psi_{\scriptscriptstyle \Sigma_{0}}\in \cl^{++}_{4,1} \simeq
\cl^+_{1,3}$. We are then lead 
to the following

\proclaim Definition. Consider the quotient set
$\{I_{\scriptscriptstyle \Sigma}\}/ \cR$ where 
$\{I_{\scriptscriptstyle \Sigma}\}$ is the set of all geometrically
equivalent minimal left 
ideals of $\cl_{1,3}$ generated by $e_{\scriptscriptstyle
\Sigma_{0}}= \frac{1}{2}(1+ E_0), \Sigma_0= (E_0, E_1,
E_2, E_3)$ [i.e., $I_{\ddot{\Sigma}}, I_{\dot{\Sigma}}\in
\{I_{\scriptscriptstyle \Sigma}\}$ then $I_{\ddot{E}}=uI_{\dot{\Sigma}}u^{-1}
\equiv I_{\dot\Sigma} u^{-1}$ for some $u\in \Spin_+ (1,3)$]. An
algebraic Dirac 
Spinor (ADS) is an element of
$\{I_{\scriptscriptstyle \Sigma}\}/ \cR$. Then if
$\Phi_{\dot{\Sigma}}\in I_{\dot{\Sigma}}, \Phi_{\ddot{\Sigma}}\in
I_{\ddot{\Sigma}}$, then $\Phi_{\ddot{\Sigma}} \simeq
\Phi_{\dot{\Sigma}}$(mod $\cR$) if and only if
$\Phi_{\ddot{\Sigma}}=\Phi_{\dot{\Sigma}} u^{-1}$, for some $u\in
\Spin_+(1,3)$. 

\noindent
We remark that (see Eq.~\ref{2.38})
$$
\Phi_{\dot{\Sigma}}=\psi_{\dot{\Sigma}} e_{\dot{\Sigma}}, \quad
\Phi_{\ddot{\Sigma}}=\psi_{\dot{\Sigma}} e_{\ddot{\Sigma}} \qquad 
\psi_{\dot{\Sigma}}, \psi_{\ddot{\Sigma}}\in \cl^+_{1,3}
$$
and since $e_{\ddot{\Sigma}}=u e_{\scriptscriptstyle \Sigma} u^{-1}$
for some $u\in 
\Spin_+(1,3)$ we get\footnote{In~\cite{REF-9,REF-10}
$2\Phi$ is called mother of all the real spinors.}
\begin{equation}
\psi_{\ddot{\Sigma}}=\psi_{\dot{\Sigma}}u^{-1}  . 
\end{equation}
Now, we quoted in Section~2.3 that for $p+q \leq 5, \ \
\Spin_+(p,q)=\{u\in \cl^+_{p,q} | u \tilde{u}=1\}$. Then for all
$\psi_{\scriptscriptstyle \Sigma} \in \cl^+_{1,3}$ such that
$\psi_{\scriptscriptstyle \Sigma} \tilde\psi_{\scriptscriptstyle
\Sigma} \neq 0$ we obtain immediately the polar form
\begin{equation} \label{2.51}
\psi_{\scriptscriptstyle \Sigma}=\rho^{1/2} e^{\beta E_{5} /2}
R_{\scriptscriptstyle \Sigma},  
\end{equation}
where $\rho\in \R^+, \beta\in \R, R_{\scriptscriptstyle \Sigma} \in
\Spin_+(1,3), E_5=E_0 E_1 
E_2 E_3$. With the above remark in mind we present the

\proclaim Definition. A Dirac-Hestenes spinor (DHS) is an
equivalence class of 
triplets $(\Sigma, u, \psi_{\scriptscriptstyle \Sigma})$,
where $\Sigma$ is an oriented orthonormal 
basis of $\R^{1,3} \subset \cl_{1,3}$,  $u\in \Spin_+(1,3)$,  
and $\psi_{\scriptscriptstyle \Sigma} \in \cl^+_{1,3}$). 
We say that $(\Sigma, u, \psi_{\scriptscriptstyle \Sigma}) \sim 
(\Sigma_0, u_0, \psi_{\scriptscriptstyle \Sigma_{0}})$ if and only if
$\psi_{\scriptscriptstyle \Sigma}=\psi_{\scriptscriptstyle
\Sigma_{0}}u^{-1}_0 u$, $\cH(u u^{-1}_0)=L$,
$\Sigma=L\Sigma_0 (\equiv u^{-1} u_0 \Sigma_0 u_0^{-1}u)$,
$u, u_0 \in \Spin_+(1,3)$, $L\in \Lie^\uparrow_+$. 
$u_0$ is arbitrary but fixed. A DHS determines a set
of vectors $X_\mu \in \R^{1,3}$, $(\mu=0,1,2,3)$ by a given
representative $\psi_{\dot{\Sigma}}$ of the DHS in the basis
$\Sigma$ by
\begin{equation} \label{2.52}
\psi: \dot{\Sigma}\to \R^{1,3}, \ \  \psi_{\dot{\Sigma}}
\dot{E}_\mu \tilde\psi_{\dot{\Sigma}}= X_\mu  \ \ (\dot{\Sigma}=
(\dot{E}_0, \dot{E}_1, \dot{E}_2, \dot{E}_3))  .
\end{equation}

\smallbreak
We give yet another equivalent definition of a DHS

\proclaim Definition. A Dirac-Hestenes spinor is an
element of the quotient set 
$\cl^+_{1,3}/ \cR$ such that given the basis $\Sigma, \dot{\Sigma}$ of
$\R^{1,3} \subset \cl_{1,3}$, $\psi_{\scriptscriptstyle \Sigma} \in
\cl^+_{1,3}$, 
$\psi_{\dot{\Sigma}}\in \cl^+_{1,3}$ then $\psi_{\dot{\Sigma}}
\sim \psi_{\scriptscriptstyle \Sigma} ({\rm mod} \cR)$ if and only if
$\psi_{\dot{\Sigma}}= \psi_{\scriptscriptstyle \Sigma} u^{-1}$, 
$\dot{\Sigma}=L\Sigma= u\Sigma u^{-1}$, $\cH(u)=L$, $u\in
\Spin_+(1,3)$, $L\in \Lie^\uparrow_+$. 

With the canonical form of a DHS given by Eq.~\ref{2.51} some features
of the hidden geometrical nature of the Dirac spinors defined above
comes to light: Eq.~\ref{2.51}
says that when $\psi_{\scriptscriptstyle \Sigma}
\tilde\psi_{\scriptscriptstyle \Sigma} \neq 0$ the 
Dirac-Hestenes spinor $\psi_{\scriptscriptstyle \Sigma}$ is
equivalent to a Lorentz rotation 
followed by a dilation and a duality mixing given by the term
$e^{\beta E_{5} /2}$, where $\beta$ is the so-called Yvon-Takabayasi
angle \cite{REF-32,REF-33} and the justification for the name duality
rotation can be found in~\cite{REF-4}. We emphasize that the
definition of the  Dirac-Hestenes spinors gives above is new. In the
past objects $\psi\in\cl^+_{1,3}$ satisfying $\psi X \tilde\psi = Y$,
for $X,Y \in \R^{1,3} \subset \cl_{1,3}$ have been  called operator
spinors (see, e.g., in~\cite{REF-34,REF-9,REF-10}). DHS 
have been used as the departure point of many interesting
results as, e.g., in~\cite{REF-4,REF-35,REF-36,REF-37}.

\subsection{Fierz Identities}

The formulation of the Fierz \shortcite{REF-13} identities using the
CDS $\Psi \in \C^4$ is well known \cite{REF-14}. Here we present
the identities for $\Psi_{\scriptscriptstyle \Sigma_{0}}
\in I_{{\scriptscriptstyle \Sigma_{0}}} \simeq
(\C \otimes \cl_{1,3})f_{\scriptscriptstyle \Sigma_{0}}$ and for the 
DHS $\psi_{\scriptscriptstyle \Sigma_{0}}\in
\cl^+_{1,3}$~\cite{REF-9,REF-10}.  Let then $\Psi \in \C^4$
be a representative of a CDS for $\R^{1,3}$
associated to the basis $\Sigma_0=\{E_0, E_1, E_2,
E_3\}$ of $\R^{1,3} \subset \cl_{1,3}$. Then
$\Psi, \Psi_{\scriptscriptstyle \Sigma_{0}}$ determines the following
so-called bilinear covariants,
\begin{eqnarray}
&& \sigma= \Psi^\dagger \gamma_0 \Psi= 4
\dgr{\tilde\Psi_{\scriptscriptstyle \Sigma_{0}}^* 
\Psi_{\scriptscriptstyle \Sigma_{0}}}_0, \nonumber\\
&& J_\mu= \Psi^\dagger \gamma_0 \gamma_\mu \Psi= 4
\dgr{\tilde\Psi_{\scriptscriptstyle \Sigma_{0}}^* E_\mu 
\Psi_{\scriptscriptstyle \Sigma_{0}}}_0, \nonumber\\
&& S_{\mu \nu}= \Psi^\dagger \gamma_0 i \gamma_{\mu \nu}\Psi= 4
\dgr{\tilde\Psi_{\scriptscriptstyle \Sigma_{0}}^* i E_{\mu
\nu} \Psi_{\scriptscriptstyle \Sigma_{0}}}_0, \nonumber \\
&& K_\mu= \Psi^\dagger \gamma_0 i \gamma_{0123} \Psi= 4
\dgr{\tilde\Psi_{\scriptscriptstyle \Sigma_{0}}^* i E_{0123} E_\mu
\Psi_{\scriptscriptstyle \Sigma_{0}}}_0, \nonumber\\ 
&& \om= -\Psi^\dagger \gamma_0  \gamma_{0123} \Psi= -4
\dgr{\tilde\Psi_{\scriptscriptstyle \Sigma_{0}}^* E_{0123}
\Psi_{\scriptscriptstyle \Sigma_{0}}}_0 ,   \label{2.53}
\end{eqnarray}
where $\dagger$ means Hermitian conjugation and $*$ complex
conjugation. We remark that the reversion in $\cl_{4,1}$
corresponds to the reversion plus complex conjugation in $\C \otimes
\cl_{1,3}$. 

All the bilinear covariants are real and have physical meaning in the
Dirac theory of the electron, but its geometrical nature appears
clearly when these bilinear covariants are formulated with the aid of
the DHS.

Introducing the Hodge dual of a Clifford number $X\in\cl_{1,3}$ by
\begin{equation}
\star X = \tilde X E_5, \qquad E_5=E_0 E_1 E_2 E_3
\end{equation}
the bilinear covariants given by Eq.~\ref{2.53} become in terms of
$\psi_{\scriptscriptstyle \Sigma_{0}}$, the representative of a DHS in the
orthonormal basis $\Sigma_0=\{E_0, E_1, E_2, E_3\}$ of $\R^{1,3} \subset
\cl_{1,3}$
\begin{equation}
\begin{array}{lll}
\psi_{\scriptscriptstyle \Sigma_{0}}\tilde\psi_{\scriptscriptstyle
\Sigma_{0}} = \sigma+\star\omega & & J=J_\mu E^\mu\\ 
\psi_{\scriptscriptstyle \Sigma_{0}}E_0\tilde\psi_{\scriptscriptstyle
\Sigma_{0}} = J & & S=\frac{1}{2} S_{\mu\nu} E^\mu E^\nu\\
\psi_{\scriptscriptstyle \Sigma_{0}}E_1 E_2\tilde\psi_{\Sigma} = S &&
K=K_\mu E^\mu\\ 
\psi_{\scriptscriptstyle \Sigma_{0}}E_3\tilde\psi_{\scriptscriptstyle
\Sigma_{0}} = K & & E^\mu=\eta^{\mu\nu} E_\mu\\ 
\psi_{\scriptscriptstyle \Sigma_{0}}E_0
E_3\tilde\psi_{\scriptscriptstyle \Sigma_{0}} = \star S & &
\eta^{\mu\nu}= \mbox{diag}(1,-1,-1,-1)\\
\psi_{\scriptscriptstyle \Sigma_{0}}E_0 E_1
E_2\tilde\psi_{\scriptscriptstyle \Sigma_{0}} = \star K &&
\end{array} 
\end{equation}

The Fierz identities are
\begin{equation}
J^2=\sigma^2+\omega^2, \quad J\cdot K=0, \quad J^2=-K^2, \quad J\wedge
K=-(\omega+\star\sigma)S
\end{equation}
\begin{equation}
\left\{\begin{array}{lll}
S\cdot J=\omega K && S\cdot K=\omega J\\
(\star S)\cdot J= - \sigma K  && (\star S)\cdot K=-\sigma J\\
S\cdot S=\omega^2-\sigma^2 && (\star S)\cdot S=-2\sigma\omega
\end{array}\right. 
\end{equation}
\begin{equation}
\left\{\begin{array}{ll}
JS=-(\omega+\star\sigma)K & KS=-(\omega+\star\sigma)J\\
SJ=-(\omega-\star\sigma)K & SK=-(\omega-\star\sigma)J\\
S^2=\omega^2-\sigma^2-2\sigma(\star\omega) &\\
S^{-1}=-S(\sigma-\star\omega)^2/(\sigma^2+\omega^2)=KSK/(\sigma^2
+\omega^2)^2 & \end{array}\right.
\end{equation}

The proof of these identities using the DHS is
almost a triviality. 

The importance of the bilinear covariants is due to the fact that
we can recover from them the CDS $\Psi_{\scriptscriptstyle
\Sigma_{0}} \in M_4 (\C)f$ or 
all other kinds of Dirac spinors defined above through an algorithm
due to Crawford (see also~\cite{REF-9,REF-10}). Indeed,
representing the images of the bilinear covariants in $\cl_{1,3}$ and
$\cl^+_{4,1} \subset \cl_{4,1}$ under the mapping $g$ (Eq.~\ref{2.25})
by the same letter we have that the following result holds true: 
let
\begin{equation}
Z_{\scriptscriptstyle \Sigma_{0}}=(\sigma+J+i S+i(\star
K)+\star\omega)\in \C \otimes \cl_{1,3}
\end{equation}
where $\sigma$, $J$, $S$, $K$, $\omega$ are the bilinear covariants of 
$\Psi_{\scriptscriptstyle \Sigma_{0}} \simeq (\C \otimes \cl_{1,3})
f_{\scriptscriptstyle \Sigma_{0}}$. Take 
$\eta_{\scriptscriptstyle \Sigma_{0}} \in  
(\C \otimes \cl_{1,3}) f_{\scriptscriptstyle \Sigma_{0}}$ such that
$\tilde\eta^*_{\scriptscriptstyle \Sigma_{0}}\Psi_{\scriptscriptstyle
\Sigma_{0}}\neq  
0$. Then $\Psi_{\scriptscriptstyle \Sigma_{0}}$ and
$Z_{\scriptscriptstyle \Sigma_{0}}\eta_{\scriptscriptstyle
\Sigma_{0}}$ differ  
by a complex factor. We have
\begin{equation}
\Psi_{\scriptscriptstyle
\Sigma_{0}}=\frac{1}{4N_{\eta_{\scriptscriptstyle \Sigma_{0}}}}  
e^{-i\al} Z_{\scriptscriptstyle \Sigma_{0}}\eta_{\scriptscriptstyle
\Sigma_{0}} 
\end{equation}
\begin{equation}
N_{\eta_{\scriptscriptstyle \Sigma_{0}}}=
\sqrt{\dgr{\tilde\eta_{\scriptscriptstyle \Sigma_{0}}^* 
Z_{\scriptscriptstyle \Sigma_{0}}\eta_{\scriptscriptstyle
\Sigma_{0}}}_0} \ , \ e^{-i\al}= 
\frac{4}{N{\eta_{\scriptscriptstyle \Sigma_{0}}}} \dgr{\tilde\eta^*\Psi}_0
\end{equation}
Choosing $\eta_{\scriptscriptstyle \Sigma_{0}} =
f_{\scriptscriptstyle \Sigma_{0}}$, we obtain 
\begin{equation}
N_{f_{\scriptscriptstyle \Sigma_{0}}} = \frac{1}{2} \sqrt{\sigma+
J\cdot E_0-S\cdot (E_1  
E_2)-K\cdot E_3} \ , \ \ \
e^{-i\al} = \psi_1/|\psi_1|  , 
\end{equation}
where $\psi_{1}$ is the first component of $\Psi_{{\scriptscriptstyle
\Sigma_{0}}}$ in the
spinorial basis $\{s_{i}\}$. 

It is easier to recuperate the CDS from its bilinear covariants
if we use the DHS $\psi_{\scriptscriptstyle \Sigma_{0}} \in
\cl_{1,3}^{+} \simeq  
(\cl^+_{4,1})^+$ since putting
\begin{equation}
\left\{\begin{array}{l}
\psi_{\scriptscriptstyle \Sigma_{0}}(1+
E_0)\tilde\psi_{\scriptscriptstyle \Sigma_{0}}=P\\ 
\psi_{\scriptscriptstyle \Sigma_{0}}(1+ E_0) E_1
E_2\tilde\psi_{\scriptscriptstyle \Sigma_{0}}=Q 
\end{array}\right.
\end{equation}
\begin{equation}
\psi_{\scriptscriptstyle \Sigma_{0}}(1+ E_0)(1+\bi  E_1  E_2)
\tilde\psi_{\scriptscriptstyle \Sigma_{0}}=(P+\bi Q)
\end{equation}
results
\begin{equation}
P=\sigma+J+\om \qquad Q=S+\star K
\end{equation}
and
\begin{equation}
Z_{\scriptscriptstyle \Sigma_{0}}
= P {\textstyle\frac{1}{2}}(1 + {\textstyle\frac{\bi}{2\sigma}} Q)^2
\end{equation}
valid for $\sigma \neq 0$, $\om \neq 0$ (for other cases
see~\cite{REF-10}).
From the above results it follows that $\Psi_{\scriptscriptstyle
\Sigma_{0}}$ can be easily 
determined from its bilinear covariant except for a ``complex" 
$ E_2  E_1$ phase factor.

\section{The Clifford Bundle of Spacetime and their Irreducible
Module Representations}

\subsection{The Clifford Bundle of Spacetime}

Let $M$ be a four dimensional, real, connected,
paracompact manifold. Let $TM$ [$T^*M$] be the tangent [cotangent]
bundle of $M$.

\proclaim Definition. A Lorentzian manifold is a pair $(M,g)$,
where $g\in\sec 
T^*M\times T^*M$ is a Lorentzian metric of signature (1,3), i.e.,
for all
$x \in M$, $T_xM\simeq T^*_x M\simeq \R^{1,3}$, where $\R^{1,3}$ is the
vector Minkowski space.

\proclaim Definition. A spacetime $\cM$ is a triple
$(M,g,\nabla)$ where $(M,g)$ 
is a time oriented and spacetime oriented Lorentzian manifold and
$\nabla$ is a linear connection for $M$ such that $\nabla g=0$. If in
addition $\bT(\nabla)=0$ and $\bR(\nabla)\neq 0$, where $\bT$ and
$\bR$ are respectively the torsion and curvature tensors, then
$\cM$ is said to be a Lorentzian spacetime. When $\nabla g=0$,
$\bT(\nabla)=0$, $\bR(\nabla)=0$, $\cM$ is called Minkowski spacetime
and will be denote by $\BM$.  When $\nabla g=0$, $\bT(\nabla)\neq 0$ 
and $\bR(\nabla)=0$ or $\bR(\nabla)\neq 0$, $\cM$ is said to be a
Riemann-Cartan spacetime.

In what follows $P_{\SO_+(1,3)}(\cM)$ denotes the principal bundle of
oriented Lorentz tetrads \cite{REF-8,REF-23}. By $g^{-1}$ we denote the
``metric" of the cotangent bundle.

It is well known that the natural operations on metric vector spaces,
such as, e.g., direct sum, tensor product,
exterior power, etc., carry over canonically to vector bundles with
metrics. Take, e.g., the cotangent bundle $T^* M$. If
$\pi : T^* M \to M$ is the canonical projection, then in each fiber
$\pi^{-1}(x)=T^*_x M\simeq\R^{1,3}$, the ``metric"
$g^{-1}$ can be used to construct a Clifford algebra $\Cl(T^*_x M)
\simeq \cl_{1,3}$. We have the

\proclaim Definition. The Clifford bundle of spacetime $\cM $
is the bundle of algebras
\begin{equation}
\Cl (\cM)=\bigcup_{x \in M} \Cl (T^*_x M)
\end{equation}

As is well known $\Cl (\cM)$ is the quotient bundle
\begin{equation}
\Cl(\cM) = \frac{\tau M}{\bJ(\cM)}
\end{equation}
where $\tau M={\displaystyle\oplus^\infty_{r=0}} T^{0,r}(M)$ 
and 
$T^{(0,r)}(M)$ is the space of $r$-covariant tensor fields,
and $\bJ(\cM)$ is the bundle of ideals whose
fibers at $x\in M$ are the two side ideals in $\tau M$ generated by
the elements of the form $a \otimes b + b \otimes a - 2g^{-1}(a,b)$ for
$a,b\in T^*\cM$.

Let $\pi_c : \Cl (\cM) \to M$ be the canonical projection of
$\Cl (\cM)$ and let $\{U_\al\}$ be an open covering of $M$. From
the definition of a fibre bundle \cite{REF-23} we know that there is
a trivializing mapping 
$\fii\al : \pi^{-1}_c(U_\al)\to U_\al\times\cl_{1,3}$ of the form
$\fii\al(p)=(\pi_c(p), \stackrel{\scriptscriptstyle
\Delta}{\fii}_\al(p))$. If 
$U_{\al\beta}=U_\al\cap U_\beta$ and $x\in U_{\al\beta}, \ \ p \in
\pi^{-1}_c(x)$, then
\begin{equation}
\stackrel{\scriptscriptstyle \Delta}{\fii}_\al(p) =
f_{\al\beta}(x)\stackrel{\scriptscriptstyle \Delta}{\fii}_\beta(p)
\end{equation}
for $f_{\al\beta}(x)\in \Aut(\cl_{1,3})$, where $f_{\al\beta} :
U_{\al\beta}\to \Aut(\cl_{1,3})$ are the transition mappings of
$\Cl (\cM)$. We know that every automorphism of $\cl_{1,3}$ is inner
and it follows that,
\begin{equation} \label{3.4}
f_{\al\beta}(x)\stackrel{\scriptscriptstyle
\Delta}{\fii}_\beta(p)= g_{\al\beta}(x) 
\stackrel{\scriptscriptstyle \Delta}{\fii}_\beta(p)
g_{\al\beta}(x)^{-1}
\end{equation}
for some $g_{\al\beta}(x)\in\cl^\star_{1,3}$, the group of invertible
elements of $\cl_{1,3}$. We can write equivalently instead of
Eq.~\ref{3.4},
\begin{equation}
f_{\al\beta}(x)\stackrel{\scriptscriptstyle \Delta}{\fii}_\beta(p)= 
\stackrel{\scriptscriptstyle \Delta}{\fii}_\beta (a_{\al\beta} p
a^{-1}_{\al\beta}) 
\end{equation}
for some invertible element $a_{\al\beta}\in\cl(T^*_x M)$.

Now, the group $\SO_+(1,3)$ has, as we know (Section~2), a
natural extension in  the Clifford algebra $\cl_{1,3}$. Indeed we
know that $\cl^\star_{1,3}$ acts naturally on $\cl_{1,3}$ as an
algebra automorphism through its adjoint representation ${\rm Ad}: u
\mapsto {\rm Ad}_{u}, \ \ {\rm Ad}_{u}(a) = u a u^{-1}$. 
Also $\Ad|_{\Spin_+(1,3)}=\sigma$ defines a group homeomorphism
$\sigma : \Spin_+(1,3) \to \SO_+(1,3)$ which is onto with kernel
$\BZ_2$. It is clear, since Ad$_{-1}=$ identity, that  
$\Ad : \Spin_+(1,3) \to \Aut(\cl_{1,3})$ descends to a
representation of $\SO_+(1,3)$. Let us call $\Ad'$ this
representation, i.e., $\Ad' : \SO_+(1,3) \to \Aut(\cl_{1,3})$.  
Then we can write $\Ad'_{\sigma(u)}a={\rm Ad}_ua=uau^{-1}$.

From this it is clear that the structure group of the Clifford bundle
$\cl(\cM)$ is reducible from $\Aut (\cl_{1,3})$ to $\SO_+(1,3)$. This
follows immediately from the existence of the Lorentzian structure
$(M,g)$ and the fact that $\cl(\cM)$ is the exterior bundle where the
fibres are equipped with the Clifford product. Thus the
transition maps of the principal bundle of oriented Lorentz tetrads
$P_{\SO_+(1,3)}(\cM)$ can be (through $\Ad'$)  taken as transition
maps for the Clifford bundle. We then have the result \cite{REF-39}
\begin{equation}
\Cl (\cM)=P_{{\SO_+(1,3)}}(\cM)\times_{\Ad'} \cl_{1,3}
\end{equation}

\subsection{Spinor Bundles}

\proclaim Definition. A spinor structure for $\cM$
consists of a principal fibre bundle 
$\pi_s : P_{\Spin_+(1,3)}(\cM) \to M$
with group $SL(2,\C) \simeq \Spin_+(1,3)$ and a map
$$
s: P_{\Spin_+(1,3)}({\cal M}) \to P_{\SO_+(1,3)}({\cal M})
$$
satisfying the following conditions
\begin{enumerate}
\item  $\pi(s(p))=\pi_s(p) \ \forall p \in P_{\Spin_+(1,3)}(\cM)$
\item $s(pu)=s(p)\cH(u) \  \forall p \in P_{\Spin_+(1,3)}(\cM)$ and
$\cH : SL(2,\C) \to \SO_+(1,3)$.
\end{enumerate}

Now, in Section~2 we learned that the minimal left (right) ideals of
$\cl_{p,q}$ are irreducible left (right) module representations of
$\cl_{p,q}$ and we define covariant and algebraic Dirac spinors as
elements of quotient sets of the type $\{I_{\scriptscriptstyle
\Sigma}\}/\R$ (sections~2.6 and 2.7) in appropriate Clifford algebras.
We defined also in Section~2 the DHS. We are now interested in defining
algebraic Dirac spinor fields (ADSF) and also Dirac-Hestenes spinor
fields (DHSF).

So, in the spirit of Section~2 the following question naturally
arises: Is it possible to find a vector	bundle
$\pi_s : S(\cM) \to M$ with the property that each fiber over
$x\in M$ is an irreducible module over $\cl(T^*_x M)$?

The answer to the above question is in general no. Indeed it is 
well known \cite{REF-40} that the necessary and sufficient conditions
for $S(\cM)$ 
to exist is that the Spinor Structure bundle 
$P_{\Spin_+(1,3)}(\cM)$ exists, which implies the vanishing of the
second Stiefel-Whitney class of $M$, i.e., $\om_2(M)=0$. For a
spacetime $\cM$ this is equivalent, as shown originally by 
Geroch \shortcite{REF-41,REF-42} that $P_{\SO_+(1,3)}(\cM)$ is a trivial
bundle, i.e., that it admits a global section. When
$P_{\Spin_+(1,3)}(\cM)$ exists we said that $\cM$ is a spin manifold.

\proclaim Definition. A real spinor bundle for $\cM$ is the
vector bundle
\begin{equation} \label{3.7}
S(\cM)=P_{\Spin_+(1,3)}(\cM) \times_\mu \bM
\end{equation}
where $\bM$ is a left (right) module for $\cl_{1,3}$ and where $\mu :
P_{\Spin_+(1,3)} \to \SO_+(1,3)$ 
is a representation  given by
left (right) multiplication by elements of $\Spin_+(1,3)$.

\proclaim Definition. A complex spinor bundle for $\cM$ is the
vector bundle 
\begin{equation}
S_c(\cM)=P_{\Spin_+(1,3)}({\cal M}) \times_{\mu_c} \bM_c
\end{equation}
where $\bM$ is a complex left (right) module for $\C\otimes\cl_{1,3}
\simeq \cl_{4,1} \simeq M_4(\C)$, and where $\mu_c : P_{\Spin_+(1,3)}
\to \SO_+(1,3)$  is a
representation  given by
left (right) multiplication by elements of $\Spin_+(1,3)$.

Taking, e.g. $\bM_c=\C^4$ and $\mu_c$ the $D^{(1/2,0)} \oplus D^{(0,1/2)}$
representation of $\Spin_+(1,3)$ in $\End(\C^4)$, we recognize
immediately the usual definition of the covariant spinor bundle of
$\cM$, as given, e.g., in~\cite{REF-31}.

Since, besides being right (left) linear spaces over
$\BH$, the left (right) ideals of $\cl_{1,3}$
are representation modules of $\cl_{1,3}$, we have the 

\proclaim Definition. $I(\cM)$ is a real spinor bundle for $\cM$ 
such that $\bM$ in Eq.~\ref{3.7} is
$I$, a minimal left (right) ideal of $\cl_{1,3}$.

In what follows we fix the ideal  taking $I=\cl_{1,3} 
\frac{1}{2}(1+E_0)=\cl_{1,3} e$. 
If $\pi_I : I(\cM) \to M$ is
the canonical projection and $\{U_\al\}$ is an open covering of $M$ we
know from the definition of a fibre bundle that there is a
trivializing mapping $\chi_\al(q) = (\pi_I(q), 
\stackrel{\scriptscriptstyle \Delta}{\chi}_\al(q))$. 
If $U_{\al\beta}=U_\al \cap U_\beta$ and $x\in
U_{\al\beta}$, $q \in \pi^{-1}_I(U_\al)$, then
\begin{equation}
\stackrel{\scriptscriptstyle \Delta}{\chi}_\al(q)=g_{\al\beta}(x)
\stackrel{\scriptscriptstyle \Delta}{\chi}_\beta(q) 
\end{equation}
for the transition maps in $\Spin_+(1,3)$.\footnote{
We start with transition maps in $\cl^\star_{p,q}$ and then by the
bundle reduction process we end with $\Spin_+(1,3)$.} Equivalently
\begin{equation}
\stackrel{\scriptscriptstyle \Delta}{\chi}_\al(q)=
\stackrel{\scriptscriptstyle \Delta}{\chi}_\beta 
(a_{\al\beta} q) 
\end{equation}
for some $a_{\al\beta}\in \Cl (T^*_x \cM)$. Thus, for the transition
maps to be in $\Spin_+(1,3)$ it is equivalent that the right action of 
$\BH e=e\BH=e\cl_{1,3}e$ be the defined in the bundle, since for
$q\in\pi^{-1}_x(x), x\in U_\al$ and $a\in\BH$ we define  $qa$ as the
unique element of $\pi^{-1}_q(x)$ such that
\begin{equation} \label{3.11}
\stackrel{\scriptscriptstyle \Delta}{\chi}_\al(qa)=
\stackrel{\scriptscriptstyle \Delta}{\chi}_\al(q)a 
\end{equation}

Naturally, for the validity of Eq.~\ref{3.11} to make sense it is
necessary that 
\begin{equation} \label{3.12}
g_{\al\beta}(x)(\stackrel{\scriptscriptstyle \Delta}{\chi}_\al(q)a)=
(g_{\al\beta}(x) 
\stackrel{\scriptscriptstyle \Delta}{\chi}_\al(q))a
\end{equation}
and Eq.~\ref{3.12} implies that the transition maps are 
$\BH$-linear.\footnote{Without the $\BH$-linear
structure there exists more general bundles of irreducible modules
for $\Cl (\cM)$ \cite{REF-43}.}

Let $f_{\al\beta} : U_{\al\beta} \to \Aut(\cl_{1,3})$ be the
transition
functions for $\Cl (\cM)$. On the intersection $U_\al\cap U_\beta\cap
U_\al$ it must hold
\begin{equation}
f_{\al\beta}f_{\beta\g} = f_{\al\g}
\end{equation}

We say that a set of {\it lifts\/} of the transition functions of
$\cl(\cM)$ is a set of elements in $\cl^\star_{1,3}, \{g_{\al\beta}\}$
such that if
$$
\begin{array}{l}
\Ad : \cl^\star_{1,3} \to \Aut(\cl_{1,3})\\
\Ad(u)X = uXu^{-1} \ , \ \forall X \in \cl_{1,3} \end{array}
$$
then $\Ad_{g_{\al\beta}}=f_{\al\beta}$ in all intersections.

Using the theory of the C\v{e}ch cohomology \cite{REF-43} it can
be shown that any set of lifts can be used to define a characteristic
class $\om(\cl(\cM)) \in \check H{}^2 (M,\BH^*)$, the
second C\v{e}ch 
cohomology group with values in $\BH^*$, the space of all non zero
$\BH$-valued germs of functions in $M$.

We say that we can coherently lift the transition maps $\cal C (\cM)$ to
a set $\{g_{\al\beta}\} \in \cl^\star_{1,3}$ if in the intersection
$U_\al\cap U_\beta\cap U_\g, \ \forall \al, \beta, \g$, we have
\begin{equation}
g_{\al\beta} g_{\beta\g}=g_{\al\g}
\end{equation}

This implies that $\om(\cl(\cM))=$ id$_{(2)}$, i.e., $M$ is
C\v{e}ch trivial and the coherent lifts can be classified by an
element of the first C\v{e}ch cohomology group
$\check H{}^1 (M,\BH^*)$.
Benn and Tucker \shortcite{REF-43} proved the important result:

\proclaim Theorem. There exists a bundle of irreducible
representation modules 
for $\cl(\cM)$ if and only if the transition maps of $\cl(\cM)$ can
be coherently lift from $\Aut(\cl_{1,3})$ to $\cl^\star_{1,3}$.

They showed also by defining the concept of equivalence classes of
coherent lifts that such classes are in one to one correspondence
with the equivalence classes of bundles of irreducible representation
modules of $\Cl(\cM)$, $I(\cM)$ and $I'(\cM)$ being equivalent if there
is a bundle isomorphism $\rho : I(\cM) \to I'(\cM)$ such that
$$
\rho(a_x q)=a_x\rho(q), \ \ \forall a_x\in\Cl (T^*_x M), \forall q\in
\pi^{-1}_I(x)
$$

By defining that a {\it spin structure} for $M$ is an equivalence
class of bundles of irreducible representation modules for
$\cl(\cM)$, represented by $I(\cM)$, Benn and Tucker showed that
this agrees with the usual conditions for $M$ to be a spin manifold.

Now, recalling the definition of a vector bundle we see that the
prescription for the construction of $I(\cM)$ is the following. Let
$\{U_\al\}$ be an open covering of $M$ with $f_{\al\beta}$ being the
transition functions for $\cl(\cM)$ and let $\{g_{\al\beta}\}$ be a
coherent lift which is then used to quotient the set $\cup_\al \ U_\al
\times I$, where e.g., $I=\cl_{1,3}\frac{1}{2}(1+E_0)$
to form the 
bundle $\cup_\al \ U_\al \times I/\cR$ where $\cR$ is the equivalence
relation 
defined as follows. For each $x\in U_\al$ we choose a minimal left
ideal $I^\al_{\Sigma(x)}$ in $\Cl (T^*_x M)$ by 
requiring\footnote{Recall the notation of
Section~2 where $\Sigma$ is an orthonormal frame, etc.}
\begin{equation}
\stackrel{\scriptscriptstyle \Delta}{\fii}_\al(I^\al_{\Sigma(x)})=I
\end{equation}
As before we introduce $a_{\al\beta}\in\Cl (T^*_x M)$ such that
\begin{equation}
\stackrel{\scriptscriptstyle
\Delta}{\fii}_\beta(a_{\al\beta})=g_{\al\beta}(x) 
\end{equation}
Then for all $X\in\cl(T^*_x\cM),  \stackrel{\scriptscriptstyle
\Delta}{\fii}_\al(X) = 
\stackrel{\scriptscriptstyle \Delta}{\fii}_\beta(a_{\al\beta} X
a^{-1}_{\al\beta})$. So, if $X\in 
I^\al_{\Sigma(x)}$ then  $a_{\al\beta} X a^{-1}_{\al\beta}$ and also
$X a^{-1}_{\al\beta}\in I^\beta_{\dot \Sigma(x)}$. Putting $Y_\al= U_\al
\times I^\al_{\Sigma(x)} \ Y = \cup_\al Y_\al$, the equivalence
relation $\cR$ is defined on $Y$ by                         
$(U_\al, x, \psi_{\scriptscriptstyle \Sigma}) \simeq  (U_\beta, x,
\psi_{\dot\Sigma})$ if and only if
\begin{equation} \label{3.17}
\psi_{\dot\Sigma} = \psi_{\scriptscriptstyle \Sigma} a^{-1}_{\al\beta}
\end{equation}

Then, $I(\cM)=Y/\cR$ is a bundle which is an irreducible module
representation of $\cC(\cM)$. We see that Eq.~\ref{3.17} captures nicely
for $a_{\al\beta}\in\Spin_+(1,3)\subset\cl^\star_{1,3}$ our discussion
of ADS of Section~2. We then have

\proclaim Definition. An algebraic Dirac Spinor Field (ADSF) is a
section of
$I(\cM)$ with $a_{\al\beta}\in\Spin_+(1,3)\subset\cl^\star_{1,3}$ in
Eq.~\ref{3.17}.

From the above results we see that ADSF are equivalence classes of
sections of $\cl(\cM)$ and it follows that ADSF can locally be
represented by a sum of inhomogeneous differential forms that lie in
a minimal left ideal of the Clifford algebra $\cl_{1,3}$ at each
spacetime point.

In Section~2 we saw that besides the ideal 
$I=\cl_{1,3}\frac{1}{2}(1+E_0)$, other ideals exist for
$\cl_{1,3}$ that are only algebraically equivalent  to this one. In
order to capture all possibilities we recall that $\cl_{1,3}$ can be
considered as a module over itself by left (or right) multiplication by
itself. We are thus lead to the

\proclaim Definition. The Real Spin-Clifford bundle  of $\cM$ is
the vector bundle
\begin{equation}
\cl_{\Spin_+(1,3)}(\cM)=P_{\Spin_+(1,3)}(\cM)\times_\ell \cl_{1,3}
\end{equation}

It is a ``principal $\cl_{1,3}$ bundle", i.e., it admits a free
action of $\cl_{1,3}$ on the right \cite{REF-7,REF-39}. There is a
natural embedding  $P_{\Spin_+(1,3)}(\cM) \subset
\cl_{{\Spin_+(1,3)}}(\cM)$ which comes from the embedding $\Spin_+(1,3)
\subset \cl^+_{1,3}$. Hence every real spinor bundle for $\cM$ can be
captured from $\cl_{{\Spin_+(1,3)}}(\cM). \ \cl_{{\Spin_+(1,3)}}(\cM)$ is
different from $\cl(\cM)$. Their relation can be discovered
remembering that the representation
$$
\Ad : \Spin_+(1,3) \to \Aut(\cl_{1,3}) \quad
\mbox{Ad}_u X = uXu^{-1} \qquad
u\in \Spin_+(1,3)
$$
is such that $\Ad_{-1}=$ identity and so $\Ad$ descends to a
representation $\Ad'$ of $\SO_+(1,3)$ which we considered above. It
follows that when $P_{\Spin_+(1,3)}(\cM)$ exists
\begin{equation}  \label{3.19}
\cl(\cM) = P_{{\Spin_+(1,3)}}(\cM) \times_{\Ad'} \cl_{{1,3}}
\end{equation}
From this it is easy to prove that indeed $S(\cM)$ is a bundle of
modules over the bundle of algebras $\cl(\cM)$.

We end this section defining the local Clifford product of
$X\in\sec\cl(\cM)$ by a section of $I(\cM)$ or
$\cl_{\Spin_+(1,3)}(\cM)$.  If $\fii\in I(\cM)$ we put 
$X\fii=\phi\in\sec I(\cM)$ and the meaning of Eq.~\ref{3.19}
is that 
\begin{equation} \label{3.20}
\phi(x)=X(x)\rho(x) \qquad \forall x \in M
\end{equation}
where $X(x)\fii(x)$ is the Clifford product of the Clifford numbers
$X(x), \fii(x) \in \cl_{1,3}$.

Analogously if $\psi\in \cl_{\Spin_+(1,3)}(\cM)$
\begin{equation}
X\psi=\xi\in \cl_{\Spin_+(1,3)}(\cM)
\end{equation}
and the meaning of Eq.~\ref{3.20} is the same as in Eq.~\ref{3.19}.

With the above definition we can ``identify" from the algebraically
point of view sections of $\cl(\cM)$ with sections of $I(\cM)$ or
$\cl_{\Spin_+(1,3)}(\cM)$.

\subsection{Dirac-Hestenes Spinor Fields (DHSF)}

The main conclusion of Section~3.2 is that a given ADSF which is a
section of $I(\cM)$ can locally be represented by a sum of
inhomogeneous differential forms in $\cl(\cM)$ that lies in a minimal
left ideal of the Clifford algebra $\cl_{1,3}$ at each point $x\in
M$. Our objective here is to define a DHSF on $\cM$. In order
to achieve our goal we need to find a vector bundle such that a DHSF
is an appropriate section.

In Section~2.7 we defined a DHS as an element of the quotient
set $\cl^+_{1,3}/ \cR$ where $\cR$ is the equivalence relation given
by Eq.~\ref{2.52}.
We immediately realize that if it is possible to define
globally on $M$ the equivalence relation $\cR$, then a DHSF can be
defined as an even section of the quotient bundle $\cl(\cM)/ \cR$.

More precisely, if $\Sigma=\{\g^a\}$, ($a=0,1,2,3$) and
$\dot{\Sigma}=\{\dot{\g}^a\}$, 
$\g^a, \dot{\g}^a \in \sec\bigw^1 (T^* M) \subset \cl(M)$ are such that
$\dot{\g}^a=R\g^a R^{-1}$, where $R\in \sec \cl^+(\cM)$ is such
that $R(x)\in \Spin_+(1,3)$ for all $x\in M$, we say that $\dot{\Sigma}
\sim \Sigma$. Then a DHSF is an equivalence class of even sections of
$\cl(\cM)$ such that its representatives
$\psi_{\scriptscriptstyle \Sigma}$ and
$\psi_{\dot{\Sigma}}$ in the basis $\Sigma$ and $\dot{\Sigma}$
define a set of $1$-forms $X^a \in \sec \Lambda^1 (T^* M)
\subset \sec \cl(\cM)$ by
\begin{equation}
X^a (x) = \psi_{\dot{\Sigma}}(x) \dot{\g}^a(x)
\tilde\psi_{\dot\Sigma}(x)=\psi_{\scriptscriptstyle \Sigma}(x)
\g^a(x) \tilde\psi_{\scriptscriptstyle \Sigma}(x) 
\end{equation}
i.e., $\psi_{\scriptscriptstyle \Sigma}$ and $\psi_{\dot{\Sigma}}$
are equivalent if and only if
\begin{equation} \label{3.24}
\psi_{\dot{\Sigma}}=\psi_{\scriptscriptstyle \Sigma} R^{-1}. 
\end{equation}
Observe that for $\dot{\Sigma}\sim \Sigma$ to be globally defined it
is necessary that the $1$-forms $\{\g^a\}$ and
$\{\dot{\g^a}\}$ are globally defined. It follows that
$P_{\SO_+(1,3)}(\cM)$, the
principal bundle of orthonormal frames must have a global section,
i.e., it must be trivial. This conclusion follows directly from our
definitions, and it is a necessary condition for the existence of a
DHSF. It is obvious that the condition is also sufficient. This
suggests the

\proclaim Definition. A spacetime $\cM$ admits a spinor structure
if and only if it is possible to define a global DHSF on it.

\noindent Then, it follows the

\proclaim Theorem. Let $\cM$ be a spacetime $(\dim M=4)$. Then the
necessary and sufficient condition for $M$ to admit a spinor structure
is that $P_{\SO_+(1,3)}(\cM)$ admits a global section.

In Sect. 3.1 we defined the spinor structure as the principal bundle
$P_{\Spin_+(1,3)}(\cM)$ and a theorem with the same statement as the
above one is known in the literature as Geroch's \shortcite{REF-41}
theorem. Geroch's deals with 
the existence of covariant spinor fields on $\cM$, but since we already
proved, e.g., that covariant Dirac spinors are equivalent to DHS, our
theorem and Geroch's one are equivalent. This can be seen more
clearly once we verify that
\begin{equation}
\frac{\Cl (\cM)}{\cR} \equiv \cl_{\Spin_+(1,3)}(\cM)
\end{equation}
where $\cl_{\Spin_+(1,3)} (\cM)=P_{\Spin_+(1,3)} \times_\ell \cl_{1,3}$
is the Spin-Clifford bundle defined in Section~3.1. To see this,
recall that a DHSF determines through Eq.~\ref{3.20} a set of
$1$-forms 
$X^a \in \sec \bigw^1(T^* M) \subset \sec \cl(\cM)$. Under
an active transformation,
\begin{equation}
X^a \mapsto \dot{X}^a=RX^a R^{-1}, \quad  R(x)\in \Spin_+(1,3), \qquad 
\forall x\in M
\end{equation}
we obtain the active transformation of a DHSF which in the
$\Sigma$-frame is given by\footnote{Observe also that in the
$\dot{\Sigma}$ we have for the representative of the actively
transformed DHSF the relation $\psi'_{\dot{\Sigma}}
= R\psi_{\scriptscriptstyle \Sigma} R^{-1}$.}
\begin{equation}
\psi_{\scriptscriptstyle \Sigma} \mapsto
\psi'_{\scriptscriptstyle \Sigma}= R \psi_{\scriptscriptstyle \Sigma} 
\end{equation}
From Eq.~\ref{3.24} it follows that the action of $\Spin_+(1,3)$ on the
typical fibre $\cl_{1,3}$ of $\cl(\cM)/\cR$ must be through left
multiplication, i.e. given $u\in \Spin_+ (1,3)$ and $X\in \cl_{1,3}$,
and taking into account that $\cl_{1,3}$ is a module over itself we
can define $\ell_u\in \End (\cl_{1,3})$ by $\ell_u(X)= ux, \forall
X\in \cl_{1,3}$. In this way we have a representation $\ell :
\Spin_+(1,3) \to \End(\cl_{1,3}), u \mapsto \ell_u$. Then we can
write,
$$
\frac{\Cl (\cM)}{ \cR} = P_{\Spin_+(1,3)}(\cM) \times_\ell \cl_{1,3}
$$

\subsection{A Comment on Amorphous Spinor Fields}

Crumeyrolle \shortcite{REF-15} gives the name of amorphous spinors fields
to ideal sections of the Clifford bundle $\cl(\cM)$. Thus an
amorphous spinor field $\phi$ is a section of $\cl(\cM)$ such that
$\phi e=\phi$,  with $e$ being an idempotent section of $\cl(\cM)$. 

It is clear from our discussion of the Fierz identities that are
fundamental for the physical interpretation of Dirac theory that
these fields cannot be used in a physical theory. The same holds true
for the so-called Dirac-K\"{a}hler
fields \cite{REF-16,REF-17,REF-18,REF-25} which are sections of
$\cl(\cM)$. These fields do not have the appropriate transformation
law under a Lorentz rotation of the local tetrad field. In particular
the Dirac-Hestenes equation written for amorphous fields is not
covariant (see Section~6).
 We think that with our
definitions of algebraic and DH spinor fields physicists can safely
use our formalism which is not only nice but extremely powerful.

\section{The Covariant Derivative of Clifford and Dirac-Hestenes
Spinor Fields}

In what follows, as in Section~3, $\cM = \dgr{M, \nabla, g}$ will denote
a general Riemann-Cartan spacetime. Since $\Cl(\cM)=\tau M /J(\cM)$ it
is clear that any linear connection defined in $\tau M$ such that
$\nabla g=0$ passes to the quotient $\tau M /J(\cM)$ and thus define an
algebra bundle connection \cite{REF-15}. In this way, the covariant
derivative of a Clifford field $A\in\sec\Cl(\cM)$ is completely
determined.

Although the theory of connections in a principal fibre bundle and
on its associate vector bundles is well described in many
textbooks, we recall below the main definitions concerning to this
theory. A full understanding of the various equivalent definitions
of a connection is necessary in order to deduce a nice formula that
permit us to calculate in a simple way the covariant derivative of
Clifford fields and of Dirac-Hestenes spinor fields (Section~4.3).
Our simple formula arises due to the fact that the Clifford algebra
$\cl_{1,3}$, the typical fibre of $\Cl(\cM)$, is an associative algebra.

\subsection{Parallel Transport and Connections in Principal and
Associate Bundles}

To define the concept of a connection on a PFB $(\bP,M,\pi,G)$ over
a four-dimensional manifold $M$ ($\dim G = n$), we first recall
that the total space $\bP$ of that PFB is itself a
$(n+4)$-dimensional manifold and each one of its fibres
$\inv\pi(x)$, $x\in M$, is a $n$-dimensional sub-manifold of $\bP$.
The tangent space $T_p\bP$, $p\in\inv\pi(x)$ is a
$(n+4)$-dimensional linear space and the tangent space
$T_p\inv\pi(x)$ of the fibre over $x$, at the same point
$p\in\inv\pi(x)$, is a $n$-dimensional linear subspace of $T_p\bP$.
It is called {\it vertical subspace\/} of $T_p\bP$ and denoted by
$V_p\bP$.

A connection is a mathematical object that governs the parallel
transport of frames along smooth paths in the base manifold~$M$.
Such a transport takes place in $\bP$, along directions specified
by vectors in $T_p\bP$, which {\sl does not\/} lie within the
vertical space $V_p\bP$. Since the tangent vectors to the paths on
on the base manifold, passing through a given point $x\in M$, span
the entire tangent space $T_xM$, the corresponding vectors $\bX\in
T_p\bP$ (in whose direction parallel transport can generally take
place in $\bP$) span a four-dimensional linear subspace of
$T_p\bP$, called {\it horizontal space\/} of $T_p\bP$ and denoted
by $H_p\bP$. The mathematical concept of a connection is given
formally by

\proclaim Definition. A connection on a PFB $(\bP,M,\pi,G)$ is
a field of vector spaces $H_p\bP \subset T_p\bP$ such that
\begin{enumerate}
\item $\pi':H_p\bP \to T_x M$, $x=\pi(p)$, is an isomorphism
\item $H_p\bP$ depends differentially on $p$
\item $H_{\tilde R_g p} = \tilde R'_g (H_p)$
\end{enumerate}

\noindent 
The elements of $H_p\bP$ are called {\it horizontal
vectors\/} and the elements of $T_p\inv\pi(x) = V_p\bP$ are called
{\it vertical vectors}. In view of the fact that $\pi:\bP\to M$ is
a smooth map of the entire manifold $\bP$ onto the base manifold
$M$, we have that $\pi'\equiv\pi_*:T\bP\to TM$ is a globally defined 
map from the entire tangent bundle $T\bP$ (over the bundle space
$\bP$) onto the tangent bundle $TM$.

If $x=\pi(p)$, then due to the fact that $x=\pi(p(t))$ for any
curve in $\bP$ such that $p(t)\in\inv\pi(x)$ and $p(0)=0$, we
conclude that $\pi'$ maps all vertical vectors into the zero vector
in $T_xM$, that is $\pi'(V_p\bP)=0$, and we have
$$
T_p\bP = H_p\bP \oplus V_p\bP, \qquad p\in \bP
$$
so that every $\bX\in T_p\bP$ can be written
$$
\bX = \bX_{\rm h} + \bX_{\rm v}, \qquad \bX_{\rm h}\in H_p\bP, \quad
\bX_{\rm v}\in V_p\bP.
$$
Therefore, if $\bX\in T_p\bP$ we get $\pi'(\bX) = \pi'(\bX_{\rm h})
= X \in T_x M$. $\bX_{\rm h}$ is then called {\it horizontal
lift\/} of $X\in T_x M$. An equivalent definition for a connection
on $\bP$ is given by

\proclaim Definition. A connection on the principal fibre bundle
$(\bP,M,\pi,G)$ is a mapping $\Gamma_p : T_x M \to T_p P$,
$x=\pi(p)$ such that 
\begin{enumerate}
\item $\Gamma_p$ is linear
\item $\pi' \circ \Gamma_p = {\rm Id}_{T_x M}$, where
${\rm Id}_{T_x M}$ is the identity mapping in $T_x M$,
and $\pi'$ is the differential of the
canonical  projection mapping $\pi : \bP \to M$
\item the mapping $p \mapsto \Gamma_p$ is differentiable
\item $\Gamma_{R_{g}p} = R'_g\Gamma_p$, $g \in G$ and $R_g$ being the
right translation in $(\bP,\pi,M,G)$. 
\end{enumerate}

\proclaim Definition. Let $C : \R \supset I \to M$,
$t \mapsto C(t)$, with $x_0=C(0)\in M$ be a curve in $M$ and let
$p_0\in \bP$ be such that $\pi(p_0)=x_0$. The {\it parallel transport\/}
of $p_0$ along $C$ is given by the curve
$\bC : \R\supset I \to \bP$, $t \mapsto \bC(t)$
defined by
$$
\frac{d}{dt} \bC(t) = \Gamma_p \frac{d}{dt} C(t)
$$
with $\bC(0)=p_0$, $\bC(t)=p_{\scriptscriptstyle\parallel}$,
$\pi(p_{\scriptscriptstyle\parallel})=x=C(t)$. 

\medbreak
We need now to know more about the nature of the vertical space
$V_p\bP$. For this, let $\hat X\in T_eG = {\got G}$ be an element of
the Lie algebra of $G$ and let $f: G \supset U_e \to \R$,
where $U_e$ is some neighborhood of the identity element of ${\got G}$.
The vector $\hat\bX$ can be viewed as the tangent to the curve
produced by the exponential map
$$
\hat\bX(f) = \frac{d}{dt} f(\exp(\hat\bX t)) |_{t=0}
$$
Then to every $u\in\bP$ we can attach to each $\hat\bX\in T_e G$ a
unique element of $V_p\bP$ as follows: Let $\cF:\bP\to\R$ be given by
$$
\hat\bX_{\rm v}(p) (\cF) = \frac{d}{dt} \cF(p \exp(\hat\bX t)) |_{t=0}
$$
By this construction we have attached to each $\hat\bX\in T_eG$ a
unique global section of $T\bP$, called fundamental field
corresponding to this element. We then have the canonical
isomorphism
$$
\hat\bX_{\rm v}(p) \leftrightarrow \hat\bX, \qquad
\hat\bX_{\rm v}(p)\in V_p\bP, \quad \hat\bX\in T_e G
$$
and we have
$$
V_p\bP \simeq {\got G}
$$
It follows that another equivalent definition for a connection is:

\proclaim Definition. A connection on $(\bP,M,\pi,G)$ is a
1-form field $\bomega$ on $\bP$ with values in the Lie algebra
${\got G}$ such that, for each $p\in\bP$,
\begin{enumerate}
\item $\bomega_p(\bX_{\rm v}) = \hat\bX$, $\bX_{\rm v}\in V_p\bP$
and $\hat\bX\in {\got G}$ are related by the canonical isomorphism
\item $\bomega_p$ depends differentially on $p$
\item $\bomega_{\tilde R_g p}(\tilde R'_g \bX) =
(\Ad_{g^{-1}}\omega_p)(\bX)$
\end{enumerate}

\noindent It follows that if $\{\cG_a\}$ is a basis of ${\got G}$
and $\{\theta^i\}$ is a basis of $T^*_p\bP$, we can write $\bomega$ as
\begin{equation}
\bomega_p = \bomega^a \otimes \cG_a = \bomega^a_i \theta^i \otimes \cG_a
\end{equation}
where $\bomega^a$ are 1-forms on $\bP$.

The horizontal spaces $H_p\bP$ can then be defined by
$$
H_p\bP = \mathop{\rm ker}(\bomega_p)
$$
and we can verify that this is equivalent to the definition of
$H_p\bP$ given in the first definition of a connection.

Now, for a given connection $\bomega$, we can associate with each
differentiable local section of $\inv\pi(U)\subset\bP$, $U\subset
M$, a 1-form with values in ${\got G}$. Indeed, let
$$
f : M \supset U \to \inv\pi(U) \subset \bP \qquad
\pi \circ f = {\rm Id}_M
$$
be a local section of $\bP$. We define the 1-form $f^*\bomega$ on
$U$ with values in $\got G$ by the pull-back of $\bomega$ by $f$.
If $X\in T_xM$, $x\in U$,
$$
(f^*\bomega)_x(X) = \bomega_{f(x)}(f'X)
$$
Conversely, we have:

\proclaim Theorem. Given $\bomega\in TM\otimes {\got G}$ and a
differentiable section of $\inv\pi(U)$, $U\subset M$, there exists
one and only one connection $\omega$ on $\inv\pi(U)$ such that
$f^*\bomega = \omega$.

\noindent It is important to keep in mind also the following
result:

\proclaim Theorem. On each principal fibre bundle
with paracompact base manifold there exists infinitely many
connections.

As it is well known, each local section $f$ determines a local
trivialization
$$
\Phi : \inv\pi(U) \to U \times G
$$
of $\pi:\bP \to M$ by setting $\inv\Phi(x,g) = f(x) g$. Conversely,
$\Phi$ determines $f$, since $f(x)=\inv\Phi(x,e)$, where $e$ is the
identity of $G$. We shall also need the following

\proclaim Proposition. Let be given a local trivialization
$(U,\Phi)$, $\Phi:\inv\pi(U)\to U\times G$, and let $f:M\supset U
\to \bP$ be the local section associated to it. Then the connection
form can be written:
\begin{equation}
(\Phi^{-1*}\bomega)_{x,g} = \inv g d g + \inv g \omega g
\end{equation}
where $\bomega = f^*\omega \in TU\otimes {\got G}$. We usually
write, for abuse of notation, $\Phi^{-1*}\bomega\equiv \bomega$.
(The proof of this proposition is trivial.)

We can now determine the nature of ${\rm span}(H_p\bP)$. Using local
coordinates $\langle x^i\rangle$ for $U\subset M$ and $g_{ij}$ for
$U_e\in G$,\footnote{For simplicity, $G$ is supposed here to be a
matrix group. The $g_{ij}$ are then the elements of the matrix
representing the element $g\in G$.} we can write
$$
\bomega = \inv g_{ij} d g_{ij} + \inv g \omega g
$$
$$
\omega = \omega^A_\mu \cG_A dx^\mu = \omega^A \otimes \cG_A
\in T_xU\otimes {\got G}
$$
and
$$
[\cG_A,\cG_B] = f_{ABC} \cG_C
$$
with $f_{ABC}$ being the structure constants of the Lie algebra
${\got G}$ of the group $G$.

Recall now that $\dim H_p\bP=4$. Let its basis be
$$
\frac{\partial}{\partial x^\mu} + d_{\mu ij}
\frac{\partial}{\partial g_{ij}}
$$
$\mu=0,1,2,3$ and $i,j=1,\ldots,n=\dim G$. Since $H_p\bP={\rm
ker}(\bomega_p)$, we obtain, by writing
$$
\bX_{\rm h} = \beta^\mu\left(
\frac{\partial}{\partial x^\mu} + d_{\mu ij}
\frac{\partial}{\partial g_{ij}} \right)
$$
that
$$
d_{\mu ij} = - \omega^A_\mu \cG_{A i k} g_{kl}
$$
where $\cG_{A i k}$ are the matrix elements of $\cG_A$.

Consider now the vector bundle $E=\bP\times_{\rho(G)}F$ associated to
the PFB $(\bP,M,\pi,G)$ through the linear representation $\rho$ of
$G$ in the vector space $F$.  Consider the local trivialization
$\fii_\al : \pi^{-1}(U_\al) \to U_\al \times G$
of $(\bP,M,\pi,G)$, $\fii_\al(p)=(\pi(p),
\stackrel{\scriptscriptstyle \Delta}{\fii}_\al(p))$ with
$\stackrel{\scriptscriptstyle \Delta}{\fii}_{\al,x}(p) :  
\pi^{-1}(x) \to G$, $x \in U_\al \in M$. Also, consider
the local trivialization
$\chi_\al :  \hbox{\boldmath$\pi$}^{-1}(U_\al) \to U_\al \times F$
of $E$ where  $\hbox{\boldmath$\pi$} : E \to M$ is the canonical
projection. We have $\chi_\al(y)
=(\hbox{\boldmath$\pi$}(Y),
\stackrel{\scriptscriptstyle \Delta}{\chi}_\al(y))$ with
$\stackrel{\scriptscriptstyle \Delta}{\chi}_{\al,x}(y) :
\hbox{\boldmath$\pi$}^{-1}(x) \to F$.
Then, for each $x\in U_{\al\beta}=U_\al \cap U_\beta$ we must have,
$$
\stackrel{\scriptscriptstyle \Delta}{\chi}_{\beta,x} \circ
\stackrel{\scriptscriptstyle \Delta}{\chi}_{\beta,x} ^{-1} =  
\rho(\stackrel{\scriptscriptstyle \Delta}{\fii}_{\beta,x} \circ
\stackrel{\scriptscriptstyle \Delta}{\fii}_{\al,x} ^{-1}) 
$$
We then have 

\proclaim Definition. The {\it parallel transport} of $v_0\in E$,
$\pi(v_0)=x_0$ along the curve
$C : \R \supset I \to M, x_0=C(0)$
from $x_0$ to $x=C(t)$ is the element
$v_{\scriptscriptstyle\parallel}\in E$ such that
\begin{enumerate}
\item $\pi(v_{\scriptscriptstyle\parallel})=x$
\item $\stackrel{\scriptscriptstyle \Delta}{\chi}_{\al,x}
(v_{\scriptscriptstyle\parallel}) =
\rho(\stackrel{\scriptscriptstyle \Delta}{\fii}_{\al,x}
(p_{\scriptscriptstyle\parallel}) \circ 
\stackrel{\scriptscriptstyle \Delta}{\fii}_{\al,x_0} ^{-1}(p_0))
\stackrel{\scriptscriptstyle \Delta}{\fii}_{\beta,x_0}(v_0)$ 
\end{enumerate}

\proclaim Definition. Let $X$ be a vector at $x_0\in M$ tangent
to the curve $C : t \mapsto C(t)$ on $M$, $x_0=C(0)$.
The {\it covariant derivative\/} of $X \in \sec E$ in the direction of
$V$ at $x_0$ is $(\nabla_V X)_{X_0} \in\sec E$ such that
\begin{equation}
(\nabla_V X)(x_0) \equiv (\nabla_V X)_{x_0} = \lim_{t\to 0} 
\frac{1}{t} (X^0_{{\scriptscriptstyle\parallel},t} - X_0)
\end{equation}
where $X^0_{{\scriptscriptstyle\parallel},t}$ is the
``vector" $X_t \equiv X(x(t))$ of a section
$X \in \sec E$ parallel transported along $C$ from $x(t)$ to $x_0$,
the unique requirement on $C$ being 
${\dis\frac{d}{dt}} C(t)\biggl|_{t=0} = V$.

In the local trivialization $(U_\al, \chi_\al)$ of $E$ we have,
\begin{equation} \label{4.4}
\stackrel{\scriptscriptstyle \Delta}{\chi}_\al
(X^0_{{\scriptscriptstyle\parallel},t}) = \rho(g_0 g^{-1}_t)
\stackrel{\scriptscriptstyle \Delta}{\chi}_{\al,x(t)}(X_t)
\end{equation}

From this last definition it is trivial to calculate the covariant
derivative of $A \in \sec \Cl(\cM)$ in the direction of $V$.
Indeed, since a {\sl spin manifold\/} for $M$ is (Section~3)
$\Cl(\cM) = P_{{\SO_+(1,3)}} \times_{\Ad'} \cl_{1,3}=
P_{{\Spin_+(1,3)}} \times_{\Ad} \cl_{1,3}$, $g_0, g^{-1}_t \in
\Spin_+(1,3)$ and $\rho$ is the adjoint representation of
$\Spin_+(1,3)$ in $\cl_{1,3}$, we can verify (just take into
account that our bundle is trivial and put $g_0=1$ for simplicity) that
that we can write
\begin{equation}
A^0_{{\scriptscriptstyle\parallel},t}
= g^{-1}_t A_t \, g_t \qquad g_t=g(x(t)) \in \Spin_+(1,3)
\end{equation}
Then,
\begin{equation}
(\nabla_V A)(x_0) = \lim_{t\to 0} \frac{1}{t}(g^{-1}_t A_t g_t-A_0)
\end{equation}

Now, as we observed in Section~2, each $g \in \Spin_+(1,3)$ is of the
form $\pm {\rm e}^{F(x)}$, where
$F\in\sec\bigw^2(T^* M)\subset \sec \Cl(\cM)$, and $F$ can be
chosen in such a way to have a positive sign in this expression,
except in the particular case where $F^{2} = 0$ and $R = -{\rm
e}^{F}$. We then  
write,\footnote{The negative sign in the
definition of $\om$ is only for convenience, in order to obtain
formulas in agreement with known results.}
\begin{equation} \label{4.7}
g_t = e^{-1/2\om t}
\end{equation}
and
\begin{equation} \label{4.8}
\om=-2g'_t g^{-1}_t|_{t=0}
\end{equation}

Using Eq.~\ref{4.8} in Eq.~\ref{4.7} gives
\begin{equation} \label{4.9}
(\nabla_V A) (x_0) = \left\{ \frac{d}{dt} A_t + \frac{1}{2} [\om,
A_t]\right\}|_{t=0}
\end{equation}

Now let $\bil{x^\mu}$ be a coordinate chart for $U \subset M, e_a =
h^\mu_a \partial_\mu, a = 0,1,2,3$ an orthonormal basis for $T U
\subset TM$.\footnote{Since $M$ is a spin manifold,
$P_{SO_+(1,3)}(\cM)$ is trivial and $\{e_a\}$, $a= 0,1,2,3$ can be
taken as a global tetrad field for the tangent bundle.}
Let $\g^a \in \sec (T^* M) \subset \sec  \Cl(\cM)$ be
the dual basis of $\{e_a\} \equiv \cB$. Let $\Sigma = \{\g^a\}$ and
$\{\g_a, a = 0,1,2,3\}$ the reciprocal basis of $\{\g^a\}$, i.e., 
\ $\g^a \cdot \g_b=\delta^a_b$ where $\cdot$ is the internal product in
$\cl_{1,3}$. We have $\g^a = h^a_\mu  dx^\mu \ , \ \g_a = h^\mu_a
\eta_{\mu \al} d x^\al$. 
\begin{eqnarray}
&& \nabla_{\partial_\mu}  \partial_\nu
= \Gamma^\al_{\mu\nu} \partial_\al, \qquad
\nabla_{\partial_\mu} (d x^\al)
= - \Gamma^\al_{\mu \beta} (d x^\beta) \label{4.10}\\
&& \nabla_{e_a}e_b  = \om^c_{ab} e_c, \qquad 
\nabla_{e_a} \g^b = - \om^b_{ac} \g^c, \qquad
\nabla_{e_a} \g_b = \om^c_{ab} \g_c \label{4.11}\\
&& \nabla_{e_\mu} e_b =  \om^c_{\mu b} e_c, \qquad
\nabla_\mu \g^b = - \om^b_{\mu c} \g^c,  \qquad
\nabla_\mu \g_b = \om^c_{\mu b} \g_c
\end{eqnarray}

From Eq.~\ref{4.10} we easily obtain $(\nabla_{\partial_\mu} \equiv
\nabla_\mu)$
\begin{equation}
(\nabla_\mu A) = \partial_\mu A
+ {\textstyle\frac{1}{2}} [\om_\mu, A]
\end{equation}
with 
\begin{equation}
\om_\mu = - 2 (\partial_\mu g) g^{-1} \in \sec\bigw^2 (T^* M) \subset
\sec \Cl(\cM)
\end{equation}
where $g \in \sec \Cl^+(\cM)$ is such that $g|_{c(t)} \equiv g_t \in
\Spin_+(1,3)$.

We observe that our formulas, Eq.~\ref{4.10} and Eq.~\ref{4.11} for the
covariant derivative of an homogeneous Clifford field preserves (as
it must be), 
its graduation, i.e., if $A_p \in \sec \bigw^p(T^* M) \subset \sec
\Cl(\cM), p = 0,1,2,3,4,$ then $[\om_\mu, A_p] \in \sec \bigw^p (T^* M)
\subset \sec \Cl(\cM)$ as can be easily verified.

Since
\begin{equation}
{\textstyle\frac{1}{2}} [\om_\mu, \g^\al] = \om_\mu \cdot \g^\al
= - \g^\al \cdot \om_\mu
\end{equation}
we have
\begin{equation}
\om_\mu = {\textstyle\frac{1}{2}} \om^{ab}_{\mu} ( \g_a \w \g_b)
\end{equation}
and we observe that
\begin{equation}
\om^{ab}_{\mu} = - \om^{b a}_\mu  
\end{equation}

For $A = A_a \g^a$ we immediately obtain
\begin{equation}
\nabla_{e_a}A_b = e_a(A_b) - \om^c_{ab}A_c
\end{equation}
which agrees with the well known formula for the derivative of a
covariant vector field.

Also we have
\begin{eqnarray}
\nabla_\mu A_a &=& \partial_\mu(A_a) - \om^{\ b}_{\mu a} A_b\nonumber\\
\nabla_\mu A_\al &=& \partial_\mu(A_\al)
- \Gamma^{\beta}_{\mu\al} A_\beta
\end{eqnarray}

From the general formula~\ref{4.9} it follows immediately the

\proclaim Proposition. The covariant derivative $\nabla_{X}$
on $\Cl(\cM)$ acts as a derivation on the algebra of sections,
i.e., for $A,B\in\sec\Cl(\cM)$ it holds
\begin{equation}
\nabla_X (AB) = (\nabla_X A) B + A (\nabla_X B)
\end{equation}

\noindent The proof is trivial.

\subsection{The Lie Derivative of Clifford Fields}

Let $V \in \sec TM$ be a vector field on $M$ which induces a local
one-parameter  transformation group $t \mapsto \fii_t$. It $\fii_{*
t}$ stands as usual to the natural extension of the tangent map $d
\fii_t$ to tensor fields, the Lie derivative $\Lie_V$ of a given tensor
field $X \in \sec T M$ is defined by
\begin{equation} \label{4.21}
(\Lie_V X) (x) = \lim_{t\to 0} \frac{1}{t}
(X_x - (\fii_{*t}(x))_x)
\end{equation}
$\Lie_V$ is a derivation in the tensor algebra $\tau \cM$. Then, we have
for $a, b \in \sec \bigw^1(T^*\cM) \subset \Cl(\cM)$.
\begin{equation}
\Lie_V( a \otimes b + b \otimes a - 2 g^{-1}(a,b)) = (\Lie_V a) \otimes
b + b \otimes (\Lie_V a) - 2 \Lie_V(g^{-1}(a,b))
\end{equation}

Since $a \otimes b + b \otimes a - 2 g^{-1}(a,b)$ belongs to $J(\cM)$,
the bilateral ideal generating the Clifford bundle $\Cl(\cM)$ we see
from Eq.~\ref{4.21} that $\Lie_V$ preservers $J(\cM)$ if and only if
$\Lie_V g =0$, i.e., $V$ induces a local isometry group and then $V$ is
a Killing vector \cite{REF-23}.

\subsection{The Covariant Derivative of Algebraic Dirac Spinor Fields}

As discussed in Section~3 ADSF are sections of the Real Spinor Bundle
$I(\cM) = P_{\Spin_+ (1,3)}(M) \times_\ell I$ where $I = \Cl_{1,3}
\frac{1}{2} (1+E_0)$. $I (\cM)$ is a sub-bundle of the Spin-Clifford
bundle $\Cl_{\Spin_+(1,3)} (\cM)$. Since both $I(\cM)$ and 
$\Cl_{\Spin_+(1,3)} (\cM)$ are vector bundles, the covariant
derivatives of a ADSF or a DHSF can be immediately calculated using
the general method discussed in Section~4.1. 

Before we calculate the covariant spinor derivative $\nabla^s_V$
of a section of $I(\cM)$ [or $\Cl_{\Spin_+(1,3)}(\cM)$] where $V \in
\sec TM$ is a vector field we must recall that $\nabla^s_V$ is a
module derivation \cite{REF-39}, i.e., if $X \in\sec \Cl (\cM)$ and 
$\fii \in \sec I(\cM)$ [or $\sec \Cl_{\Spin_+ (1,3)} (\cM)]$ then 
it holds:

\proclaim Proposition. Let $\nabla$ be the connection in
$\Cl(\cM)$ to which $\nabla^s$ is related. Then,
\begin{equation}
\nabla^s_V (X \fii) = (\nabla_V X) \fii + X(\nabla^s_V \fii) 
\end{equation}

\noindent
The proof of this proposition is trivial once we derive an explicit
formula to compute $\nabla^s_V(\fii)$, $\fii\in\sec I(\cM) \subset
\sec \Cl_{\Spin_{+}(1,3)}(\cM)$.

Let us now calculate the covariant derivative
$\nabla^s_v$ in the direction of $v$, a vector at
$x_0 \in M$ of $\phi \in\sec I(\cM) \subset
\sec \Cl_{\Spin_+(1,3)} (\cM)$.

Putting $g_0 = 1 \in \Spin_+ (1,3)$ we have using the general
procedure 
\begin{equation} \label{4.24}
\phi^0_{{\scriptscriptstyle\parallel},t} = g^{-1}_t \phi_t 
\end{equation}
where $\phi^0_{{\scriptscriptstyle\parallel},t}$ is the
``vector" $\phi_t = \phi(x(t))$ of a
section $\phi \in \sec I(\cM) \subset \sec \Cl_{\Spin_+(1,3)} (\cM)$
parallel transported along $C : \R \supset I \to M$, 
$t \mapsto C(t)$ from $x(t) \equiv C(t)$ to $x_0 = C(0)$,
${\displaystyle\frac{d}{dt}} C(t) \biggl|_{t=0} = v$

Putting as in Eq.~\ref{4.8} $g_t = e^{-1/2 wt}$, we get by using
Eq.~\ref{4.4} 
\begin{equation} \label{4.25}
(\nabla^s_v \phi) (x_0) = \left( \frac{d}{dt} \phi_t + \frac{1}{2}
\om \phi_t \right) \biggl|_{t=0}
\end{equation}
If $\{\g^a\}$ is an orthogonal field of 1-forms, $\g^a \in\sec
\bigw^1(T^*M) \subset \sec  \Cl(\cM)$ dual to the orthogonal frame
field $\{e_a\}$, $e_a \in\sec TM$, $g(e_a, e_b) = \eta_{ab}$ and if
$\{\g_a\}$ is the reciprocal frame of $\{\g^a\}$, i.e., $\g^a \cdot \g_b
= \delta^a_b$  ($a, b = 0, 1, 2, 3$) then for Eq.~\ref{4.25} we get 
\begin{equation} \label{4.26}
\nabla^s_{e_a} \phi = e_a (\phi)
+ {\textstyle\frac{1}{2}} \om_a \phi
\end{equation}
with 
\begin{equation}
\om_a = {\textstyle\frac{1}{2}} \om^{bc}_a \g_b \w \g_c
\end{equation}
and we recognize the 1-forms $\omega_a$
as being $\omega_a=\omega(e_a)$ where $\omega=f^*\bomega$,
$f:M\to U\times G$ is the global section used to write
Eq.~\ref{4.24}. The Lie algebra of $\Spin_{+}(1,3)$ is, of course,
generated by the ``vectors'' $\{\g_a\w\g_b\}$.
\begin{equation}
\nabla_{e_a} \g^b = -\om^{bc}_a \g_c
\end{equation}
If $\dgr{x^\mu}$ is a coordinate chart for $U \subset M$ and $\g^a =
h^a_\mu dx^\mu$, $a,\mu = 0,1,2,3$, we also obtain 
\begin{equation}
\nabla^s_\mu \phi = \partial_\mu (\phi)
+ {\textstyle\frac{1}{2}} \om_\mu \phi,
\qquad
\om_\mu = {\textstyle\frac{1}{2}} \om^{bc}_\mu \g_b \w \g_c
\end{equation}
Now, since $\phi \in\sec I(\cM) \subset  \sec  \Cl_{\Spin_+(1,3)} (\cM)$
is such that $\phi e_{\scriptscriptstyle \Sigma} = \phi$ with
$e_{\scriptscriptstyle \Sigma} =  
\frac{1}{2} (1+\g^0)$ it follows from $\nabla^s_{e_a}\phi
=\nabla^s_{e_a} (\phi e_{\scriptscriptstyle \Sigma})$ that 
\begin{equation}
e_{\scriptscriptstyle \Sigma} \nabla^s_{e_a}
e_{\scriptscriptstyle \Sigma} = 0
\end{equation}
Now, recalling Eq.~\ref{2.30} we have a spinorial basis for
$I(\cM)$ given
by $\beta^s = \{s^A\}$, $A = 1, 2, 3, 4$, $s^A \in\sec I(\cM)$ with 
\begin{equation}
s^1 = e_{\scriptscriptstyle \Sigma} = {\textstyle\frac{1}{2}} (1+\g^0), 
\quad s^2 = - \g^1 \g^3 e_{\scriptscriptstyle \Sigma}, \quad 
s^3 = \g^3 \g^0 e_{\scriptscriptstyle \Sigma}, \quad 
s^4 = \g^1 \g^0 e_{\scriptscriptstyle \Sigma} .  
\end{equation}
Then as we learn in Section~2, $\phi = \phi_A s^A$ where $\phi_A$ are
formally complex numbers. Then 
\begin{eqnarray}
\nabla^s_{e_a} \phi & = & e_a (\phi)
+ {\textstyle\frac{1}{2}} \om_a \phi \nonumber \\
& = & \left[e_a (\phi_A)
+ {\textstyle\frac{1}{2}} \om_a \phi_A\right]s^A \nonumber \\
& = & \left(e_a (\phi_A)
+ {\textstyle\frac{1}{2}} [\om_a]^B_A \phi_B\right)s^A 
\label{4.32}
\end{eqnarray}
with
\begin{equation}
\om_a s^A = [\om_a]^A_B s^B
\end{equation}
\begin{eqnarray} \label{4.34}
\nabla^s_{e_a} \phi & = & \nabla^s_{e_a} (\phi_A s^A) \nonumber \\
& = & e_a (\phi_A)s^A + \phi_A \nabla^s_{e_a} s^A
\end{eqnarray}
From Eq.~\ref{4.32} and Eq.~\ref{4.34} it follows that 
\begin{equation}
\nabla^s_{e_a} s^A = {\textstyle\frac{1}{2}} [\om_a]^A_B s^B
\end{equation}
We introduce the dual space $I^*(\cM)$ of $I(M)$ where $I^*(M) = 
P_{\Spin_+(1,3)}(\cM) \times_r I$ where here the action of	
$\Spin_+(1,3)$ on the typical fiber is on 
the right. A basis for $I^*(\cM)$ is then $\rho_s = \{s_A\}$,
$A = 1, 2, 3, 4$, $s_A \in \sec I^*(\cM)$ such that 
\begin{equation}
s_A (s^B) = \delta^B_A
\end{equation}
A simple calculation shows that 
\begin{equation}
\nabla^s_{e_a} s_A = - \frac{1}{2} [\om_a]^B_A s_B 
\end{equation}
Since $\Cl(\cM) = I^*(\cM) \otimes I(\cM)$ (the ``tensor-spinor
space'')
is spanned by the basis $\{s^A \otimes s_B\}$ we can write 
\begin{equation}
\g_a s_A = [\g_a]^B_A s_B
\end{equation}
with 
\begin{equation}
[\g_a]^B_A = \g^B_{aA} \equiv \g_a (s^B, s_A)
\end{equation}
being the matricial representation of $\g_a$. It follows that 
\begin{equation}
\nabla^s_{e_b} \g_a (s^B, s_A) = e_b ([\g_a]^B_A) - \om^c_{ba} \g^B_{cA}
+ \frac{1}{2} \om^B_{b C} \g^C_{bA} - \frac{1}{2} \om^C_{bA} \g^B_{aC}
\end{equation}
Now, 
\begin{equation} \label{4.41}
(\frac{1}{2} \om^B_{bc} \g^C_{aA}
- \frac{1}{2} \om^C_{bA} \g^B_{aC}) s^A = (\g_a \cdot \om_b) s^B
\end{equation}
and from $\om_b = \frac{1}{2} \om^{cd}_b \g_c \w \g_d$, we get 
\begin{equation} \label{4.42}
(\g_a \cdot \om_b)s^B = (-\om^c_{ba} \g^B_{cA})s^A
\end{equation}
From Eq.~\ref{4.41} and Eq.~\ref{4.42} we obtain 
\begin{equation} \label{4.43}
{\textstyle\frac{1}{2}} \om^B_{bC} \g^C_{aA}
- {\textstyle\frac{1}{2}} \om^C_{bA} \g^B_{aC}
= - \om^d_{ba} \g^B_{dA}
\end{equation}
and then 
\begin{equation} \label{4.44}
\nabla^s_{e_b} [\g_a]^B_A = e_b ([\g_a]^B_A) = 0
\end{equation}
since according to a result obtained in Section~2.6 $[\g_a]^B_C$ are
constant matrices.
Eq.~\ref{4.43} agrees with the result presented, e.g., in~\cite{REF-23}.
Also from $\om_a = \frac{1}{2} \om^{bc}_a \g_b \w \g_c$
it follows 
\begin{equation}
\om^A_{aB} = {\textstyle\frac{1}{2}} \om^{bc}_a [\g_b, \g_c]^A_B  . 
\end{equation}

We can also easily obtain the following results: Writing 
\begin{equation}
\nabla^s_{e_a}\phi \equiv (\nabla^s_{e_a} \phi_A)s^A
\end{equation}
it follows that
\begin{equation}
\nabla^s_{e_a} \phi_A = e_a (\phi_A) + \frac{1}{8} \om^b_{ac} [\g_b,
\g^c]^B_A \phi_B
\end{equation}
and
\begin{equation} \label{4.48}
\nabla^s_{e_a} \phi^A = e_a (\phi^A) - \frac{1}{8}\om^b_{ac} [\g_b,
\g^c]^A_B \phi^B
\end{equation}
Eq.~\ref{4.48} agrees exactly with the result presented, e.g., by 
Choquet-Bruhat et al \shortcite{REF-23} for the components of the
covariant  derivative of a CDSF $\psi \in \sec P_{\Spin_+(1,3)}(\cM) 
\times_\rho\C^4$. It is important to emphasize here that the condition
given by Eq.~\ref{4.44},
namely $\nabla^s_{e_b} [\g_a]^B_A = 0$ holds true but this does not
imply that $\nabla_{e_b} \g^a = 0$, i.e., $\nabla$ need not be the so
called connection of parallelization of the 
$\cM = \langle M, g, \nabla\rangle$, which as well known has zero
curvature but non zero torsion \cite{REF-44}.

The main difference between $\nabla^s$ acting on sections of
$I(\cM)$ or of $\Cl_{\Spin_+(1,3)} (\cM)$ and $\nabla$ acting on
sections of $\Cl (\cM)$ is that, for $\phi \in \sec I(\cM) {\rm or}
\sec \Cl_{\Spin_+(1,3)}(\cM)$ and $A \in \sec \Cl(\cM)$, we must have
\begin{equation}
\nabla^s_{e_a} (A \phi) = (\nabla_{e_a} A) \phi + A 
(\nabla^s_{e_a} \phi)  , 
\end{equation}
and of course $\nabla$ cannot be applied to sections of $I(\cM)$ or
of $\Cl_{\Spin_+(1,3)} (\cM)$.

\subsection{The Representative of the Covariant Derivative of a
Dirac-Hestenes Spinor Field in $\Cl (\cM)$}

In Section~3.2 we defined a DHSF $\psi$ as an even section of
$\Cl_{\Spin_+ (1,3)} (\cM)$. Then, by the same procedure used in
Section~4.3 we get\footnote{The meaning of $e_a, \g^b$, etc.\ is as
before.}  
\begin{equation} \label{4.50}
\nabla^s_{e_a} \psi = e_a (\psi)
+ {\textstyle\frac{1}{2}} \om_a \psi \qquad
\nabla^s_{e_a} \tilde{\psi} = e_a (\tilde{\psi})
- {\textstyle\frac{1}{2}} \tilde{\psi} \om_a
\end{equation}
and as before
\begin{equation}
\om_a = {\textstyle\frac{1}{2}} \om^{bc}_a \g_b \w \g_c
\in \sec \Cl (\cM)
\end{equation}

Now, let $\bgamma^a \in \sec \Cl_{\Spin_+ (1,3)} (\cM)$ 
such that
$\bgamma^a  \bgamma^b + \bgamma^b \bgamma^a = 2 \eta^{ab}$,
$(a,b = 0,1,2,3)$, and let us
calculate $\nabla^{s}_{e_a} (\psi \bgamma^b)$. Using
Eq.~\ref{4.26} we have,
\begin{equation} \label{4.52}
\nabla^s_{e_a} (\psi \bgamma^b) = e_a (\psi \bgamma^b)
+ {\textstyle\frac{1}{2}} \om_a \psi
\bgamma^b = (\nabla^s_{e_a} \psi) \bgamma^b
\end{equation}
On the other hand, 
\begin{equation} \label{4.53}
\nabla^s_{e_a} (\psi \bgamma^b) = (\nabla^s_{e_a} \psi) 
\bgamma^b + \psi (\nabla^s_{e_a} \bgamma^b)
\end{equation}
Comparison of Eq.~\ref{4.52} and Eq.~\ref{4.53} implies that 
\begin{equation} \label{4.54}
\nabla^s_{e_a} \bgamma^b = 0
\end{equation}
The matrix version of Eq.~\ref{4.54} is Eq.~\ref{4.44}.

We know that if $\psi, \tilde{\psi} \in \sec  \Cl^+_{\Spin_+(1,3)}
(\cM)$ then $\psi \bgamma^a \tilde{\psi} = 
\bX^a$ is such that $\bX^a (x) \in \R^{1,3}\  \forall x \in M$. Then,
\begin{equation} \label{4.55}
\nabla^s_{e_a} (\psi \bgamma^b \tilde{\psi})
= (\nabla^s_{e_a} \psi) \bgamma^b \tilde{\psi}
+ \psi \bgamma^b (\nabla^s_{e_a} \tilde{\psi}) 
\end{equation}
and $\nabla^s_{e_a} (\psi \bgamma^b \tilde{\psi})(x) \in \R^{1,3}, \
\forall x \in M$.

We are now prepared to find the representative of the covariant
derivative of a DHSF in $\Cl (\cM)$. We recall that $\psi$ is an
equivalence class of even sections of $\Cl(\cM)$ such that in the
basis $\Sigma = \{\g^a\}, \g^a \in \sec \bigw^1(T^* M) \subset \sec 
\Cl(\cM)$ the representative of $\psi$ is $\psi_{\scriptscriptstyle
\Sigma} \in \Cl^+ (\cM)$ 
and the representative of $\bX^a$ is $X^a \in \sec
\bigw^1(T^*M)\subset \sec 
 \Cl(\cM)$ such that 
\begin{equation}
X^a = \psi_{\scriptscriptstyle \Sigma} \g^a
\tilde{\psi}_{\scriptscriptstyle \Sigma} 
\end{equation}
Let $\nabla$ be the connection acting on sections of $\Cl (\cM)$. Then, 
\begin{eqnarray}
&& \nabla_{e_a} (\psi_{\scriptscriptstyle \Sigma} \g^b
\tilde{\psi}_{\scriptscriptstyle \Sigma}) = \left\{e_a 
(\psi_{\scriptscriptstyle \Sigma}) + 
\frac{1}{2} [\om_a, \psi_{\scriptscriptstyle \Sigma}]\right\}\g^b
\tilde{\psi}_{\scriptscriptstyle \Sigma} \nonumber \\ 
&&+ \psi_{\scriptscriptstyle \Sigma} (\nabla_{e_a} \g^b)
\tilde{\psi}_{\scriptscriptstyle \Sigma} + \psi_{\scriptscriptstyle
\Sigma} \g^b 
\left\{e_a (\psi_{\scriptscriptstyle \Sigma}) + \frac{1}{2} [\om_a,
\tilde{\psi}_{\scriptscriptstyle \Sigma}]\right\} = \nonumber \\ 
&& = \left[e_a (\psi_{\scriptscriptstyle \Sigma}) + \frac{1}{2} \om_a
\psi_{\scriptscriptstyle \Sigma}\right] 
\g^b \psi_{\scriptscriptstyle \Sigma} + 
\psi_{\scriptscriptstyle \Sigma} \g^b \left[e_a
(\psi_{\scriptscriptstyle \Sigma}) - \frac{1}{2} 
\tilde{\psi}_{\scriptscriptstyle \Sigma}\om_a\right]. \label{4.57}
\end{eqnarray}

Comparing Eq.~\ref{4.55} and Eq.~\ref{4.57} we see that the following
definition suggests by itself

\proclaim Definition. 
\begin{eqnarray}
&& (\nabla^s_{e_a} \psi)_{\scriptscriptstyle \Sigma} \equiv
\nabla^s_{e_a} \psi_{\scriptscriptstyle \Sigma} = e_a 
(\psi_{\scriptscriptstyle \Sigma}) + \frac{1}{2} \om_a
\psi_{\scriptscriptstyle \Sigma}	\nonumber \\ 
&& (\nabla^s_{e_a} \tilde{\psi})_{\scriptscriptstyle \Sigma} \equiv
\nabla^s_{e_a} 
\tilde{\psi}_{\scriptscriptstyle \Sigma} = e_a
(\tilde{\psi}_{\scriptscriptstyle \Sigma}) - \frac{1}{2} 
\tilde{\psi}_{\scriptscriptstyle \Sigma} \om_a \\
&& (\nabla^s_{e_a} \bgamma^b)_{\scriptscriptstyle \Sigma}
\equiv \nabla^s_{e_a} \g^b = 0 \nonumber 
\end{eqnarray}
where $(\nabla^s_{e_a} \psi)_{\scriptscriptstyle \Sigma},
(\nabla^s_{e_a} \tilde{\psi})_{\scriptscriptstyle \Sigma}, 
(\nabla^s_{e_a} \bgamma^b)_{\scriptscriptstyle\Sigma}
\in \sec \Cl(\cM)$ are representatives of
$\nabla^s_{e_a}\psi$ (etc...) in the basis $\Sigma$ in $\Cl(\cM)$

Observe that the result $\nabla^s_{e_a} \g^b = 0$ is compatible with
the result $\nabla^s_{e_a} [\g_a]^B_A = 0$ obtained in Eq.~\ref{4.43}
and is an important result in order to write the Dirac-Hestenes equation
(Section~6).

\section{The Form Derivative of the Manifold and the Dirac and
Spin-Dirac Operators} 

Let $\cM = \dgr{M, g, \nabla}$ be a Riemann-Cartan manifold
(Section~4), and let $\Cl(\cM), I(\cM)$ and $\Cl_{\Spin_+(1,3)}(\cM)$ be
respectively the Clifford, Real Spinor and Spin Clifford bundles.
Let $\nabla^s$ be the spinorial connection acting on sections of
$I(\cM)$ or $\Cl_{\Spin_+(1,3)}(\cM)$. Let also $\{e_a\}, \{\g^a\}$
with the same meaning as before and for convenience when useful we
shall denote the Pfaff derivative by $\partial_a \equiv e_a$ . 

\proclaim Definition. Let $\Gamma$ be a section of $\Cl (\cM),
I(\cM)$ or  
$\Cl_{\Spin_+(1,3)}(\cM)$.  The form derivative of the manifold is a
canonical first order differential operator $\partial : \Gamma 
\mapsto \Gamma$ such that 
\begin{eqnarray}
\partial \Gamma & = & (\g^a \partial_a)\Gamma \nonumber \\
& = & \g^a \cdot (\partial_a (\Gamma)) + \g^a \w (\partial_a (\Gamma)) 
\end{eqnarray}
for $\g^a \in \sec \ \Cl(\cM)$ . 

\proclaim Definition. The Dirac operator acting on sections of
$\Cl(\cM)$ is a canonical first order differential operator
$\bpartial : A \mapsto \bpartial A$, $A \in \sec  \Cl(\cM)$, such that 
\begin{equation}
\bpartial A = (\g^a \nabla_{e_a})A 
= \g^a \cdot (\nabla_{e_a} A) + \g^a \w (\nabla_{e_a} A) 
\end{equation}

\proclaim Definition. The Spin-Dirac
operator\footnote{In~\cite{REF-39} this 
operator (acting on sections of $I(\cM)$) is called simply Dirac
operator, being the generalization of the operator originally
introduced by Dirac. See also~\cite{REF-11} for comments on the use
of this terminology.} acting on sections of $I(\cM)$ or
$\Cl_{\Spin_+(1,3)} (\cM)$ is a canonical first order differential
operator $\bD : \Gamma \to \bD\Gamma$ ($\Gamma \in \sec \ I(\cM))$
[or $\Gamma \in \sec  \Cl_{\Spin_+(1,3)}(\cM)]$ such that 
\begin{eqnarray}
\bD\Gamma & = & (\g^a \nabla^s_{e_a})\Gamma \nonumber \\
& = & \g^a \cdot  (\nabla^s_{e_a} \Gamma)
+ \g^a \w (\nabla^s_{e_a} \Gamma)
\end{eqnarray}
The operator $\bpartial$ is sometimes called the Dirac-K\"ahler operator
when $\cM$ is a Lorentzian manifold \cite{REF-17}, 
i.e., $\bT(\nabla) = 0$, $\bR(\nabla) = 0$, where $\bT$ and $\bR$ are
respectively the torsion and Riemann tensors. In this case we can show
that
\begin{equation}
\bpartial = d - \delta
\end{equation}
where $d$ is the differential operator and $\delta$ the Hodge
codifferential operator. In the spirit of section 4, we use the
convention that the  representative of $\bD$ (acting on sections of 
$\Cl_{\Spin_+(1,3)}(\cM))$ in $\Cl(\cM)$ will be also denote by 
\begin{equation}
\bD = \g^a \nabla^s_{e_a}
\end{equation}

\section{The Dirac-Hestenes Equation in Minkowski Spacetime} 

Let $\cM = \dgr{M, g, \nabla}$ be the Minkowski spacetime,
$\Cl(\cM)$ be the Clifford bundle of $\cM$ with typical fiber
$\cl_{1,3}$, and let
$\Psi \in \sec P_{\Spin_+(1,3)}(\cM) \times_\rho \C^4$
(with $\rho$ the $D^{(1/2,0)} \oplus D^{(0, 1/2)}$ representation of
$\SL (2, \C) \simeq \Spin_+(1,3)$. Then,
the Dirac equation for the charged fermion field $\Psi$ in
interaction with the electromagnetic field $A$
is \cite{REF-30} ($\hbar = c =1$)
\begin{equation} \label{6.1}
\ubi\g^\mu (i \partial_\mu - e A_\mu) \Psi = m \Psi \;\; {\rm or} \;\; i
\bD \psi - \ubi\g^\mu A_\mu \Psi = m \Psi 
\end{equation}
where $\ubi\g^\mu \ubi\g^\nu + \ubi\g^\nu \ubi\g^\mu =
2\eta^{\mu\nu}, \ubi\g^\mu$ being the Dirac matrices given by
Eq.~\ref{2.31} and $A = A_\mu dx^\mu \in\sec \bigw^1 (T^*M)$ 

As showed, e.g., in~\cite{REF-7} this equation is equivalent to the
following equation satisfied by $\phi \in \sec I(\cM)$ $[\phi
e_{\scriptscriptstyle \Sigma} 
= \phi, e_{\scriptscriptstyle \Sigma} = \frac{1}{2} (1+\g^0),
\bgamma^\mu 
\bgamma^\nu + \bgamma^\nu \bgamma^\mu =  2 \eta^{\mu\nu},
\bgamma^\mu \in\sec \Cl_{\Spin_+(1,3)}(\cM)]$,
\begin{equation} \label{6.2}
\bD\phi \bgamma^2 \g^1 - e A \phi = m \phi , 
\end{equation}
where $\bD$ is the Dirac operator on $I(\cM)$ and
$A \in \sec \bigw^1 (T^*M) \subset \sec \Cl(\cM)$ . 

Since, as discussed in Section~3, each $\phi$ is an equivalence class
of sections of $\Cl(\cM)$ we can also write an equation equivalent
to Eq.~\ref{6.2} for $\phi_{\scriptscriptstyle \Sigma} =
\phi_{\scriptscriptstyle \Sigma} e_{\scriptscriptstyle \Sigma}$,
$\phi_{\scriptscriptstyle \Sigma}, 
e_{\scriptscriptstyle \Sigma} \in \sec \Cl(\cM)$,
$e_{\scriptscriptstyle \Sigma} = \frac{1}{2} 
(1+\g^0), \g^\mu \g^\nu + \g^\nu \g^\mu = 2\eta^{\mu\nu}, \g^\mu \in
\sec \Cl(\cM)$, and $\g^\mu = dx^\mu$ for the global coordinate
functions $\dgr{x^\mu}$.
In this case the Dirac operator $\bpartial=\g^\mu\nabla_\mu$ is
equal to the form derivative $\partial = \g^\mu\partial_\mu$
and we have
\begin{equation}
\partial \phi_{\scriptscriptstyle \Sigma} \g^2 \g^1 - e
A\phi_{\scriptscriptstyle \Sigma} = m \phi_{\scriptscriptstyle
\Sigma} \g^0 
\end{equation}
Since each $\phi_{\scriptscriptstyle \Sigma}$ can be written
$\phi_{\scriptscriptstyle \Sigma} = \psi_{\scriptscriptstyle \Sigma} 
e_{\scriptscriptstyle \Sigma}, (\psi_{\scriptscriptstyle \Sigma}
\in\sec \ \Cl^+(\cM) $ being the representative of 
a DHSF) and $\gamma^{0}e_{\Sigma} = e_{\Sigma}$, we can write the
following equation for $\psi_{\scriptscriptstyle \Sigma}$ that is 
equivalent to Dirac equation \cite{REF-7,REF-9,REF-10}
\begin{equation} \label{6.4}
\partial \psi_{\scriptscriptstyle \Sigma} \g^2 \g^1 - e A
\psi_{\scriptscriptstyle \Sigma}
= m \psi_{\scriptscriptstyle \Sigma} \g^0 
\end{equation}
which is the so called Dirac-Hestenes equation \cite{REF-2,REF-3}. 

Eq.\ref{6.4} is covariant under passive (and active) Lorentz
transformations, in the following sense: 
consider the change from the Lorentz frame $\Sigma = \{\g^\mu=dx^\mu\}$
to the frame $\dot\Sigma = \{\dot\g^\mu=d\dot{x}^\mu\}$ with
$\dot\g^\mu = R^{-1} \g^\mu R$ and $R \in \Spin_+ (1,3)$ being constant.
Then the representative of the Dirac-Hestenes spinor changes as
already discussed in Section~3 from
$\psi_{\scriptscriptstyle \Sigma}$ to $\psi_{\dot\Sigma} =
\psi_{\scriptscriptstyle \Sigma} R^{-1}$. Then we have 
$\partial = \g^\mu \partial_\mu = \dot\g^\mu \partial/\partial \dot
x^\mu$ where 
$\dgr{x^\mu}$ and $\dgr{\dot x^\mu}$ are related by a Lorentz
transformation and  
\begin{equation}
\partial \psi_{\scriptscriptstyle \Sigma} R^{-1}R \g^2 R^{-1}R \g^1
R^{-1} - e A 
\psi_{\scriptscriptstyle \Sigma} R^{-1} = m 
\psi_{\scriptscriptstyle \Sigma} R^{-1}R \g^0 R^{-1} ,
\end{equation}
i.e., 
\begin{equation}
\partial \psi_{\dot\Sigma} \dot\g^2 \dot\g^1 - e A \psi_{\dot\Sigma} = m
\psi_{\dot\Sigma} \dot\g^0
\end{equation}
Thus our definition of the Dirac-Hestenes spinor fields as an
equivalence class of even sections of $\Cl(\cM)$ solves directly the
question raised by Parra \shortcite{REF-45} concerning the covariance of
the Dirac-Hestenes equation. 

Observe that if $\nabla^s$ is the spinor covariant derivative acting on
$\psi_{\scriptscriptstyle \Sigma}$ (defined in Section~4.4) we can
write Eq.~\ref{6.4} in intrinsic form, i.e., without the need of
introducing a chart for $\cM$ as follows 
\begin{equation}
\g^a \nabla^s_a \psi_{\scriptscriptstyle \Sigma} \g^2 \g^1 - e A
\psi_{\scriptscriptstyle \Sigma} = m \psi_{\scriptscriptstyle \Sigma}
 \g^0 
\end{equation}
where $\g^a$ is now an orthogonal basis of $T^*M$, and not necessarily
it is $\g^a = dx^a$ for some coordinate functions $x^a$. 

It is well-known that Eq.~\ref{6.1} can be derived from the principle of
stationary action through variation of the following action 
\begin{equation}
S (\Psi) = \int d^4 x \cL
\end{equation}
\begin{equation}
\cL = - \frac{i}{2} (\g^\mu \partial_\mu \Psi^+) \Psi
+ \frac{i}{2} \Psi^+ (\ubi\g^\mu \partial_\mu \bar\Psi)
- m \Psi^+ \bar\Psi - e A_\mu \Psi^+ \ubi\g_\mu \bar\Psi
\end{equation}
with $\Psi^+ = \Psi^* \ubi\g^0$.

In the next section we shall present the rudiments of the multiform
derivative approach to Lagrangian field theory (MDALFT) developed
in~\cite{REF-24} --- see also~\cite{REF-46} -- and we apply this formalism
to obtain the 
Dirac-Hestenes equation on a Riemann-Cartan spacetime.

\section{Lagrangian Formalism for the Dirac-Hestenes Spinor Field on
a Riemann-Cartan Spacetime}

In this section we apply the concept of multiform (or multivector)
derivatives first 
introduced by Hestenes and Sobczyk \shortcite{REF-34} (HS) to present a
Lagrangian formalism for the Dirac-Hestenes spinor field DHSF on a
Riemann-Cartan spacetime. In Section~7.1 we briefly present our
version of the multiform derivative approach to Lagrangian field
theory for a Clifford field $\phi \in \sec \Cl(\cM)$ where $\cM$
is Minkowski spacetime. In Section~7.2 we present the theory for the
DHSF on Riemann-Cartan spacetime.

\subsection{Multiform Derivative Approach to Lagrangian Field Theory}

We define a Lagrangian density for $\phi \in \sec \Cl(\cM)$ as a
mapping 
\begin{equation} \label{7.1}
\cL : (x, \phi(x), \bpartial \w \phi(x), \bpartial \cdot \phi (x))
\mapsto \cL (x, \phi(x), \bpartial \w \phi(x),
\bpartial \cdot \phi (x))\in \bigw^4 (T^*M) \subset \Cl(\cM)
\end{equation}
where $\bpartial$ is the Dirac operator acting  on 
sections of\footnote{An example of a Lagrangian of the form given by
Eq.~\ref{7.1} appears, e.g., in the theory of the gravitational field
in Minkowski spacetime \cite{REF-47}.  In~\cite{REF-13a} we present
further mathematical results, derived in the Clifford bundle formalism.
Those results are important for the gravitational theory and other
field theories.} $\Cl(\cM)$, and by the above
notation we mean an arbitrary multiform function of $\phi$,
$\bpartial \wedge \phi$ and $\bpartial \cdot \phi$. 

In this section we shall perform our calculations using an
orthonormal and coordinate basis for the tangent (and cotangent)
bundle. If $\langle x^\mu\rangle$ is a global Lorentz chart, then
$\g^\mu=dx^\mu$ and  $\bpartial = \g^\mu\nabla_\mu =
\g^\mu\partial_\mu = \partial$, so that the Dirac operator
($\bpartial$) coincides with the form derivative ($\partial$)
of the manifold.

We introduce also for $\phi$ a Lagrangian $L(x, \phi(x), \partial \w
\phi(x), \partial \cdot \phi(x)) \in \bigw^0 (T^*M) \subset \Cl(\cM)$
by 
\begin{equation}
\cL (x, \phi (x), \partial \w \phi(x), \partial \cdot \phi (x))
= L(x, \phi(x), \partial \w \phi(x), \partial \cdot \phi (x)) \tau_g 
\end{equation}
where $\tau_g \subset\sec \bigw^4 (T^*M)$ is the volume form,
$\tau_g = dx^0 \w dx^1 \w dx^2 \w dx^3$ for $\dgr{x^\mu}$ a global
Lorentz chart.

In what follows we suppose that $\cL[L]$ does not depend explicitly of
$x$ and we write $L(\phi, \partial \w \phi, \partial \cdot  \phi)$
for the Lagrangian. Observe that 
\begin{equation}
L(\phi, \partial \w \phi, \partial \cdot \phi)
= \dgr{L(\phi, \partial \w \phi, \partial \cdot \phi)}_0 
\end{equation}
As usual, we define the action for $\phi$ as 
\begin{equation}
S(\phi) = \int_U L(\phi, \partial \w \phi, \partial \cdot \phi)\tau_g
\qquad  U \subseteq M
\end{equation}
The field equations for $\phi$ is obtained from the principle of
stationary action for $S(\phi)$. Let $\eta \in\sec \ \Cl(\cM)$
containing the same grades as $\phi \in \sec \ \Cl(\cM)$. We say
that $\phi$ is stationary with respect to $L$ if 
\begin{equation} \label{7.5}
\frac{d}{dt} S(\phi + t \eta)\biggl|_{t=0} = 0
\end{equation}
But, recalling HS \shortcite{REF-34} we see that Eq.~\ref{7.5} is just the
definition of the multiform derivative of $S(\phi)$ in the direction
of $\eta$, i.e., we have using the notation of HS
\begin{equation}
\eta \ast \partial_\phi S(\phi) = \frac{d}{dt} S(\phi + t\eta)
\biggl|_{t=0} 
\end{equation}
Then, 
\begin{equation} \label{7.7}
\frac{d}{dt} S(\phi+t\eta)\biggl|_{t=0} = \int \tau_g \frac{d}{dt} \{
L[(\phi + t\eta), \partial \w (\phi+t \eta), \partial \cdot (\phi +
t\eta)]\}\biggl|_{t=0}
\end{equation}
Now 
\begin{eqnarray}
&& \frac{d}{dt} \{[L(\phi+t\eta), \partial \w (\phi+t\eta),
\partial \cdot (\phi + t\eta)]\}_{t=0} \nonumber \\[1ex]
&&\qquad= \eta \ast \partial_\phi L
+ (\partial \w \eta) \ast \partial_{\partial \w \phi} L
+ (\partial \cdot \eta) \ast \partial_{\partial \cdot \phi} L 
\label{7.8}
\end{eqnarray}
Before we calculate (\ref{7.8}) for a general
$\phi \in\sec \ \Cl (\cM)$, let us suppose that
$\phi = \dgr{\phi}_r$, i.e., it is homogeneous.
Using the properties of the multiform derivative \cite{REF-34}
we obtain after some algebra the following fundamental formulas,
$(\eta = \dgr{\eta}_r)$ 
\begin{eqnarray}
&& \eta \ast \partial_{\phi_r} L
= \eta \cdot \partial_{\phi_r} L \\
&& (\partial \w \eta) \ast \partial_{\partial \w \phi_r} L
= \partial \cdot [\eta \cdot (\partial_{\partial \w \phi_r} L)]
- (-1)^r \eta \cdot [\partial \cdot (\partial_{\partial \w \phi_r} L)]\\
&& (\partial \cdot \eta) \ast \partial_{\partial \cdot\phi_r}  L
= \partial \cdot [\eta \cdot (\partial_{\partial \cdot\phi_r} L)]
+ (-1)^r \eta\cdot [\partial \w (\partial_{\partial \cdot \phi_r} L)]
\end{eqnarray}
Inserting Eq.~7.9 into Eq.~\ref{7.8} and then in Eq.~\ref{7.7}
we obtain, imposing
${\displaystyle\frac{d}{dt}} S (\phi_r + t\eta) = 0$, 
\begin{eqnarray}
&& \int_U \{ \eta \cdot [ \partial_{\phi_r} L - (-1)^r \partial \cdot
(\partial_{\partial \w \phi_r} L) + (-1)^r \partial \w
(\partial_{\partial \cdot \phi_r} L)] \} \tau_g \nonumber \\
&&\qquad + \int_U \partial \cdot
[\eta \cdot (\partial_{\partial \w \phi_r} L
+ \partial_{\partial \cdot \phi_r} L)] \tau_g = 0 \label{7.10}
\end{eqnarray}
The last integral in Eq.~\ref{7.10} is null by Stokes theorem if we
suppose as usual that $\eta$ vanishes on the boundary of $U$.

Then Eq.~\ref{7.10} reduces to
\begin{equation}
\int_U \{\eta \cdot [\partial_{\phi_r} L - (-1)^r \partial \cdot
(\partial_{\partial \w \phi_r} L) + (-1)^r \partial \w
(\partial_{\partial \cdot \phi_r} L)]\} \tau_g = 0 
\end{equation}
Now since $\eta= \dgr{\eta}_r$ is arbitrary and $\partial_{\phi_r} L$,
$\partial \cdot  (\partial_{\partial \w \phi_r} L)$, $\partial \w
(\partial_{\partial \cdot\phi_r}  L)$ are of grade $r$ we get 
\begin{equation} \label{7.12}
\dgr{\partial_{\phi_r} L
- (-1)^r \partial \cdot (\partial_{\partial \w \phi_r} L)
+ (-1)^r (\partial_{\partial \cdot \phi_r} L)}_r = 0
\end{equation}
But since $\partial_{\phi_r} \dgr{L}_0 = \dgr{\partial_{\phi_r} L}_r =
\partial_{\phi_r} L, \partial_{\partial \w \phi_r} L =
\dgr{\partial_{\partial \w \phi_r} L}_{r+1}$, etc Eq.~\ref{7.12}
reduces to 
\begin{equation} \label{7.13}
\partial_{\phi_r} L - (-1)^r \partial \cdot
(\partial_{\partial \w \phi_r} L)
+ (-1)^r \partial \w (\partial_{\partial \cdot \phi_r} L) = 0 
\end{equation}
Eq.~\ref{7.13} is a {\it multiform Euler-Lagrange equation}. 
Observe that as $L = \dgr{L}_0$ the equation has the graduation of
$\phi_r \in \sec \bigw^r (T^*M) \subset\sec \Cl(\cM)$. 

Now, let $X \in \sec \ \Cl(\cM)$ be such that
$X = \sum^4_{s=0} \dgr{X}_r$ and $F(x) = \dgr{F(x)}_0$.
From the properties of the multivectorial derivative we can easily
obtain
\begin{eqnarray}
\partial_X F(x) & = & \partial_X \dgr{F(x)}_0 \nonumber \\
& = &  \sum^4_{s=0} \partial_{\dgr{X}_s} \dgr{F(x)}_0 = \sum^4_{s=0}
\dgr{\partial_{\dgr{X}_s} F(X)}_0 
\end{eqnarray}
In view of this result if $\phi = \sum^4_{r=0}
\dgr{\phi}_r \in \sec  \Cl(\cM)$ we get as {\it Euler-Lagrange
equation} for $\phi$ the following equation 
\begin{equation} \label{7.15}
\sum_r [\partial_{\dgr{\phi}_r} L - (-1)^r \partial \cdot
(\partial_{\partial \w \dgr{\phi}_r} L) + (-1)^r \partial \w
(\partial_{\partial\cdot\dgr{\phi}_{r}} 
L)] = 0
\end{equation}
We can write Eq.~\ref{7.13} and Eq.~\ref{7.15} in a more convenient
form if we take into account that
$A_r \cdot B_s = (-1)^{r(s-1)} B_s \cdot A_r (r \leq s)$ and
$A_r \w B_s = (-1)^{rs} B_s \w A_r$. Indeed, we now have for $\phi_r$
that 
\begin{eqnarray}
&& \partial \cdot (\partial_{\partial \w \phi_r} L) \equiv \partial \cdot
(\partial_{\partial \w \phi_r} L)_{r+1} = (-1)^r (\partial_{\partial
\w \phi_r} L)_{r+1} \cdot  \stackrel{\leftarrow}{\partial} \\
&& \partial \w (\partial_{\partial \cdot \phi_r} L) \equiv \partial \w
(\partial_{\partial \cdot  \phi_r} L)_{r-1}
= (-1)^r (\partial_{\partial \cdot
\phi_r} L)_{r+1} \w \stackrel{\leftarrow}{\partial} 
\end{eqnarray}
where $\stackrel{\leftarrow}{\partial}$ means that the internal
and exterior products are to be done on the right.
Then, Eq.~\ref{7.15} can be written as
\begin{equation} \label{7.18}
\partial_\phi L - (\partial_{\partial \w \phi} L) \cdot
\stackrel{\leftarrow}{\partial} - (\partial_{\partial \cdot \phi} L)
\w \stackrel{\leftarrow}{\partial} = 0
\end{equation}
We now analyze the particular and important case where 
\begin{equation}
L(\phi, \partial \w \phi, \partial \cdot \phi)
= L (\phi, \partial \w \phi + \partial \cdot \phi)
= L(\phi, \partial \phi) 
\end{equation}
We can easily verify that
\begin{eqnarray}
\partial_{\partial \cdot \phi} L(\partial \phi) =
\dgr{\partial_{\partial\phi} L(\partial \phi)}_{r-1} \\
\partial_{\partial \w \phi} L(\partial \phi) = \dgr{\partial_{\partial
\phi} L(\partial\phi)}_{r+1}
\end{eqnarray}
Then, Eq.~\ref{7.18} can be written 
\begin{eqnarray}
&&\partial_\phi L - \dgr{\partial_{\partial\phi}L}_{r+1} \cdot
\stackrel{\leftarrow}{\partial} -
\dgr{\partial_{\partial\phi}L}_{r-1} 1
\stackrel{\leftarrow}{\partial} \nonumber\\
&&\qquad = \partial_\phi L -
\dgr{(\partial_{\partial_\phi}L)\cdot\stackrel{\leftarrow}{\partial}}_
r - \dgr{(\partial_{\partial\phi}L)\w
\stackrel{\leftarrow}{\partial}}_r \nonumber\\
&&\qquad = \dgr{\partial_\phi L - (\partial_\phi L)\cdot
\stackrel{\leftarrow}{\partial} - (\partial_{\partial\phi}L) \w
\stackrel{\leftarrow}{\partial}}_r = 0 \nonumber\\
&&\qquad = \dgr{\partial_\phi L - (\partial_\phi L)
\stackrel{\leftarrow}{\partial}}_r = 0 
\end{eqnarray} 
from where it follows the very elegant equation
\begin{equation} \label{7.23}
\partial_\phi L - (\partial_{\partial\phi} L) 
\stackrel{\leftarrow}{\partial} = 0 ,
\end{equation}
also obtained in~\cite{REF-46}. 

As an example of the use of Eq.~\ref{7.23} we write the Lagrangian in
Minkowski space for a Dirac-Hestenes spinor field represented in the
frame $\Sigma = \{\g^\mu\}, [\g^\mu \g^\nu + \g^\nu \g^\mu = 2
\eta^{\mu\nu}, \g^\mu \in\sec \ \w^1 (T^*M)( \subset\sec \ \Cl
(\cM)]$ by $\psi \in\sec \ \Cl(\cM)^+$ in interaction with the
electromagnetic field $A \in\sec \ \w^1 (T^*M) \subset\sec \
\Cl(\cM)$. We have\footnote{Note that we are omitting, for sake of
simplicity, the reference to the basis $\Sigma$ in the notation for
$\psi$.}, 
\begin{equation}
L = L_{DH} = \dgr{(\partial \psi \g^2 \g^1 - m \psi \g^0)\g^0 \tilde\psi
- e A \psi \g^0 \tilde\psi}_0 
\end{equation}
Then 
\begin{equation}
\partial_{\tilde\psi} L = (\partial \psi \g^2 \g^1 - m \psi \g^0)
\g_0 -	e A \psi \g^0  \qquad {\rm and} \qquad
\partial_{\partial\tilde\psi} L = 0
\end{equation}
and we get the Dirac-Hestenes equation
\begin{equation}
\partial \psi \g^2 \g^1 - e A \psi = m \psi \g^0
\end{equation}

Also since $\dgr{A\psi \g^0 \tilde\psi}_0 = \dgr{\psi\g^0\tilde\psi
A}_0$ we have 
\begin{eqnarray}
&& \partial_\psi L = - m \tilde\psi - e \g^0 \tilde\psi A \\
&& \partial_{\partial\psi} L = \g^{210} \tilde\psi  , \ \ \ (\g^{210} =
\g^2\g^1\g^0). 
\end{eqnarray}
Now, 
$$
(\partial_{\partial\psi} L) \stackrel{\leftarrow}{\partial}  = (\g^{210}
\tilde\psi) \stackrel{\leftarrow}{\partial} 
$$
and from the above equations we get 
$$
-m \tilde\psi - e \g^0 \tilde\psi A - (\g^{210} \tilde\psi) 
\stackrel{\leftarrow}{\partial}  = 0
$$
and this gives again, 
$$
\partial \psi \g^2 \g^1 - e A \psi = m \psi \g^0 
$$
Another Lagrangian that also gives the DH equation is, as can be
easily verified, 
\begin{equation} \label{7.29}
L_{DH}' = \dgr{ {\textstyle\frac{1}{2}} \partial \psi \g^{210}
\tilde\psi - {\textstyle\frac{1}{2}} \psi \g^{210} \tilde\psi
\stackrel{\leftarrow}{\partial} - m \psi \tilde\psi
- e A \psi \g^0 \psi }_0
\end{equation}

\subsection{The Dirac-Hestenes Equation on a Riemann-Cartan
Spacetime}

Let $\cM = \langle M, g, \nabla\rangle$ be a Riemann-Cartan
spacetime (RCST), i.e., $\nabla g=0,  \bT(\nabla)\neq 0, 
\bR(\nabla)\neq
0$. Let $\Cl(\cM)$ be the Clifford bundle of spacetime with typical
fibre $\cl_{1,3}$ and let $\psi\in\sec\Cl^+(\cM)$ be the
representative of a Dirac-Hestenes spinor field in the basis
$\Sigma=\{\g^a\}, [\g^a\in\sec\bigw^1(T^*M) \subset\sec \Cl(\cM),
\g^a\g^b+\g^b\g^a=2\eta^{ab}]$ dual to the basis $\cB=\{e_a\},
e_a\in\sec TM, a,b=0,1,2,3$.

To describe the ``interaction" of the DHSF $\psi$ with the
Riemann-Cartan spacetime we invoke the principle of minimal coupling.
This consists in changing $\partial=\g^a\partial_a$ in the Lagrangian 
given by Eq.~\ref{7.29} by
\begin{equation}
\g^a\partial_a\psi \longmapsto \g^a
\nabla^s_{e_a}\psi
\end{equation}
where $\nabla^s_{e_a}$ is the spinor covariant derivative of the DHSF
introduced in Section~4.4, i.e.,
\begin{equation}
\nabla^s_{e_a}\psi = e_a(\psi)+\frac{1}{2} \omega_a\psi.
\end{equation}

Let $\dgr{x^\mu}$ be a chart for $U\subset M$ and let be $\partial_a
\equiv e_a=h^\mu_a\partial_u$ and $\g^a=h^a_\mu dx^\mu$, with
$h^a_\mu h^v_a=\delta^\mu_v, h^a_\mu h^\mu_b=\delta^a_b$.

We take as the action for the DHSF $\psi$ on a RCST,
\begin{equation}
S(\psi) = \int_U 
\dgr{\frac{1}{2} \bD\psi\g^{210}\tilde{\psi}-
\frac{1}{2}\psi\g^{210}\tilde{\psi}\stackrel{\leftarrow}{\bD}
-m\psi\tilde{\psi}}_0 h^{-1}
dx^0\wedge dx^1\wedge dx^2\wedge dx^3
\end{equation}
where $\bD=\g^a\nabla^s_{e_a}$ is the operator Dirac operator made with the
spinor connection acting on sections of $\Cl(\cM)$ and
$h^{-1} = [\det(h^\mu_a)]^{-1}$. The Lagrangian $L=\dgr{L}_0$ is then
\begin{eqnarray}
&&L=h^{-1}\dgr{\frac{1}{2} \bD\psi\g^{210}\tilde{\psi}-
\frac{1}{2}\psi\g^{210}\tilde{\psi}
\stackrel{\leftarrow}{\bD}-m\psi\tilde{\psi}}_0 = 
\nonumber\\
&&=h^{-1}\dgr{\frac{1}{2}[\g^a(\partial_a+\frac{1}{2}\om_a\psi)\g^{210}
\tilde{\psi}-\psi\g^{210}(\partial_a\tilde{\psi}-\frac{1}{2}
\tilde{\psi} \om_a)\g^a]-m\psi\tilde{\psi}}_0 
\end{eqnarray}

As in Section~7.2 the principle of stationary action gives
\begin{eqnarray}
&& \partial_{\tilde\psi}L-(\partial_{\partial_{\tilde\psi}}
L)\stackrel{\leftarrow}{\partial} = 0  \nonumber\\
&& \partial_\psi L -
(\partial_{\partial_\psi}L)\stackrel{\leftarrow}{\partial} = 0  .
\label{7.34}
\end{eqnarray}

To obtain the equations of motion we must recall that
\begin{equation} 
(\partial_{\partial_\psi}L)\stackrel{\leftarrow}{\partial}
= \partial_\mu (\partial_{\partial_\mu \psi} L)
\end{equation}
and
\begin{equation}
\partial_{\partial_\mu \psi} L
= h_a^\mu \partial_{\partial_a \psi} L  .
\end{equation}
Then Eqs.~\ref{7.34} become
\begin{eqnarray}
&& \partial_\psi L - \partial_\mu(h^\mu_a) \partial_{\partial_a \psi} L
- \partial_a(\partial_{\partial_a \psi} L) = 0  ,\nonumber \\
&& \partial_{\tilde\psi} L - \partial_\mu(h^\mu_a) 
\partial_{\partial_a \tilde\psi} L - 
\partial_a(\partial_{\partial_a \tilde\psi} L) = 0 . \label{7.37}
\end{eqnarray}

Now, taking into account that $[e_a, e_b]=c^d_{ab}e_d$ and that
$\partial_a h/h = h^a_\mu\partial_a h^\mu_a$ we get
\begin{equation} \label{7.38}
\partial_\mu h^\mu_a = -c^b_{ab} + \partial_a \ln{h}
\end{equation}
and Eqs.~\ref{7.37} become
\begin{eqnarray}
&& \partial_\psi L-[\partial_a+\partial_a \ln{h} -
c^b_{ab}]\partial_{\partial_a \psi} L =0  ,\nonumber\\ 
&& \partial_{\tilde\psi} L-[\partial_a+\partial_a \ln{h} -
c^b_{ab}]\partial_{\partial_a \tilde\psi} L =0 . 
\end{eqnarray}

Let us calculate explicitly the second of Eqs.~\ref{7.37}. We have,
\begin{eqnarray}
&&\partial_{\tilde\psi}=h^{-1}[\frac{1}{2}
\g^a(\nabla_{e_a}\psi)\g^{210}
+ \frac{1}{4} \om_a\g^a\psi\g^{210}-m\psi]  , \label{7.40}\\
&&\partial_{\partial_a \tilde\psi} L = h^{-1}(-\frac{1}{2}
\g^a\psi\g^{210}).  \label{7.41}
\end{eqnarray}
Then,
\begin{eqnarray}
\partial_a(\partial_{\partial_a \tilde\psi} L)&=& 
(\partial_a \ln{h^{-1}}) 
h^{-1}(-\frac{1}{2} \g^a\psi\g^{210})-h^{-1}\frac{1}{2}
\g^a\partial_a\psi\g^{210}  = \nonumber\\
&=& -(\partial_a \ln{h})\partial_{\partial_a \tilde\psi} L - 
h^{-1}\frac{1}{2} \g^a\partial_a\psi\g^{210} .
\end{eqnarray}
 
Using Eq.~\ref{7.38} and Eq.~\ref{7.40} in the second of
Eqs.~\ref{7.37} we obtain
$$
\frac{1}{2} (\bD\psi)\g^{210}+
\frac{1}{4}
\om_a\g^a\psi\g^{210}-m\psi+\frac{1}{2}\g^a\partial_a\g^{210}
-\frac{1}{2} c^b_{ab} \g^a\psi\g^{210}=0
$$
or
$$
\bD\psi\g^{210}-\frac{1}{4}(\g^a \om_a-\om_a\g^a)\psi\g^{210}
- m\psi-\frac{1}{2} c^b_{ab} \g^a\psi\g^{210}=0  . 
$$
Then
\begin{equation} \label{7.43}
\bD\psi\g^{210}-\frac{1}{2}(\g^a \cdot\om_a)\psi\g^{210}
-\frac{1}{2} c^b_{ab} \g^a\psi\g^{210}- m\psi=0  . 
\end{equation}
But
\begin{equation}
\g^a \cdot \om_a=\om^b_{ba}\g^a
\end{equation}
and since $\om^b_{ab}=0$ because it is $\om^{bc}_a=-\om^{cb}_a$ we have
\begin{equation} \label{7.45}
\g^a \cdot \om_a = (\om^b_{ba}-\om^b_{ab})\g^a   . 
\end{equation}

Using Eq.~\ref{7.45} in Eq.~\ref{7.43} we obtain
$$
\bD\psi\g^{210}-\frac{1}{2}[\om^b_{ba}
-\om^b_{ab}+c^b_{ab}]\g^a\psi\g^{210} - m\psi=0  . 
$$
Recalling the definition of the torsion tensor, $T^c_{ab} =
\om^c_{ba}-\om^c_{ba}+c^c_{ab}$, we get
\begin{equation} \label{7.46}
(\bD + {\textstyle\frac{1}{2}} T)\psi\g^1\g^2 + m\psi\g^0=0, 
\end{equation}
where $T=T^b_{ab} \g^a$.

Eq.~\ref{7.46} is the Dirac-Hestenes equation on Riemann-Cartan
spacetime.  Observe that if $\cM$ is a Lorentzian spacetime
($\nabla g=0$, $\bT(\nabla)=0$, $\bR(\nabla)\neq 0$)
then Eq.~\ref{7.46} reduces to
\begin{equation}
\g^a(\partial_a+\frac{1}{2} \om_a)\psi\g^1\g^2+m\psi\g^0=0  , 
\end{equation}
that is exactly the equation proposed by Hestenes \shortcite{REF-48} as
the equation for a spinor field in a gravitational field
modeled as a Lorentzian spacetime $\cM$. Also, Eq.~\ref{7.46} is the
representation in $\Cl(\cM)$ of the spinor equation proposed by Hehl
et al \shortcite{REF-25} for a covariant Dirac spinor field $\Psi \in
P_{\Spin_+(1,3)} \times_\rho \C^4$ on a Riemann-Cartan spacetime. The
proof of this last statement is trivial. Indeed, first we multiply
$\psi$ in Eq.~\ref{7.46} by the idempotent field 
$\frac{1}{2} (1+\g^0)$ thereby obtaining an equation for the
representative of the Dirac algebraic spinor field in $\Cl(\cM)$.
Then we translate the  equation in $I(\cM) =  
P_{\Spin_+(1,3)} \times_\ell I$, from where taking a matrix
representation with the techniques already discussed in Section~2 we
obtain as equation for  $\Psi \in P_{\Spin_+(1,3)} \times_\rho \C^4$,
\begin{equation}
i\left(\ubi\g_a \nabla_a^s\Psi
- {\textstyle\frac{1}{2}} T\Psi\right)-m\Psi=0 \qquad  i=\sqrt{-1} 
\end{equation}
with $T=T^b_{ab}\ubi\g^a$, $\ubi\g^a$ being the Dirac matrices
(Eq.~\ref{2.31}).

We must comment here that Eq.~\ref{7.46} looks like, but it is indeed
very different from an equation proposed by Ivanenko and
Obukhov \shortcite{REF-26} as a generalization of the so called
Dirac-K\"{a}hler (-Ivanenko) equation for a Riemann-Cartan spacetime.
The main differences in the equation given in~\cite{REF-26} and our
eq(7.46) is that in~\cite{REF-26} $\Psi\in\sec\Cl(\cM)$ 
whereas in our approach $\psi_a \in \Cl^+(\cM)$ is only the
representative of the Dirac-Hestenes spinor field in the basis
$\Sigma=\{\g^a\}$ and also \cite{REF-26} use $\nabla_{e_a}$ instead of
$\nabla^s_{e_a}$.

Finally we must comment that Eq.~\ref{7.46} have played an important
role in our recent approach to a geometrical equivalence of Dirac and
Maxwell equations \cite{REF-4,REF-28a}	and also to the double
solution interpretation of Quantum Mechanics
\cite{REF-4,REF-49,REF-50}.

\section{Conclusions}

We presented in this paper a thoughtful and rigorous study of the
Dirac-Hestenes Spinor Fields (DHSF), their Covariant Derivatives and
the Dirac-Hestenes Equations on a Riemann-Cartan manifold $\cM$.

Our study shows in a definitive way that Covariant  Spinor Fields
(CDSF) can be represented by DHSF that are equivalence classes of even
sections of the Clifford Bundle $\Cl(\cM)$, i.e., spinors are
equivalence classes of a sum of even differential forms. We clarified
many misconceptions and misunderstanding appearing on the earlier
literature concerned with the representation of spinor fields by
differential forms. In particular we proved that the so-called 
Dirac-K\"{a}hler  spinor fields that are sections of $\Cl(\cM)$ and
are examples of amorphous spinor fields (Section~4.3.4) cannot be
used for representation of the field of fermionic matter. With
amorphous spinor fields the Dirac-Hestenes equation is not covariant.

We presented also an elegant and concise formulation of Lagrangian
theory in the Clifford bundle and use this powerful method to derive
the Dirac-Hestenes equation on a Riemann-Cartan spacetime.

\acknowledgements{The authors are grateful to CNPq and CAPES for
financial support.}


\begin{thebibliography}{}


\bibitem[\protect\citeauthoryear{Ablamowicz {\it et al.}}{1991}]{REF-27}
R. Ablamowicz, P. Lounesto and J. Maks,
``Conference Report: Second Workshop on Clifford Algebra and their
Applications in Mathematical Physics,''
{\it Found. Phys.} {\bf 21}, 735--748 (1991). 

\bibitem[\protect\citeauthoryear{Becher}{1981}]{REF-18} P. Becher,
``Dirac fermions on the lattice -- a local approach without
spectrum degeneracy,''
{\it Phys. Lett.} {\bf B104}, 221--225 (1981). 

\bibitem[\protect\citeauthoryear{Becher and Joos}{1982}]{REF-19}
P. Becher and H. Joos,
``The Dirac-K\"ahler equation and fermions on the lattice,''
{\it Zeits. f\"ur Phys.} {\bf C15}, 343--365 (1982). 

\bibitem[\protect\citeauthoryear{Benn and Tucker}{1987}]{REF-11}
I. M. Benn and R. W. Tucker,
{\it An Introduction to Spinors and Geometry with Applications in
Physics\/} (Adam Hilger, Bristol and New York, 1987). 

\bibitem[\protect\citeauthoryear{Benn and Tucker}{1988}]{REF-43}
I. M. Benn and R. W. Tucker,
``Representing spinors with differential forms,''
in A. Trautman and G. Furlan (eds.)
{\it Spinors in Physics and Geometry\/}
(World Scientific, Singapore, 1988).

\bibitem[\protect\citeauthoryear{Bishop and Goldberg}{1980}]{REF-44}
R. L. Bishop and S. I. Goldberg,
{\em Tensor Analysis on Manifolds\/}
(Dover Publ. Inc., New York, 1980). 

\bibitem[\protect\citeauthoryear{Bjorken and Drell}{1964}]{REF-30}
J. D. Bjorken and S. Drell,
{\it Relativistic Quantum Mechanics\/}
(McGraw-Hill, New York, 1964). 

\bibitem[\protect\citeauthoryear{Blaine Lawson and Michelson}{1989}]{REF-39}
H. Blaine Lawson, Jr. and M. L. Michelsohn,
{\em Spin Geometry\/}
(Princeton Univ. Press., Princeton, 1989). 

\bibitem[\protect\citeauthoryear{Blau}{1985}]{REF-12}
M. Blau,
``Connections on Clifford bundles and the Dirac Operator,''
{\it Lett. Math. Phys.} {\bf 13}, 83--86 (1985). 

\bibitem[\protect\citeauthoryear{Choquet-Bruhat}{1968}]{REF-31}
Y. Choquet-Bruhat,
{\it G\'eom\'etrie Diff\'erentielle et Syst\`emes Ext\'erieurs\/}
(Dunod, Paris, 1968). 

\bibitem[\protect\citeauthoryear{Choquet-Bruhat {\it et al.}}{1982}]{REF-23}
Y. Choquet-Bruhat, C. Dewitt-Morette and M.  Dillard-Bleick,
{\em Analysis, Manifolds and Physics}, revised edition,
(North-Holland Publ. Co., Amsterdam, 1982). 

\bibitem[\protect\citeauthoryear{Crawford}{1985}]{REF-14}
J. Crawford,
``On the algebra of Dirac bispinor densities: factorization and
inversion theorems,''
{\it J. Math. Phys.} {\bf 26},
1439--1441 (1985). 

\bibitem[\protect\citeauthoryear{Crumeyrole}{1991}]{REF-15}
A. Crumeyrolle,
{\em Orthogonal and Sympletic Clifford Algebras\/} 
(Kluwer Acad. Publ., Dordrecht, 1991). 

\bibitem[\protect\citeauthoryear{Fierz}{1937}]{REF-13}
M. Fierz,
``Zur Fermischen theorie des $\beta$-zerfalls,''
{\it Z. Phys.} {\bf 104}, 553--565 (1937).

\bibitem[\protect\citeauthoryear{Figueiredo {\it et al.}}{1990}]{REF-5}
V. L. Figueiredo, W. A. Rodrigues, Jr. and E. C. Oliveira,
``Covariant, algebraic, and operator spinors,''
{\it Int. J. Theor. Phys.} {\bf 29}, 371--395 (1990).

\bibitem[\protect\citeauthoryear{Figueiredo {\it et al.}}{1990a}]{REF-6}
V. L. Figueiredo, W. A. Rodrigues, Jr. and E. C. Oliveira,
``Clifford algebras and the hidden geometrical nature of spinors,''
{\em Algebras, Groups and Geometries}, {\bf 7}, 153--198 (1990). 

\bibitem[\protect\citeauthoryear{Geroch}{1968}]{REF-41}
R. Geroch,
``Spin structure of space-times in general relativity I,''
{\it J. Math. Phys.} {\bf 9}, 1739--1744 (1968).

\bibitem[\protect\citeauthoryear{Geroch}{1970}]{REF-42}
R. Geroch,
``Spin structure of space-times in general relativity II,''
{\it J. Math. Phys.} {\bf 11}, 343--348 (1970). 

\bibitem[\protect\citeauthoryear{Graf}{1978}]{REF-17}
W. Graf,
``Differential forms as spinors,''
{\it Ann. Inst H. Poincar\'e}, {\bf 29}, 85--109 (1978). 

\bibitem[\protect\citeauthoryear{Hehl and Datta}{1971}]{REF-25}
F. W. Hehl and B. K. Datta,
``Nonlinear spinor equation and asymmetric connection in general
relativity,''
{\it J. Math. Phys.} {\bf 12}, 1334--1339 (1971). 

\bibitem[\protect\citeauthoryear{Hestenes}{1967}]{REF-2}
D. Hestenes, ``Real spinor fields,''
{\it J. Math. Phys.} {\bf 8}, 798--808 (1967). 

\bibitem[\protect\citeauthoryear{Hestenes}{1976}]{REF-3}
D. Hestenes,
``Observables, operators, and complex numbers in the Dirac theory,''
{\it J. Math. Phys.} {\bf 16}, 556--571 (1976). 

\bibitem[\protect\citeauthoryear{Hestenes}{1985}]{REF-48}
D. Hestenes,
``Spinor approach to gravitational motion and precession,''
{\it Int. J. Theor. Phys.} {\bf 25}, 59--71 (1985). 

\bibitem[\protect\citeauthoryear{Hestenes and Sobczyk}{1984}]{REF-34}
D. Hestenes and G. Sobczyk,
{\it Clifford Algebra to Geometrical Calculus\/}
(D. Heidel, Publ. Co., Dordrecht, 1984). 

\bibitem[\protect\citeauthoryear{Ivanenko and Obukhov}{1985}]{REF-26}
D. Ivanenko and Yu. N. Obukhov,
``Gravitational interaction of fermion antisymmetric connection in
general relativity,''
{\it Anallen der Physik} {\bf 17}, 59--70 (1985). 

\bibitem[\protect\citeauthoryear{K\"ahler}{1962}]{REF-16}
E. K\"{a}hler,
``Der innere differentialkalk\"ul,''
{\it Rendiconti di Matematica e delle sue Applicazioni\/}
{\bf 21}, 425--523 (1962). 

\bibitem[\protect\citeauthoryear{Lasenby {\it et al.}}{1993}]{REF-46}
A. Lasenby, C. Doran, and S. Gull,
``A multivector derivative approach to lagrangian field theory,''
{\em Found. Phys.} i{\bf 23}, 1295--1327 (1993).

\bibitem[\protect\citeauthoryear{Lichnerowicz}{1964}]{REF-21}
A. Lichnerowicz,
``Champs spinoriels et propagateurs en relativit\'e g\'en\'erale,''
{\it Ann. Inst. H. Poincar\'e} {\bf 13}, 233--290 (1964). 

\bibitem[\protect\citeauthoryear{Lichnerowicz}{1984}]{REF-22}
A. Lichnerowicz,
``Champ de Dirac, champ du neutrino et transformations C, P, T sur
un espace temps courbe,''
{\it Bull. Soc. Math. France} {\bf 92}, 11--100 (1984). 

\bibitem[\protect\citeauthoryear{Lounesto}{1981}]{REF-29}
P. Lounesto,
``Scalar product of spinors and an extension of Brauer-Wall
groups,''
{\it Found. Phys.} {\bf 11},  721--740 (1981). 

\bibitem[\protect\citeauthoryear{Lounesto}{1993}]{REF-9}
P. Lounesto,
``Clifford algebras and Hestenes spinors,''
{\it Found. Phys.} {\bf 23}, 1203--1237 (1993). 

\bibitem[\protect\citeauthoryear{Lounesto}{1993a}]{REF-10}
P. Lounesto,
``Clifford algebras, relativity and quantum mechanics,''
in P. Letelier and W. A. Rodrigues, Jr.  (eds.)
{\it Gravitation: The Spacetime Structure}, Proc. SILARG VIII,
\'Aguas de Lind\'oia, Brazil, 1993, pp.~49--80,
(World Scientific Publ. Co., Singapore, 1994). 

\bibitem[\protect\citeauthoryear{Milnor}{1963}]{REF-40}
J. Milnor,
``Spin structures on manifolds,''
{\it L'Enseignement Mathematique} {\bf 9}, 198--203 (1963).

\bibitem[\protect\citeauthoryear{Parra}{1992}]{REF-45}
J. M. Parra,
``Relativistic invariance of Dirac's equation revisited,''
in {\it Proc. of the Relativity Meeting\/}
(Bilbo, Spain, World Scient. Publ. Co., 1992).

\bibitem[\protect\citeauthoryear{Pav\v{s}i\v{c} {\it et al.}}{1993}]{REF-35}
M. Pav\v{s}i\v{c}, E. Recami, W. A. Rodrigues, Jr., 
D. Maccarrone, F. Raciti and G. Salesi,
``Spin and electron structure,''
{\it Phys. Lett.} {\bf 31B}, 481--488 (1993). 

\bibitem[\protect\citeauthoryear{Porteous}{1969}]{REF-28}
I. Porteous,
{\it Topological Geometry\/} 
(van Nostrand, London, 1969). 

\bibitem[\protect\citeauthoryear{Rapoport}{1994}]{REF-28a}
D. L. Rapoport, W. A. Rodrigues, Jr., Q. A. G. de~Souza, and J. Vaz, Jr.,
``The Riemann-Cartan-Weyl geometry generated by a Dirac-Hestenes spinor field,''
{\it Algebras, Groups and Geometries\/} {\bf 11}, 23--35 (1995).

\bibitem[\protect\citeauthoryear{Rodrigues and Figueiredo}{1990}]{REF-8}
W. A. Rodrigues, Jr. and V. L. Figueiredo,
``Real spin-Clifford bundle and the spinor structure of the
spacetime,''
{\it Int. J. Theor. Phys.} {\bf 29}, 413--424 (1990). 

\bibitem[\protect\citeauthoryear{Rodrigues and Oliveira}{1990}]{REF-7}
W. A. Rodrigues, Jr. and E. C. Oliveira,
``Dirac and Maxwell equations in the Clifford and spin-Clifford
bundles,''
{\it Int. J. Theor. Phys.} {\bf 29}, 397--412 (1990). 
 
\bibitem[\protect\citeauthoryear{Rodrigues and Souza}{1993}]{REF-47}
W. A. Rodrigues, Jr. and Q. A. G. de Souza,
``The Clifford bundle and the nature of the gravitational field,''
{\it Found. Phys.} {\bf 23}, 1465--1490 (1993).

\bibitem[\protect\citeauthoryear{Rodrigues {\it et al.}}{1993}]{REF-36}
W. A. Rodrigues Jr, J. Vaz, Jr. and E. Recami,
``About Zitterbewegung and electron structure",
{\em Phys. Lett.} {\bf B 318}, 623--628 (1993). 

\bibitem[\protect\citeauthoryear{Rodrigues {\it et al.}}{1993a}]{REF-49}
W. A. Rodrigues, Jr., J. Vaz, Jr. and E. Recami,
``Free Maxwell equations, Dirac equation and non dispersive
De Broglie wave packets,'' in G. Lochak and P. Lochak (eds.)
{\it Courants, Am\'ers, \'Ecueils en Microphysique}, pp.~379--392
(Fond. Louis de Broglie, Paris, 1993). 

\bibitem[\protect\citeauthoryear{Rodrigues {\it et al.}}{1994}]{REF-24}
W. A. Rodrigues, Jr., Q. A. G. de Souza and J. Vaz, Jr.,
``Lagragian formulation in the Clifford bundle of the Dirac-Hestenes
equation on a Riemann-Cartan manifold", in P. Letelier and
W. A. Rodrigues Jr (eds.),
{\it Gravitation: The Spacetime Structure},
Proc. SILARG VIII, \'Aguas de Lind\'oia, Brazil, 1993, pp.~522--531
(World Scientific Publ. Co., Singapore, 1994). 

\bibitem[\protect\citeauthoryear{Souza and Rodrigues}{1993}]{REF-13a}
Q. A. G. de Souza and W. A. Rodrigues, Jr.,
``The Dirac operator and the structure of Riemann-Cartan-Weyl spaces,''
in P. Letelier and W. A. Rodrigues, Jr. (eds.),
{\it Gravitation: The Spacetime Structure}, Proc. SILARG VIII,
\'Aguas de Lind\'oia, Brazil, 1993, pp.~177--210
(World Scientific Publ. Co., Singapore, 1994).  

\bibitem[\protect\citeauthoryear{Takabayasi}{1957}]{REF-33}
T. Takabayasi,
``Relativistic hydrodynamics of the Dirac matter,''
{\it Suppl. Progr. Theor. Phys.} {\bf 4}, 1--80 (1957). 

\bibitem[\protect\citeauthoryear{Vaz and Rodrigues}{1993}]{REF-4}
J. Vaz, Jr. and W. A. Rodrigues, Jr.,
``Equivalence of the Dirac and Maxwell equations and quantum
mechanics,''
{\it Int. J. Theor. Phys.} {\bf 32}, 945--958 (1993).

\bibitem[\protect\citeauthoryear{Vaz and Rodrigues}{1993a}]{REF-50}
J. Vaz, Jr. and W. A. Rodrigues, Jr.,
``A basis for double solution theory,''
in R. Delanghe, F. Brackx and H. Serras (eds.),
{\it Clifford algebras and their applications in
mathematical physics},  pp.~345--351
(Kluwer Acad. Publ., Dordchet, 1993).

\bibitem[\protect\citeauthoryear{Vaz and Rodrigues}{1994}]{REF-37}
J. Vaz, Jr. and W. A. Rodrigues, Jr.,
``Zitterbewegung and the electromagnetic field of the electron,''
{\em Phys. Lett.} {\bf B 319}, 203--208 (1994). 

\bibitem[\protect\citeauthoryear{Vaz and Rodrigues}{1995}]{REF-MD}
J. Vaz, Jr. and W. A. Rodrigues, Jr., 
``Maxwell and Dirac theories as an already unified theory'', 
preprint hep-th 9511181, to appear in the proceedings of 
the International Conference on the Theory of the Electron, 
J.Keller and Z. Oziewicz (eds.), UNAM, Mexico (1995). 


\bibitem[\protect\citeauthoryear{Yvon}{1940}]{REF-32}
J. Yvon,
``Equations de Dirac-Madelung,''
{\it J. Phys. et de Radium} {\bf 8}, 18--30 (1940). 


\end{thebibliography}
\end{document}